\pdfoutput=1

\documentclass[british,a4paper,12pt]{phdthesis}
\usepackage[latin1]{inputenc}
\usepackage[toc,page]{appendix}

\usepackage[intoc]{nomencl}
\newcommand*{\nom}[2]{#1\nomenclature{#1}{#2}}

\makenomenclature

\ifx\undefined\pdfoutput
\usepackage[dvips]{graphicx}
\usepackage[dvips]{color}
\DeclareGraphicsExtensions{.eps}
\else
\usepackage[pdftex]{graphicx}
\usepackage[pdftex]{color}
\usepackage{thumbpdf}
\DeclareGraphicsExtensions{.png,.pdf,.jpg}
\fi
\usepackage[british]{babel}
\usepackage{subfigure}
\usepackage{amsmath}
\usepackage{mathpple} 
\usepackage{latexsym}
\usepackage[draft]{varioref}
\usepackage[english,vario]{fancyref}
\usepackage{setspace}
\usepackage{url}
\usepackage{algorithm}
\usepackage{algorithmic}
\usepackage{enumerate}

\usepackage{amssymb}
\usepackage{graphicx}
\usepackage{units}
\usepackage{amsmath}
\usepackage{fancyref}

\usepackage{upgreek}
\usepackage{listings}
\usepackage{courier}
\usepackage{lscape}

\DeclareRobustCommand{\gobblefour}[4]{}

\usepackage{makeidx}
\makeindex

\usepackage{bibentry}
\newcommand{\bibverse}[1]{\begin{verse} \bibentry{#1}. \end{verse}}

\makeatletter
 \renewcommand*{\@fancyref@page@ref}{%
  \let\vref@space\space
  \@ifnextchar[
  \@vpageref{\@vpageref[\unskip]}%
}%
\makeatother

\title{Study of Peer-to-Peer Network Based Cybercrime Investigation: Application on Botnet Technologies}

\author{Mark Scanlon, B.A. (Hons.), M.Sc.}

\authoraddress{
         UCD Centre for Cybersecurity and Cybercrime Investigation,\\
	School of Computer Science and Informatics,\\
         University College Dublin, Belfield, Dublin 4, Ireland.
     }

\date{\today}

\submission{A thesis submitted to University College Dublin\\for the degree of Ph.D. in the College of Science}

\department{School of Computer Science and Informatics} \hod{Mr. John Dunnion, M.Sc.} \supervision{Prof. M-Tahar Kechadi, Ph.D.}
\submissionmonth{October} \submissionyear{2013}

\begin{document}

\nobibliography*

\maketitle 

\begin{dedication}
  \vspace{-1.5cm}

This thesis is dedicated to my wife, Joanne, who has supported, encouraged and motivated me throughout the last nine years and has been especially patient and thoughtful throughout my research. This thesis is also dedicated to my parents, Philomena and Larry Scanlon.

\end{dedication}

\tableofcontents 

\begin{acknowledgements}
  \vspace{-1.5cm}
With no doubt, the work on this thesis has been the most challenging endeavour I have undertaken so far. I am thankful to my supervisor, Prof. M-Tahar Kechadi, for his guidance and encouragement. I would like to thank the staff and students in the School of Computer Science and Informatics, University College Dublin for providing me with the opportunity to learn, facilities to perform my research, and a motivating environment that carried me forward through my course work. My gratitude goes to my friends Alan Hannaway, Cormac Phelan, John-Michael Harkness, Michael Whelan, Alex Cronin, Pat Tobin, Jason Farina and Dr. Pavel Gladyshev for many interesting and developing discussions, presentations and collaborations. Many thanks to all my immediate friends for their constant encouragement and support.

This work was co-funded by the Irish Research Council (formally the Irish Research Council for Science, Engineering and Technology) and Intel Ireland Ltd., through the Enterprise Partnership Scheme. Amazon Web Services also generously contributed to this research with grants funding the costs involved in experimentation conducted on their cloud infrastructure, including Elastic Compute Cloud (EC2) and Relational Database Service (RDS).

\end{acknowledgements}

\cleardoublepage
\addcontentsline{toc}{chapter}{\listtablename}
\listoftables

\cleardoublepage
\addcontentsline{toc}{chapter}{\listfigurename}
\listoffigures

\printnomenclature[2.5 cm]

\begin{abstract}
\onehalfspacing
\setstretch{1.5}
 \vspace{-1.5cm}
The scalable, low overhead attributes of Peer-to-Peer (\nom{P2P}{Peer-to-Peer}) Internet protocols and networks lend themselves well to being exploited by criminals to execute a large range of cybercrimes. The types of crimes aided by P2P technology include copyright infringement, sharing of illicit images of children, fraud, hacking/cracking, denial of service attacks and virus/malware propagation through the use of a variety of worms, botnets, malware, viruses and P2P file sharing. This project is focused on study of active P2P nodes along with the analysis of the undocumented communication methods employed in many of these large unstructured networks. This is achieved through the design and implementation of an efficient P2P monitoring and crawling toolset. 
%

The requirement for investigating P2P based systems is not limited to the more obvious cybercrimes listed above, as many legitimate P2P based applications may also be pertinent to a digital forensic investigation, e.g, voice over IP, instant messaging, etc. Investigating these networks has become increasingly difficult due to the broad range of network topologies and the ever increasing and evolving range of P2P based applications. In this work we introduce the Universal P2P Network Investigation Framework (\nom{UP2PNIF}{Universal P2P Network Investigation Framework}), a framework which enables significantly faster and less labour intensive investigation of newly discovered P2P networks through the exploitation of the commonalities in P2P network functionality. In combination with a reference database of known network characteristics, it is envisioned that any known P2P network can be instantly investigated using the framework, which can intelligently determine the best investigation methodology and greatly expedite the evidence gathering process. A proof of concept tool was developed for conducting investigations on the BitTorrent network. A Number of investigations conducted using this tool are also outlined.

\end{abstract}

\begin{listofpublications}
\onehalfspacing
\setstretch{1.2}
\begin{itemize}
\item Peer-Reviewed International Journal Publications

\bibverse{scanlon2011investigating}

\item Peer-Reviewed International Conference Publications

\bibverse{scanlon2013case}

\bibverse{scanlon2013universal}

\bibverse{scanlon2013bag}

\newpage

\bibverse{scanlonfuturenet}

\bibverse{scanlon2010week}


\bibverse{scanlon}

\item International Conference Posters

\bibverse{scanlon2010poster}

\end{itemize}
\end{listofpublications}
\onehalfspacing
\setstretch{1.5}


\chapter{Introduction}
\label{ch:introduction}

\section{Background}
\label{background}
In June 1999, the control that the content producing industry (composed of movie producers, TV show producers, musicians, writers, etc.) had over their traditional distribution model was permanently changed due to the release, and subsequent rise in popularity, of Napster by Shawn Fanning \cite{giesler2003anthropology}. Napster brought the relatively new concept of Internet file sharing into the mainstream. It facilitated its users in sharing music with millions of other users around the world. The ease of use, vast library of available content, perceived anonymity and zero cost model enabled Napster to grow rapidly. It's rise in popularity also coincided with the release of new portable devices capable of playing digital audio files, MP3 players \cite{van2005introducing}. The difference in user difficulty between converting store bought CDs into a suitable format when compared to performing a search for the song's title and double clicking the version you wanted was significant. At its peak, it enabled over 25 million users to share more than 80 millions digital songs with each other \cite{greenfeld2000meet}. This was not the first implementation of Peer-to-Peer (P2P) technology, but it certainly was the first to gather attention. It enabled regular computer users with Internet connections to perform copyright infringement on a scale incomparable to physical copying of tapes and CDs.

P2P technologies are most known for unauthorised distribution of copyrighted content but the merits of P2P have been exploited by other criminals with more sinister intentions. The ever increasing proliferation of computers has resulted in a new breed of high-tech, highly skilled, computer savvy criminals emerging. For the lesser skilled criminal, a large underground market creating and selling software packages to enable the online execution of a range of crimes has emerged. As this phenomenon continues, an increasing number of ``offline" crimes are being aided by computers, e.g., fraud, identity theft, phishing, terrorism, child sexual exploitation, etc. As a result, digital forensic investigators and law enforcement in general are playing catchup in an attempt to gain the necessary expertise to combat these crimes. Looking to always be one step ahead of the law, criminals are continually looking for more advanced methods of conducting their crimes. With the advent of ``botnets", i.e., large distributed networks of compromised machines, criminals are now able to take advantage of far superior distributed processing power, bandwidth and other resources than a single machine could ever afford them. These botnets also award the criminal a relative degree of anonymity if the botnet itself is entirely decentralised, i.e., no central server or single point of penetration, such as a P2P botnet. Each compromised node in a P2P botnets is obliged to forward on received commands and queries to other known active nodes in the network. The scalable and minimal investment attributes of P2P and similar distributed Internet protocols lend themselves well to being exploited by criminals to execute a range of cybercrimes. These crimes not only include those offline examples previously mentioned, but also new computer targeted crimes, such as distributed denial of service (\nom{DDoS}{Distributed Denial of Service}) attacks, virus/malware propagation, etc.

\section{Research Problem}
\label{sec:problem} 
Much of the existing research into P2P cybercrimes relies on packet sniffing as the primary method for collecting information. This method involves setting up a honeypot, as outlined in greater detail in Section \ref{ch4:hardware}, and deliberately infecting the machine with the required malware. The downside of this type of investigation is that the system is reliant on recording typical network communication to find out information about the system being investigated. Any single node on a P2P network may never communicate with every other node, as each node generally maintains a list in the order of 5--10 other known active nodes. The motivation for the research detailed in this thesis is to design and test a new methodology for investigating P2P networks. This methodology involves emulating and multiplying regular client usage resulting in the distributed capability of crawling an entire network. 
\\
\\

The objectives of this research are as follows:
\begin{enumerate}
\item Provide an insight into the technical requirements of the design and implementation of a forensically sound P2P crawling and investigation tool; collecting of digital evidence and the counter-detection measures that may need to be employed.
\item Demonstrate the application of a P2P network crawling system as a plausible option for forensic investigation.
\item Design an architecture for such a system. It should be forensically verifiable, cost effective, expandable, reliable and widely compatible with current computer hardware and network capacities.
\item Prototype the system and perform experimental analysis to measure the viability of the system for both documented and undocumented networks.
\item Draw some recommendations about future use of these technologies.
\end{enumerate}

\section{Contribution of this Work}
\label{sec:Contribution-of-this-work}
Many of the tools available in the field of digital network forensic investigations are based upon the deployment of packet sniffing or deep packet inspection devices and software, which are outlined further in Sections \ref{packetsniffing} and \ref{dpi} . These methods can result in a huge volume of data to be analysed by the forensic investigator. ``Typically, only a small fraction of the examined data is of interest in an investigation" \cite{gao}. The existing techniques are concentrated around the procedures that should be implemented after the physical confiscation of the computer equipment. The research outlined as part of this thesis results in a system capable of quickly implementing the communication protocol of any given P2P network, resulting in more focused data collection. The data collected can be partially processed at the point of collection, eliminating the need to store, index and analyse irrelevant information. 

The contribution of this research can be summarised as follows:

\begin{itemize}
\item Design of a forensically sound P2P network investigation system, which can be used for the collection of court-admissible evidence or used for system monitoring. The system also enables the user to conduct a cloud based investigation. This results in the forensic investigators being able to spend more time analysing evidence, as opposed to being in the field collecting it. The design approach can be extended to defining how to best deal with the issues of cost, speed, compatibility and redundancy of the data while ensuring that the process is reproducible and reliable.
\item Proof of the viability of the system through experimentation of all the necessary components. Each component of the system was individually tested to ensure the forensic integrity of the data collected.
\item Performance results from testing ``real-world" scenarios where such a system may be used, i.e., collecting evidence from a live P2P network investigation.
\item Outline a new forensically sound method for storing remote network captured P2P evidence.
\end{itemize}

\section{Limitations of this Work}
\label{sec:limitations-of-this-work}

With such a large variety of P2P networks and P2P based cybercrimes, a number of limitations for the scope of this research were introduced: 

\begin{enumerate}
\item To conduct comprehensive testing across every known P2P network was deemed too large a task for the purposes of this work due to time and resource constraints.
\item As a proof of concept for the viability of the system designed as part of this work, it was deemed acceptable to perform testing and investigation of unauthorised file-sharing occurring on P2P networks. The methodology and techniques outlined are equally applicable to the investigation of any P2P based cybercrimes.
\end{enumerate}

\section{Structure of the Thesis}
\label{sec:Structure-of-the-thesis}
This thesis is organised as follows: 

\begin{itemize}
 \item After introducing the context and highlighting the main goals of the project in this Chapter, in Chapters \ref{ch2digitalforensicinvestigations}, \ref{ch3peertopeerfilesharing}, and \ref{ch4botnetinvestigation}, we present literature reviews of related research work and software tools relevant to the areas of Digital Forensics, P2P File-sharing and Botnet Investigation respectively. These chapters outline some of the tools, systems, architectures, and best practices associated with the corresponding fields from a technical, and legal perspective.
 
 \item Chapter \ref{ch5designandarch} presents the architecture and design of the universal P2P network investigation framework capable of expansion to deal with any P2P network investigation. We also outline the design considerations which should be incorporated into a framework of this nature. Chapter \ref{ch6forensicinvestigationofbittorrent} presents the results from a proof of concept investigation tool developed for the investigation of the BitTorrent file-sharing P2P network. The results of comprehensive experiments carried out to prove the viability of such a framework. This testing phase incorporated the testing of each individual component of the system to ensure forensic integrity and ultimately, court admissible evidence. 

 \item Chapter \ref{ch7conclusion} summarises and concludes this research. This chapter also outlines scenarios where the technology developed can be adapted and reused for additional purposes. Guidelines for further developments to the presented work are also outlined and discussed.
\end{itemize}


\chapter{Digital Forensic Investigation; State of the art}
\label{ch2digitalforensicinvestigations}

\section{Introduction}
\label{ch2:introduction}

``A forensics expert must have the investigative skills of a detective, the legal skills of a lawyer, and the computing skills of the criminal." \cite{yasinsac2001policies}. This chapter outlines some of the digital network evidence acquisition, investigation software, and hardware tools commonly used by forensic investigators in law enforcement and private investigations such as ForNet, Wireshark, Security Incident and Event Management Software (\nom{SIEM}{Security Incident and Event Management}), Network Forensic Analysis Tools (\nom{NFAT}{Network Forensic Analysis Tool}), and  Deep-Packet Inspection (\nom{DPI}{Deep Packet Inspection}). Current commercial, research and open-source tools are discussed specifying their benefits and designs. Common digital evidence storage formats are also discussed, outlining the cross-compatibility between the tools available and the associated formats. Best practices associated with the field of digital forensics from a technical, cryptographical and legal perspective are also discussed.

\section{Computer Forensic Investigation}
\label{ch2:evidence}
Generally speaking, the goal of a digital forensic investigation is to identify digital evidence relative to a specific cybercrime. Investigations rarely rely entirely on digital evidence to prosecute the offender, instead relying on a case built from physical evidence, digital evidence, witness testimony and cross-examination. However, when dealing solely with digital evidence, there are three major phases \cite{carrier-open}:

\begin{enumerate}
    \item \emph{Acquisition Phase} -- The acquisition phase is concerned with capturing the state of a digital system for later analysis. This is similar to the collection of physical evidence from a crime scene, e.g., taking photographs, collecting fingerprints, fibres, blood samples, tire patterns, etc. During this phase in a digital investigation, it is typically very difficult to tell which evidence is relevant to the case, so the goal of this phase is to collect all possible digital evidence (including any data on removable storage devices, network traffic, logs, etc.). 
    
    \item \emph{Analysis Phase} -- After a successful and complete acquisition of the system state from a suspect computer, the data acquired needs to be analysed to identify pieces of evidence. The analysis of evidence is carried out on an exact copy of the original evidence. This copy is verified against the original through the use of a hashing algorithm, as outlined in more detail in Section \ref{ch2:hashfunctions}. Carrier \cite{carrier-open} defines three major categories of evidence a digital investigator needs to discover when conducting his analysis:
    \begin{itemize}
        \item Inculpatory Evidence -- This is any evidence which supports a given theory.
        \item Exculpatory Evidence -- This is any evidence which contradicts a given theory.
        \item Evidence of Tampering -- This is any evidence which cannot be related to any theory currently under investigation, but shows that the system was tampered with to avoid identification.
    \end{itemize}
    The procedure followed during this phase includes examining file and directory contents (including recovered deleted content) to draw verifiable conclusions based on any evidence that is collected.
    \item \emph{Presentation Phase} -- The steps performed in the previous two phases are the same regardless of the type of investigation being conducted, e.g., corporate, law enforcement or military. However, the presentation phase will be different depending on corporate policy or local law. This phase presents the conclusions and their corresponding evidence that the digital investigator has deduced. In a court settings, the lawyers must first evaluate the evidence to confirm that it is court admissible.
\end{enumerate}

\subsection{Network Forensic Investigation}
\label{ch2:networkforensics}

The 2006 National Institute of Standards and Technology's (\nom{NIST}{National Institute of Standards and Technology}) special publication ``Guide to Integrating Forensic Techniques Into Incident Response"  \cite{Nist:2012:NSP:2331550} outlines a number of best practices and legal considerations for forensic investigators working with network data. The NIST guide outlines the typical sources of network evidence and tools that should be used during the evidence collection phase of an investigation:

\begin{itemize}
\item Firewall and router logs -- These devices are normally configured to record suspicious activity.
\item Packet Sniffing -- This allows the investigator to monitor, in real-time, the activity on the network.
\item Intrusion Detection Systems (\nom{IDS}{Intrusion Detection System}) -- Some larger networks may employ IDS to capture packets related to suspect activity.
\item Remote Access Servers -- this includes devices such as VPN gateways and modem servers that facilitate connections between networks.
\item Security Event Management Software -- These tools aid in analysis of logs files, typically produced by IDS tools, firewalls, and routers.
\item Network Forensic Analysis Tools -- These tools allow a reconstruction of events by visualising and replaying network traffic within a specified period.
\item Other Sources -- These include Internet Service Provider (\nom{ISP}{Internet Service Provider}) records, client/server applications, hosts' network configureration and connections, and Dynamic Host Configuration Protocol (\nom{DHCP}{Dynamic Host Configuration Protocol}) records.
\end{itemize}

A number of tools capable of collecting and analysing some of the above evidence are outlined in Section \vref{packetsniffing}.

\section{Network Investigation Tools}
\label{packetsniffing}

While the area of Computer Forensics and Cybercrime Investigation is relatively new among the more traditional computer security models, there is a small number of companies and open-source tools dedicated to forensic investigations. There are numerous free packet sniffing software tools available. A number of these tools are discussed in the following subsections:

\subsection{TCPDump/WinDump}
\label{ch2:tcpdump}
\begin{figure}[!h]
\centering
\includegraphics[width=0.95\textwidth]{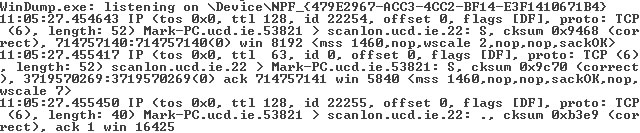}
\caption{Example Frame Capture of SSH Session Using WinDump.}
\label{Figurewindump}
\end{figure}

TCPDump and WinDump are the Unix and Windows equivalent command line network software analysers developed in the 1990s. The tools run on a local machine and are capable of capturing all the network traffic over ethernet or wireless connections. They have the ability to display in a semi-coherent fashion the captured traffic frame by frame and allow the analysis of the data. As its name might suggest, TCPDump focuses mainly on the \nom{TCP}{Transmission Control Protocol}/\nom{IP}{Internet Protocol} protocol \cite{fuentes2005ethereal}. An example capture of an \nom{SSH}{Secure Socket Handling} session using WinDump can be seen in Figure \ref{Figurewindump}.

\subsection{Ethereal}
\label{ch2:ethereal}
\begin{figure}[!h]
\centering
\includegraphics[width=0.95\textwidth]{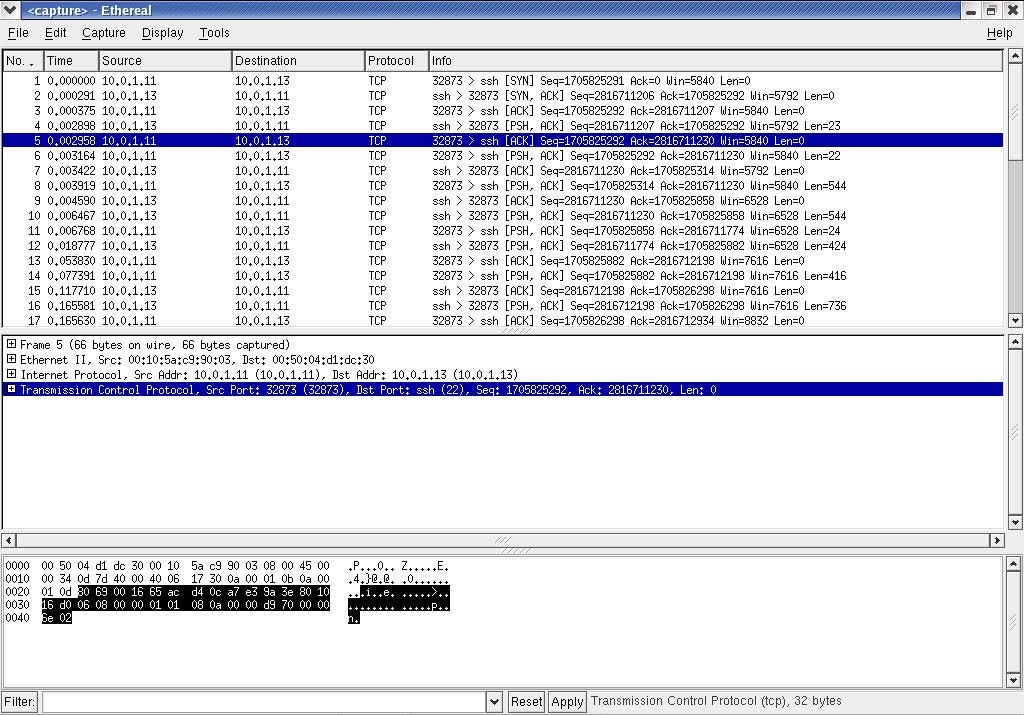}
\caption{Example Frame Capture of SSH Session Using Ethereal.}
\label{Figureethereal}
\end{figure}

Ethereal is another free tool available for both Unix and Windows. It is more user friendly than TCPDump as it has a graphical user interface (\nom{GUI}{Graphical User Interface}) to assist its users. Ethereal also provides a large number of protocol decoding options; more than 400 in total \cite{fuentes2005ethereal}. It allows the forensic investigator to analyse data collected on a packet basis or protocol basis. An example capture of an SSH session using Ethereal and its presentation in the GUI can be seen in Figure \ref{Figureethereal}.

\subsection{Network Forensic Analysis Tools}
\label{NFAT}
NFATs are intelligent packet analysis tools capable of identifying firewall circumvention \cite{corey2002network}. For example, corporate firewalls may block access to their staff from using instant messaging at work. Yahoo Messenger normally operates on port 5050, but when this port is blocked it will automatically switch to port 23 (usually reserved for telnet) \cite{hindocha2003malicious}. While this port change might bypass a firewall rule in place, an NFAT would still be able to identify the network usage as being Yahoo Messenger due to packet analysis. NFATs are not designed as a replacement for firewalls or IDS software, but are designed to work in conjunction with them. Typically NFATs will rely on another piece of software to capture the traffic, e.g., TCPDump.

\subsection{Security Incident and Event Manager Software}
\label{SIEM}

SIEM software is a combination of the formally different software categories of Security Incident Management Software and Security Event Management Software and takes a different investigative approach to the ``on-the-fly'' analysis tools outlined above. SIEM software is focused on importing security event information for a number of network traffic related sources, e.g., IDS logs, firewall logs, etc. \cite{Nist:2012:NSP:2331550}. It operates on an ``after the fact'' basis whereby it analyses copies of the logs attempting to identify suspicious network activity events by matching IP addresses, timestamps and other network traffic characteristics. An open source example of this software is called OSSIM \cite{ossim}.


\section{Packet Inspection Hardware}
\label{dpi}

In the regular operation of Network Interface Cards (\nom{NIC}{Network Interface Card}s), the devices only accept incoming packets that are specifically addressed to its IP address. However when a NIC is placed in promiscuous mode, it will accept all packets that it sees, regardless of their intended destinations. Packet sniffing hardware generally operates on this principle, with configuration available to capture all packets or only those with specific characteristics, e.g., certain TCP ports, certain source or destination IP addresses, etc.  \cite{Nist:2012:NSP:2331550}. This style of network traffic capture can be used in combination with software sampling optimisation techniques in order to reduce the overall size of the data to be investigated \cite{khalife2013performance}.

The current standard hardware device used for digital evidence acquisition in the forensic laboratory is the Forensic Recovery of Evidence Device (\nom{FRED}{Forensic Recovery of Evidence Device}). This machine incorporates a selection of equipment tailored for digital investigations available from Digital Intelligence \cite{fred}. Each FRED workstation contains a collection of write-blocked (read-only) ports including Serial Advance Technology Attachment (\nom{SATA}{Serial Advance Technology Attachment}), Integrated Drive Electronics (\nom{IDE}{Integrated Drive Electronics}), Small Computer System Interface (SCSI), Universal Serial Bus (USB) and FireWire. However in order to perform network evidence capture, the workstation incorporates a standard 10/100/1000Mb ethernet card due to the requirement for any NIC to both send and receive packets. This NIC is capable of collecting network evidence when used in conjunction with one of the software tools outlined above.

\section{Evidence Storage Formats}
\label{ch2:storage}

There is currently no universal standard for the format that digital evidence and any case related information is stored. This is due to the fact that there are no state or international governmental policies to outline a universal format. Many of the vendors developing forensic tools have their own proprietary evidence storage format. With such a small target market (mainly law enforcement), it sometimes makes business sense for them to try to lock their customers into a proprietary format. This results in their users being more likely to buy only their software in the future as it will be compatible with their existing evidence. There have been a number of attempts at creating open formats to store evidence and its related metadata. The following subsections describe the most common evidence storage formats.

\subsection{Common Digital Evidence Storage Format}
\label{ch2:cdesf}

The Common Digital Evidence Storage Format (CDESF) Working Group was created as part of the Digital Forensic Research Workshop (DRFWS) in 2006. The goal of this group was to create an open data format for storing digital forensic evidence and its associated metadata from multiple sources, e.g., computer hard drives, mobile Internet devices, etc. \cite{cdesf}. The format which the CDESF working group were attempting to create would have specified metadata capable of storing case-specific information such as case number, digital photographs of any physical evidence and the name of the digital investigator conducting the investigation. In 2006, the working group produced a paper outlining the advantages and disadvantages of various evidence storage formats \cite{dfrws2006}.

Due to resource restrictions, the CDESF working group was disbanded in 2007 before accomplishing their initial goal.

\subsection{Raw Format}
\label{ch2:raw}

According to the CDESF Working Group, ``the current de facto standard for storing information copied from a disk drive or memory stick is the so-called ``raw" format: a sector-by-sector copy of the data on the device to a file" \cite{evidencestandards}. The raw format is so-called due to the fact that it is simply a file containing the exact sector-by-sector copy of the original evidence, e.g., files, hard disk/flash memory sectors, network packets, etc. Raw files are not compressed in any manner and as a result, any deleted or partially overwritten evidence that may lay in the slackspace of a hard disk is maintained. All of the commercial digital evidence capturing tools available today have the capability of creating raw files. 


\subsection{Advanced Forensic Format}
\label{ch2:aff}

The Advanced Forensic Format (AFF) is an open source, extensible format created by S. Garfinkel in Basis Technology in 2006 \cite{aff}. The AFF format has a major emphasis on efficiency and as a result it is partitioned into two layers; the disk representation layer which defines segment name used for storing all data associated with an image and the data storage layer which defines how the image is stored (binary or XML) \cite{containers}. The format specifies three file extensions; *.aff, *.afd and *.afm. *.aff files store all data and metadata in a single file, *.afd files store the data and metadata in multiple small files, and *.afm files store the data in a raw format and the metadata is stored in a separate XML file \cite{containers}.

\subsection{Generic Forensic Zip}
\label{ch2:gfzip}

Generic Forensic Zip (gfzip) is an open source project. Its goal is to create a forensically sound compressed digital evidence format based on AFF \ref{ch2:aff} \cite{gfzip}. Due to the fact that it is based upon the AFF format, there is limited compatibility between the two in terms of segment based layout. One key advantage that gfzip has over the AFF format is that gfzip seeks to maintain compatibility with the raw format, as described in Section \ref{ch2:raw}. It achieves this by allowing the raw data to be placed first in the compressed image \cite{containers}.

\subsection{Digital Evidence Bag (QinetiQ)}
\label{ch2:debq}

The method for traditional evidence acquisition involves a law enforcement officer collecting any relevant items at the crimescene and storing the evidence in bags and seals. These evidence bags may then be tagged with any relevant case specific information, such as \cite{debq}:
\begin{itemize}
\item Investigating Agency / Police Force
\item Exhibit reference number
\item Property reference number
\item Case/Suspect name
\item Brief description of the item
\item Date and time the item was seized/produced
\item Location of where the item was seized/produced
\item Name of the person that is producing the item as evidence
\item Signature of the person that is producing the item
\item Incident/Crime reference number
\item Laboratory reference number
\end{itemize}

Physical evidence containers, such as evidence bags, are trusted due to the well understood and practised process called ``chain of custody" \cite{debwet}.

Digital Evidence Bag (\nom{DEB}{Digital Evidence Bag}) is a digital version of the traditional evidence bag, created by Philip Turner in 2005 \cite{debq}. DEB is based on an adaptation of existing storage formats, with potentially infinite capacity. The data stored in a DEB is stored in multiple files, along with metadata containing the information that would traditionally be written outside on an evidence bag. There are currently no tools released that are compatible with the QinetiQ DEB format.

\subsection{Digital Evidence Bag (WetStone Technologies)}
\label{ch2:debwet}

In 2006, C. Hosmer, from WetStone Technologies Inc. \cite{wetstone}, published a paper outlining the design of a Digital Evidence Bag (DEB) format for storing digital evidence \cite{debwet}. This format is independent from the Digital Evidence Bag outlined in Section \ref{ch2:debq}. The format emerged from a research project funded by the U.S. Air Force Research Laboratory. The motivation for this format was similar to that described in Section \ref{ch2:debq}, i.e., to metaphorically mimic the plastic evidence bag used by crime scene investigators to collect physical evidence such as fibres, hairs, etc. This format will be released publicly when complete.

\subsection{EnCase Format}
\label{ch2:encaseformat}

The EnCase format for storing digital forensic is proprietary to the evidence analysis tool of the same name \cite{garber2001encase}. It is by far the most common evidence storage option used by law enforcement and private digital investigation companies \cite{containers}. Because of the proprietary nature of the format, along with the lack of any open formal specification from Guidance Software \cite{guidance}, much remains unknown about the format itself. Some competitors to Guidance Software have attempted to reverse engineer the format to provide an element of cross-compatibility with their tools \cite{aff}. EnCase stores a disk image as a series of unique compressed pages. Each page can be individually retrieved and decompressed in the investigative computer's memory as needed, allowing a somewhat random access to the contents of the image file. The EnCase format also has the ability to store metadata such as a case number and an investigator \cite{aff}.

\section{Evidence Handling}
\label{ch2:handling}

When analysing physical evidence, the commonly used procedure is known as the ``chain of custody" \cite{debq}. The chain of custody commences at the crime scene where the evidence is collected, when the investigating officer collects any evidence s/he finds and places it into an evidence bag. This evidence bag will be sealed to avoid any contamination from external sources and signed by the officer and will detail some facts about the evidence, e.g., description of evidence, location, date and time it was found etc. The chain of custody will then be updated again when the evidence is checked into the evidence store. When it comes to analysing the evidence, it will be checked out to the analysts' custody and any modification to the evidence required to facilitate the investigation, e.g., taking a sample from a collected fibre to determine its origin or unique properties. Each interaction with the evidence will be logged and documented. 

The procedures outlined above for physical evidence need to be slightly modified for evidence acquisition and analysis. Due to the fact that digital evidence is analysed on forensic workstations, most of the above sequences can be automated into concise logging of all interactions. During a digital investigation, there is no requirement to modify the existing evidence in any way. This is because all analysis is conducted on an image of the original source and any discovered evidence can be extracted from this image, documented and stored separately to both the original source and the copied image. It is imperative when dealing with all types of evidence that all procedures used are reliable, reproducible and verifiable. In order for evidence to be court admissible, it must pass the legal criteria for the locality that the court case is being heard, as outlined in greater detail in Section \ref{ch2:legal}.

\subsection{What does ``Forensically Sound" really mean?}
\label{ch2:sound}

Many of the specifications for digital forensic acquisition and analysis tools, storage formats and hash functions state that the product in question is ``forensically sound" or that the product works with the digital evidence in a ``forensically sound manner", without specifying exactly what the term means. In 2007, E. Casey published a paper in the Digital Investigation Journal entitled ``What does ``forensically sound" really mean ?" \cite{forensicallysound}. 

In the paper, Casey outlined some of the common views of forensic professionals with regard to dealing with digital forensic evidence. Purists state that any digital forensic tools should not alter the original evidence in any way. Others point out that the act of preserving certain types of evidence necessarily alters the original, e.g., a live memory evidence acquisition tool must be loaded into memory (altering the state of the volatile memory and possibly overwriting some latent evidence) in order to run the tool and capture any evidence contained in the memory. Casey then goes onto to explain how some traditional forensic process require the alteration of some of the evidence in order to collect the required information. For example, collecting DNA evidence requires taking a sample from some collected evidence, e.g., a hair. Subsequently, the forensic analysis of this evidentiary sample (DNA profiling) is destructive in its nature which further alters the original evidence.

Casey summarises that from a forensic standpoint, evidence acquisition and handling should modify the evidence as little as possible and when modification is unavoidable, it should be well documented and considered in the final analytical results. ``Provided the acquisition process preserves a complete and accurate representation of the original data, and its authenticity and integrity can be validated, it is generally considered forensically sound" \cite{forensicallysound}.

\section{Cryptographic Hash Functions}
\label{ch2:hashfunctions}

Cryptographic hash functions are deterministic procedures which operate by taking a block of data or a file as input and output a fixed length digital fingerprint or cryptographic hash value/sum. The data input to a hash function is commonly referred to as the ``message``, while the hash sum produced is referred to as the digest.

The ideal collision resistant cryptographic hash function (h) has four main properties, defined by B. Preneel as part of his Ph.D. thesis in 1993 \cite{chf}:
\begin{enumerate}
\item The description of h must be publicly known and should not require any secret information for its operation.
\item The argument/message X can be of arbitrary length and the result h(X) has a fixed length of n bits (with n $\geq$ 128).
\item Given h and X, the computation of h(X) must be ``easy".
\item The hash function must be ``one-way'' in the sense that given a Y, it is infeasible to find a message X such that h(X) = Y , i.e., it should be impractical to modify a message without changing its hash. It should also be infeasible given X and h(X) to find a message X' $\neq$ X such that h(X') = h(X), i.e., it should not be possible to have two different messages with the same hash.
\item The hash function must be collision resistant: this means that one should not find two distinct messages that hash to the same result. It also should not be feasible to find a message X that has a given hash sum h(X).
\end{enumerate}


\subsection{Collision Resistance}
\label{ch2:collisionresistance}

The measure of the unlikelihood of two different inputs to a hashing function returning the same hash sum is known as the collision resistance of the hash function. Generally speaking, the larger the internal state size that the hashing function has to operate with, the better the collision resistance of that function.


In 2005, Wang and Yu published a paper outlining their attempts to break a number of specified hash functions, entitled ``How to Break MD5 and Other Hash Functions" \cite{wangmd5}. In this paper they described a method for engineering two files which, when hashed using MD5, would result in having the same hash sum. In their experiments, they created two different files, F1 and F2, by reverse engineering them to have the specific bits in the specific file locations required for the hashing function to produce an identical hash sum so far. It is important to note that there is no documented evidence that, if given a specific file F1, that anyone is capable of engineering a second file F2 that has the same hash sum. As a result of this paper, the United States Computer Emergency Readiness Team (US-CERT), part of the United States' Department of Homeland Security,  published a vulnerability note stating that MD5 should be considered cryptographically broken and unsuitable for further use and that most United States governmental applications will be required to move to the SHA-2 family of hashing functions by 2010 \cite{certmd5}.

To date, no collisions have been found in any of the SHA-2 family of hashing functions.

\subsection{Avalanche Effect}
\label{ch2:avalanche}

The avalanche effect of a cryptographic hashing function refers to a desirable property whereby should the input file be modified slightly \cite{zhangzheng}, e.g., changing a single bit of the file, the resultant hash sum produced changes significantly. The term ``avalanche effect" used to describe this property was created by H. Feistel in 1975 \cite{feistel}. Table \ref{Figurehashtable} shows a sample set of common hashing functions along with sample hash sums they produce for two slightly different input files showing the influence the avalanche effect has on each function.

\begin{table}
\label{ch2:hashchoice}
\begin{center}
\begin{tabular}{| p{2.1cm} | p{1.2cm} | p{4.1cm} | p{4.1cm} | p{1cm} |}
\hline
\bf{Hash Algorithm}  & \bf{Length in bits}   & \bf{\emph{Sentence 1:} The quick brown fox jumps over the lazy dog} & \bf{\emph{Sentence 2:} The quick brown fox jumps over the lazy cog} & \bf{Diff \%}\\
\hline
Adler32     & 32    & 
    \texttt{5BDC0FDA} & 
    \texttt{5BD90FD9} & 25.0\%\\
\hline
CRC32       & 32    & 
    \texttt{414FA339} & 
    \texttt{4400B5BC} & 87.5\%\\
\hline
Haval       & 128   & 
    \texttt{713502673D67E5FA 557629A71D331945} & 
    \texttt{4C9409BE8321D982 72D9252F610FBB5B} & 93.8\%\\
\hline
MD2     & 128   & 
    \texttt{03D85A0D629D2C44 2E987525319FC471} & 
    \texttt{6B890C9292668CDB BFDA00A4EBF31F05} & 93.8\%\\
\hline
MD4     & 128   & 
    \texttt{1BEE69A46BA81118 5C194762ABAEAE90} & 
    \texttt{B86E130CE7028DA5 9E672D56AD0113DF} & 93.8\%\\
\hline
MD5     & 128   & 
    \texttt{9E107D9D372BB682 6BD81D3542A419D6} & 
    \texttt{1055D3E698D289F2 AF8663725127BD4B} & 100\%\\
\hline
RipeMD128   & 128   & 
    \texttt{3FA9B57F053C053F BE2735B2380DB596} & 
    \texttt{3807AAAEC58FE336 733FA55ED13259D9} & 93.8\%\\
\hline
RipeMD160   & 160   & 
    \texttt{37F332F68DB77BD9 D7EDD4969571AD67 1CF9DD3B} & 
    \texttt{132072DF69093383 5EB8B6AD0B77E7B6 F14ACAD7} & 95.0\%\\
\hline
SHA-1       & 160   & 
    \texttt{2FD4E1C67A2D28FC ED849EE1BB76E739 1B93EB12} & 
    \texttt{DE9F2C7FD25E1B3A FAD3E85A0BD17D9B 100DB4B3} & 95.0\%\\
\hline
SHA-256     & 256   & 
    \texttt{D7A8FBB307D78094 69CA9ABCB0082E4F 8D5651E46D3CDB76 2D02D0BF37C9E592} & 
    \texttt{E4C4D8F3BF76B692 DE791A173E053211 50F7A345B46484FE 427F6ACC7ECC81BE} & 95.3\%\\
\hline
SHA-384     & 384   & 
    \texttt{CA737F1014A48F4C 0B6DD43CB177B0AF D9E5169367544C49 4011E3317DBF9A50 9CB1E5DC1E85A941 BBEE3D7F2AFBC9B1} & 
    \texttt{098CEA620B0978CA A5F0BEFBA6DDCF22 764BEA977E1C70B3 483EDFDF1DE25F4B 40D6CEA3CADF00F8 09D422FEB1F0161B} & 95.8\%\\
\hline
SHA-512     & 512   & 
    \texttt{07E547D9586F6A73 F73FBAC0435ED769 51218FB7D0C8D788 A309D785436BBB64 2E93A252A954F239 12547D1E8A3B5ED6 E1BFD7097821233F A0538F3DB854FEE6}& 
    \texttt{3EEEE1D0E11733EF 152A6C29503B3AE2 0C4F1F3CDA4CB26F 1BC1A41F91C7FE4A B3BD86494049E201 C4BD5155F31ECB7A 3C8606843C4CC8DF CAB7DA11C8AE5045} & 96.1\%\\
\hline

\end{tabular}
\end{center}
\caption[Example hash sums from popular hash functions]{Example hash sums for a small file containing the sentences outlined. The percentage difference shows the difference in the hash sums produced. While each character of a hash is hexadecimal, i.e., 1 of 16 possible values, it is notable that some hashing functions have differences greater than the expected maximum difference, i.e., $>$93.8\%. This is due to a more pronounced avalanche effect in the hashing function.}
\label{Figurehashtable}
\end{table}

\subsection{Overview of Common Hashing Algorithms}
\label{ch2:hashoverview}

While there are hundreds, if not thousands, of hashing functions in existence, the list of commonly used functions is significantly shorter. This is due to the fact that NIST and the National Security Agency (\nom{NSA}{National Security Agency}) in the United States have prioritised the standardisation of hashing functions. The most popular hashing functions, outlined below, are all based on the message digest principle. The message digest principle was designed by Ronald Rivest \cite{rivest1992md5} and constitutes a hash function taking in a message of arbitrary length and producing a fixed length message digest (hash value/sum) based on that input.

\subsubsection{MD Family}
\label{ch2:md}

The Message Digest (\nom{MD}{Message Digest}) algorithm family of hash functions were all created by Ronald Rivest, a professor in Massachusetts Institute of Technology, along with some collaboration from others. The family contains six iterations of the algorithms; MD, MD2 (1988), MD3 (1989), MD4(1990), MD5 (1991) and MD6 (2008.) From the original iteration up as far as MD5, the algorithms all produced 128-bit message digests. These MD hash values are expressed as 32 hexadecimal digits, as can be seen in Table 2.1. MD6 is based on a variable length message digest size to improve performance for smaller inputs, and as a result the message digest can be anywhere in the range from 0 - 512 bits in length.

MD5 is a popular hash function used in numerous applications. Most of the tools available to the digital investigator rely on a combination of the CRC32 and the MD5 hash functions for maintaining data integrity \cite{dfrws2006}.

MD6 was entered into the competition for the SHA-3 Family of hash functions. However, in July 2009, the algorithm was withdrawn from the competition because in order for it to be fast enough to compete, the design would have had to compromise its resistance to differential attacks.

\subsubsection{SHA-0 and SHA-1 Family}
\label{ch2:sha1}

The first specification of the Secure Hashing Algorithm (\nom{SHA}{Secure Hashing Algorithm}) family of hashing functions was published in 1993 by the US National Institute for Standards and Technology. This early specification is now known as the SHA-0 function. SHA-0 was withdrawn from use by the US National Security Agency in 1995 and was replaced by a modified version of the function; SHA-1. Both SHA-0 and SHA-1 produce 160-bit hash sums and they have a maximum input message size of $2^{64} - 1$ bits (or 2048 petabytes).

X. Wang, Y.L. Yin and H. Yu produced a paper entitled ``Finding Collisions in the Full SHA-1" in 2005 \cite{wangsha1}. This paper outlined the first attack on the SHA-1 hash function. The authors successfully found collisions on the SHA-1 function. They achieved this by first finding near-collisions. They then were able to discover full collisions based on the analysis of the near collisions. They conclude that although the SHA-1 family of hash functions has message expansion, it does not offer enough avalanche effect in terms of differing inputs.

\subsubsection{SHA-2 Family}
\label{ch2:sha2}

The SHA-2 Family consists of the following hash functions: SHA-224, SHA-256, SHA-384, and SHA-512. The number in the name of the hash function represents the output message digest size in bits. H. Gilbert and H. Handschuh produced a journal paper entitled ``Security Analysis of SHA-256 and Sisters" in 2004 \cite{gilbert} which published their results from the analysis of the SHA-2 family of hash functions. They found that the attacks that have broken the SHA-1 family no longer are applicable to the SHA-2 family.

The SHA-224 and SHA-256 have the same maximum input file size of $2^{64} - 1$ bits (or 2048 petabytes) as with the SHA-1 Family, while the SHA-384 and SHA-512 have a maximum of $2^{128} - 1$ bits (or 3.78 x $10^{22}$ petabytes).

\subsubsection{SHA-3 Family}
\label{ch2:sha3}

NIST, part of the Department of Commerce, held a five year development competition to decide on which hashing function to choose for the third iteration of the SHA Family. As part of the competition, NIST accepted over 60 entries into the first round of testing. This number was reduced down to 14 accepted into the second round which was announced in August 2009 \cite{sha3}. The remaining candidates in the second round are BLAKE \cite{aumasson2008sha}, Blue Midnight Wish \cite{thomsen2010pseudo}, CubeHash \cite{bernstein2008cubehash}, ECHO \cite{benadjila2009sha}, Fugue \cite{halevi2009hash}, Gr\o stl \cite{gauravaram2008grostl}, Hamsi \cite{kuccuk2009hash}, JH \cite{wu2011hash}, Keccak \cite{bertoni2009keccak}, Luffa \cite{de2008hash}, Shabal \cite{bresson2008shabal}, SHAvite-3 \cite{biham2008shavite}, SIMD \cite{leurent2009simd} and Skein \cite{schneier2011skein}. The winner of the hashing function,  Keccak, was announced in November 2012 after evaluation of the final round entries \cite{nistsha3}. Keccak uses a ``sponge construction" with no explicit maximum limit for file size and for produces a variable length hash.


\section{Court Admissible Evidence}
\label{ch2:legal}

Since the United States leads the way with the implementation of many standards in relation to evidence handling and the court admissibility of evidence, many other countries look to the procedures outlined by the United States in this area when attempting to create their own legal procedures \cite{commons}. As a result, much of the information available regarding the admissibility of digital forensic evidence into court cases is specifically tailored to the Unites States, but will influence law makers across the globe. Carrier \cite{carrier-open} states that in order for evidence to be admissible into a United States legal proceeding, the scientific evidence (a category which digital forensic evidence falls under in the U.S.) must pass the so-called ``Daubert Test" (see Section \ref{ch2:daubert} below). The reliability of the evidence is determined by the judge in a pre-trail ``Daubert Hearing". The judge's responsibility in the Daubert Hearing is to determine whether the methodologies and techniques used to identify the evidence was sound, and as a result, whether the evidence is reliable.

\subsection{Daubert Test}
\label{ch2:daubert}


The ``Daubert Test" stems from the United States Supreme Court's ruling in the case of Daubert vs. Merrell Dow Pharmaceuticals (1993) \cite{daubert}. The Daubert process outlines four general categories that are used as guidelines by the judge when assessing the procedure(s) followed when handling the evidence during the acquisition, analysis and reporting phases of the investigation, \cite{carrier-open} and \cite{daubert}:

\begin{enumerate}
\item \emph{Testing} -- Has the procedure been tested? Testing of any procedure should include testing of the number of false negatives, e.g., if the tool displays filenames in a given directory, then all file names must be shown. It should also incorporate testing of the number of false positives, e.g. if the tool was designed to capture digital evidence, and it reports that it was successful, then all forensic evidence must be exactly copied to the destination. NIST have a dedicated group working on Computer Forensic Tool Testing (CFTT) \cite{cftt}.
\item \emph{Error Rate} -- Is there a known error rate of the procedure? For example, accessing data on a disk formatted in a documented file format, e.g., \nom{FAT}{File Allocation Table}32 or ext2, should have a very low error rate, with the only errors involved being programming errors on behalf of the developer. Acquiring evidence from an officially undocumented file format, e.g., \nom{NTFS}{New Technology File System}, may result in unknown file access errors occurring, in addition to the potential programming error rate.
\item \emph{Publication} -- Has the procedure been published and subject to peer review? The main condition for evidence admission under the predecessor to the Daubert Test, the Frye Test, was that the procedure was documented in a public place and undergone a peer review process. This condition has been maintained in the Daubert Test \cite{carrier-open}. In the area of digital forensics, there is only one major peer-reviewed journal, the International Journal of Digital Evidence.
\item \emph{Acceptance} -- Is the procedure generally accepted in the relevant scientific community? For this guideline to be assessed, published guidelines are required. Closed source tools have claimed their acceptance by citing the large number of users they have. The developers of these tools do not cite how many of their users are from the scientific community, or how many have the ability to scientifically assess the tool. However, having a tool with a large user base can only prove acceptance of the tool; it cannot prove the acceptance of the undocumented procedure followed when using the tool.
\end{enumerate}

In 2005, The House of Commons Science and Technology Committee in the United Kingdom published a report entitled ``Forensic Science on Trial" \cite{commons}. In this report they outline numerous standards to be used across the field of forensics. As part of this report, the admissibility of expert evidence is discussed. As it stood in the UK when the report was written, the judge of any given case had the role of the ``gate-keeper" for any evidence s/he would admit into his/her court. It was determined that judges are not well-placed to determine the scientific validity without input from scientists, especially due to the absence of an agreed protocol for assessment. The main recommendation to come from the report is that the Forensic Science Advisory Council should develop a ``gate-keeping" test for expert evidence, built in partnership with judges, scientists and other key players from the criminal justice system and that it should be built upon the US Daubert Test \cite{commons}.

\section{Legal Considerations of Network Forensics}
\label{legal}

Collecting network traffic can pose legal issues. Deploying a packet sniffing or deep packet inspection device, such as those outlined above, can result in the (intentional or incidental) capture of information with privacy and security implications, such as passwords or e-mail content, etc. As privacy has become a greater concern for regular computer users and organisations, many have become less willing to cooperate or share any information with law enforcement. For example, most ISPs will now require a court order before providing any information related to suspicious activity on their networks \cite{Nist:2012:NSP:2331550}. In Europe, continental legal systems operate on the principle of free introduction and free evaluation of evidence and provide that all means of evidence, irrespective of the form they assume, can be admitted into legal proceedings \cite{karyda2007internet}.


One aspect of the use of search and seizure warrants in an Internet environment concerns the geographical scope of the warrant issued by a judge or a court authorising the access to the digital data. In the past, the use of computer-generated evidence in court has posed legal difficulties in common law countries, and especially in Australia, Canada, the United Kingdom and the USA. The countries are characterised by an oral and adversarial procedure. Knowledge from secondary sources is regarded as ``hearsay evidence", such as other persons, books, records, etc., and in principle is inadmissible. However, digital evidence has become widely admissible due to several exceptions to this hearsay rule \cite{karyda2007internet}.

\section{Summary}
\label{ch2:summary}
This chapter describes some foundations behind the system described in Chapter \ref{ch5designandarch}. It outlined some of the tools, formats, tests and procedures used for the acquisition and analysis of digital forensic evidence. This chapter also outlined some network focused forensic tools and systems developed for aiding digital forensic investigations. Traditionally, in order for a digital forensic investigation to begin, the investigator must physically visit the crime scene and collect any suspect computer equipment. This equipment will then be brought back to the forensic laboratory. When investigating network crimes, the procedure is somewhat different. The forensic investigator may need to install a physical deep packet inspection device onto the suspect's Internet connection (assuming a warrant is granted to do so). This will then typically be left in situ for a predetermined amount of time and then taken away for analysis. This DPI device will generally contain all of the suspects network traffic for the investigation duration. Analysis, and subsequent detection of any incriminating evidence, can only begin at this stage. As a result of this offline analysis, it may be some time before an arrest can be made.



\chapter{Peer-to-Peer File-Sharing}
\label{ch3peertopeerfilesharing}

\section{Introduction}
\label{ch3:introduction}

P2P networks can be used in a number of ways including distributed computing, collaboration and communication, but perhaps they are best known for their use in file-sharing \cite{fox2001peer}. In 1999, three influential P2P systems were launched attracting significant interest in the Internet technology; the Napster music sharing system, the Freenet anonymous data store and the SETI@home distributed volunteer-based scientific computing project \cite{rodrigues2010peer}.

In 2008, Cisco estimated that P2P file sharing accounted for 3,384 petabytes per month of global Internet traffic, or 55.6\% of the total usage. Cisco forecast that P2P traffic will account for 9,629 petabytes per month globally in 2013 (approximately 30\% of total global usage) \cite{cisco2008}. While the volume of P2P traffic is set to almost triple from 2008-2013, its proportion of total Internet traffic is set to decrease due to the rising popularity of media streaming sites and one-click file hosting sites (often referred to as ``cyberlockers'') such as Rapidshare, Mega, Mediafire, etc. Cisco estimate that P2P file transfer will decline over the next few years to 5,755 petabytes per month by 2017 \cite{cisco2012}. The decline is accounted for due to the rise in streaming services and traditional server based file-sharing. BitTorrent is the most popular P2P protocol used worldwide and accounts for the biggest proportion of Internet traffic when compared to other P2P protocols. The most recent measurement data from Ipoque GmbH. has measured BitTorrent traffic to account for anything from 20-70\% of total Internet usage in 2009, depending on the specific geographical area concerned \cite{schulze2009internet}. With the evolution towards employing encrypted traffic, these measurement statistics have been upwardly estimated over the measured traffic.

\subsection{Financial Impact on Content Producing Industry}
\label{ch3:impact}

The content producing industries report that revenue figures are steadily declining as a result of online piracy. The International Federation of the Phonographic Industry's (\nom{IFPI}{International Federation of the Phonographic Industry}) Digital Music Report 2011 states that legitimate digital music distribution is up 1000\% from 2004 to 2010, although total global recorded music revenues are down 31\% over the same period \cite{moore2011ifpi}. The report cites Internet piracy as having a significant impact on their sales. The report cites a study from 2010 entitled ``Piracy, Music and Movies: A Natural Experiment'' which estimates that physical sales would be up 72\% with the abolishment of piracy in Sweden \cite{adermon2010piracy}.

The 2012 Digital Music report states that 28\% of Internet users are accessing at least one unlicensed site monthly and that approximately half of those users are using P2P networks \cite{moore2012ifpi}. In 2006, Zentner \cite{zentner2006measuring} summarises that downloading MP3 files online reduces the probability of buying music by 30\%. In 2008, the Motion Picture Association of America (\nom{MPAA}{Motion Picture Association of America}) reported that Internet piracy cost the film industry \$7 billion that year \cite{serbin2012graduated}. 

While the figures outlined above are provided by the content producing industry, the figures of total physical and digital sales in comparison to illegal downloads are not available or provided by the industry for independent verification. However, unauthorised distribution of copyrighted content must have an impact on the profits of the industry as a whole. As a result of these financial losses incurred by the content producing industry, there has been a significant push for technological and legislative measures to deter users from choosing the pirated option. A number of these measures are outlined in Sections \ref{ch3:legislation} and \ref{ch3:p2pprocess}. 

Research has been conducted into the decision of an average consumer to legally purchase digital content versus the decision to illegally download the content. In his 2005 paper, Fetscherin evaluated the choices made by consumers in legally or illegally downloading movies on the Kazaa P2P network \cite{fetscherin2005movie}. He found that the majority of users prefer to download movies legally, but that a significant number will always opt for the free option. The factors affecting consumer behaviour in this legal/illegal choice are, in order, the risk of being caught, the price, the perceived value of the original and the availability of high quality copies.

In 2013, the European Commission's Institute for Prospective Technological Studies published a working paper analysing the impact of illegal downloading and legal streaming on the legal purchases of digital music \cite{eupiracy}. The results were based on over 16,000 European consumers from all European Commission (\nom{EC}{European Commission}) countries. It was found that Internet users do not view illegal downloading as a substitute for legal digital music. An increase of 10\% of clicks on illegal downloading websites was found to lead to a 0.2\% increase in clicks on legal purchase websites.

\section{Legislative Response to Online Piracy}
\label{ch3:legislation}

While not merely limited to P2P file-sharing, there has been significant effort globally to create new legislative measures to combat online copyright infringement. A number of these provisions are outlined below \cite{carrier2013sopa}:

\begin{enumerate}
\item Stop Online Piracy Act (\nom{SOPA}{Stop Online Piracy Act}) -- This United States act states that if a website is deemed to be dedicated to the theft of U.S. property then it should be blocked by various Internet companies. These include ISPs, search engines, payment providers and advertising services. Each must prevent access to the site for their customers and cease operation with the site and its owners. This act has facilitated the significant seizure of Internet domain names by the U.S. government since 2012.
\item PROTECT IP Act (\nom{PIPA}{Preventing Real Online Threats to Economic Creativity and Theft (PROTECT) of Intellectual Property Act}) -- In the United States, the ``Preventing Real Online Threats to Economic Creativity and Theft of Intellectual Property Act of 2011" allows the Attorney General to sue operators of Internet sites dedicated to infringing activities.
\item Anti-Counterfeiting Trade Agreement (\nom{ACTA}{Anti-Counterfeiting Trade Agreement}) -- This is a treaty signed between the United States, 
Australia, Canada, Korea, Japan, New Zealand, Morocco, and Singapore. The treaty specifies that laws should be created to make those responsible for copyright infringement on a commercial scale criminally liable.
\item Trans-Pacific Partnership Agreement (\nom{TPP}{Trans-Pacific Partnership}) -- This is a trade agreement negotiated between the United States, Australia, Brunei, Chile, Malaysia, New Zealand, Peru, Singapore, and Vietnam. This broadly defined agreement states that its member countries must legislate for significant wilful copyright infringement for financial gain.
\item Graduated Response -- France, New Zealand, South Korea, Taiwan and the United Kingdom have implemented various ``three strikes'' laws, with Ireland and the United States having a voluntary system put in place \cite{serbin2012graduated}. The model requires the rights holders to monitor for unauthorised online infringing activity and reporting the corresponding IP addresses to the ISP involved. The ISP can identify the customer and send them a notification. Repeat infringers risk bandwidth throttling, protocol blocking or account suspension.
\end{enumerate}

As deduced by Carrier \cite{carrier2013sopa}, the language used in the existing treaties, acts and agreements  outlined above is quite vague. The resulting lack of clarity enables biased interpretations, which in turn promotes a litigation based business model. Carrier claims this will ultimately stem innovation. While it's clear that there is a need for legislation to protect content producers, the current iterations leave much open to various interpretation.

\section{Peer-to-Peer File-sharing System Design}
\label{ch3:p2pdesign}

When designing and implementing a P2P file-sharing system, developers must make a number of key decisions regarding the network and its purposes:

\begin{enumerate}
\item{Centralised/Decentralised/Hybrid -- A centralised network can offer a simpler design whereas a decentralised design can offer a more robust network. Hybrid networks are much more complex to implement but can offer many of the advantages of both centralised and decentralised systems.}
\item{Open Source/Proprietary -- Making the network design and client open source promotes an active development community but may result in compatibility issues across numerous clients with different update cycles resulting in newer features taking some time to roll out across all clients.}
\item{Encrypted/Unencrypted -- Encrypting all network communication can help eliminate some of the packet sniffing investigation methods deployed by IT administrators and investigators but can decrease the performance of the overall system.}
\end{enumerate}

\subsection{Centralised Design}
\label{ch3:centralised}

In a centralised P2P network design, there are one or more central servers which puts users in contact with each other. When a new user wishes to join the network, s/he registers with a known server which, in turn, is able to supply the user with a list of other known active peers currently on the network. Depending on the specific centralised design, the server itself may index the entire system, i.e., maintain an active list of users and the content they are sharing, or help contribute to a distributed hash table (\nom{DHT}{Distributed Hash Table}) maintained by the server. The latter option passes much of the querying load onto the connected peers. Hashing is used to prevent the accidental downloading of incorrect content, e.g., two files on the network with the same name but with different content.

\begin{figure}[t]
\centering
\includegraphics[width=1\textwidth]{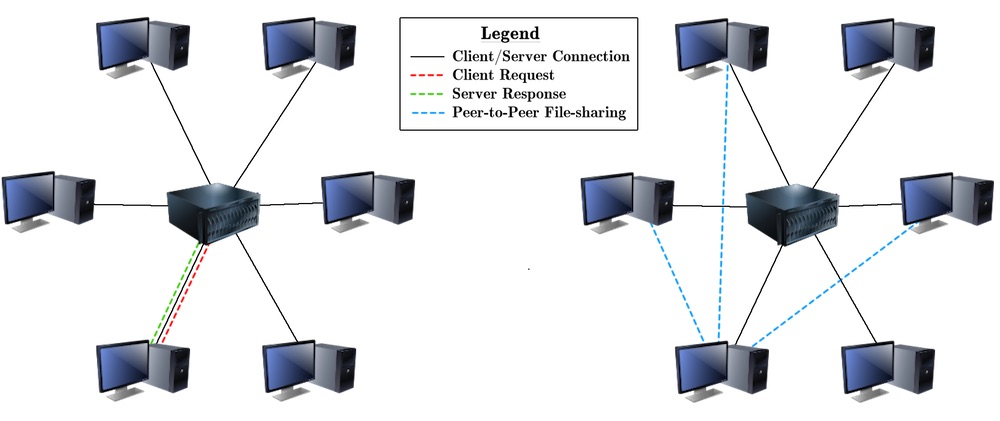}
\caption{Centralised P2P system overview.}
\label{Figurecentralised}
\end{figure}

A sample centralised usage scenario is shown in Figure \ref{Figurecentralised}, whereby the server records each user's shared files. On the left hand side of the figure, the P2P client issues a request to the server for any users sharing a specific piece of content. The server responds with a list (IP addresses and port numbers) of active nodes on the network sharing that content. Once the user chooses one of these files to download, the server no longer has any further part in the interaction. The user's client software will connect directly to the remote peer and download the file directly. In most modern systems built using this design, the user may download part of the desired file from multiple other peers simultaneously. This is shown on the right hand side of the figure and results in a faster throughput of the download by distributing the workload. Once all the required parts are complete, the file is combined into the original content and is immediately available to the user. By default, many P2P systems automatically make any newly downloaded files available on the network for other peers.

The advantage of a centralised design lies in its efficiency of conducting queries and the resultant small traffic footprint devoted to querying. However, the most significant downside to a centralised design is that there is a single point of failure. If the central server is disrupted or removed from the network, the entire network ceases to function and is rendered useless.

\subsection{Decentralised Design}
\label{ch3:decentralised}

Decentralised P2P network design removes the single point of failure from the centralised design outlined above. It achieves this by enforcing that each node in the network simultaneously plays the role of client and server. When a node receives a query request, that request is passed on to all known nodes and the results are passed back to the source of the query. Each node in the network maintains a small active list of current connections and helps to contribute to a DHT. To remove any loss to the DHT as a result of a node going offline, each distributed part will exist on multiple nodes.

The primary advantage of a decentralised system is that there is no single point of failure. Removing any single node from the network will have no significant impact on the entire system. However, due to its decentralised design, each query will take much longer to complete as the query needs to be passed directly from node to node before any response comes back to the source. As a result, the larger the network becomes, the longer the time required to conduct a complete search of the entire network will become. This results in a much larger querying traffic footprint in order to keep the network functional. There are a number of solutions to this problem, i.e., limiting the number of hops a query can be passed along, including a specified query timeout, etc. However, each of these solutions result in a partial search of the network.

\begin{figure}[t]
\centering
\includegraphics[width=1\textwidth]{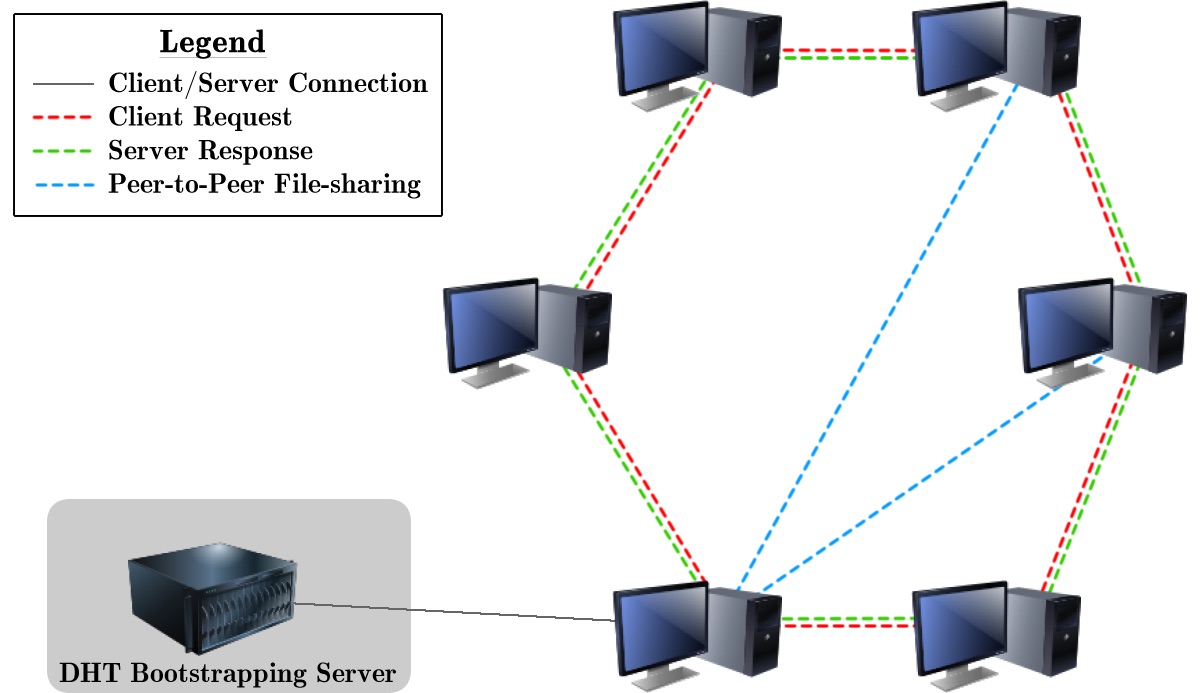}
\caption{Decentralised P2P system overview.}
\label{Figuredecentralised}
\end{figure}

A sample usage scenario is outlined in Figure \ref{Figuredecentralised} showing how each node acts as both a client and a server. The query can be seen passing from node to node until a ``hit'' is found. Depending on the design, this query hit might get passed back through the same sequence of steps the query took or alternatively may be relayed directly back to the querying node. The file transfer occurs in a similar direct fashion as exists in a centralised P2P network. When a new node wishes to join the network, it needs to bootstrap onto the DHT in some manner. Depending on the implementation of the network, there may be a hardcoded list of always-on DHT servers or a list of always-on nodes. These servers/nodes can provide a new client with a list of active nodes to bootstrap onto the network.

\subsection{Hybrid Design}
\label{ch3:hybrid}

Hybrid P2P networks take on a number of the features from both centralised and decentralised networks in an attempt to overcome the limitations of each. This type of network will employ a large number of centralised servers to prevent the network becoming dysfunctional if an individual server should be taken offline. In practice, these servers are actually regular peers on the network. Static hybrid networks allow a peer to specify that they would like to become a server, or ``supernode'' in the configuration of the client software, e.g., in eDonkey. Dynamic hybrid network clients, e.g., Limewire, FastTrack and Gnutella, can automatically promote any peer to become a supernode dependant on specified criteria such as uptime, bandwidth capability, latency etc.

\begin{figure}[t]
\centering
\includegraphics[width=1\textwidth]{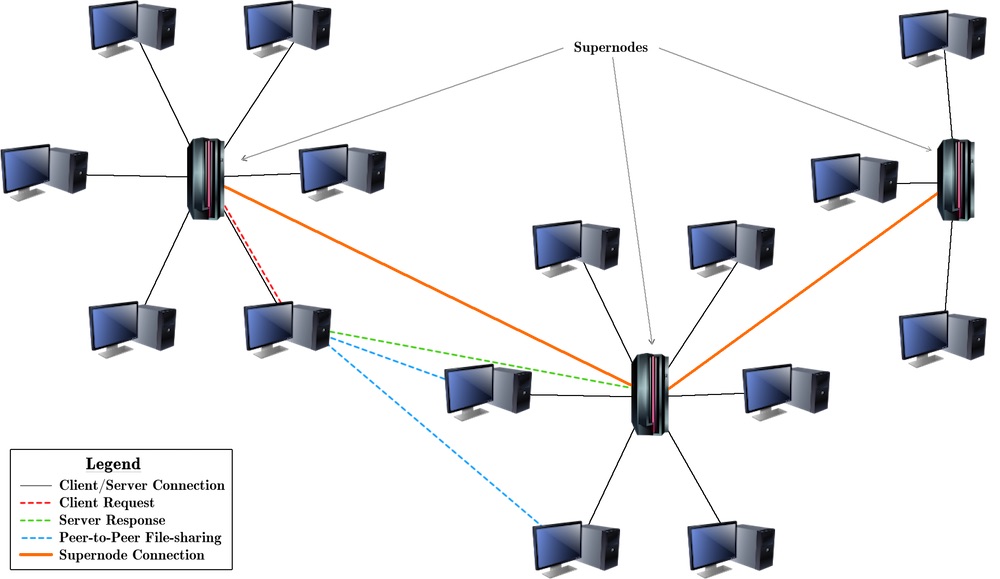}
\caption{Hybrid P2P system overview.}
\label{Figurehybrid}
\end{figure}

Querying a hybrid network can involve varying query distribution options. There are two main query propagation mechanisms employed by hybrid P2P networks \cite{ioannidis2008design}:

\begin{enumerate}
\item Random Walk -- Employing this mechanism, a file search query is sent randomly to known neighbours in the network. If this node can resolve the query, i.e., has a file matching the request, a query hit is sent back to the source of the request. If it does not resolve the query, it randomly passes this query onto another neighbour from its list and the process iterates until either a hit is returned or a timeout kills the query.
\item Expanding Ring -- This query propagation option can be thought of as a sequence of flooding searches in which the time-to-live (TTL) is increased at each iteration. A simple flooding search is conducted whereby a query is sent to all known neighbours. Each neighbour will propagate the received query onto all of its active neighbours (it will only forward on each specific query once). Again, if any node can resolve the query, it will directly reply to the origin of the query with a hit. The query dies when the TTL reaches zero. This propagation mechanism is outlined in Figure \ref{Figurehybrid}, where a query is sent from a node to a supernode. If this supernode cannot resolve the query, it is forwarded onto other supernodes and a query hit is returned directly to the source node. File transfer then commences directly between the nodes.
\end{enumerate}

\section{Peer-to-Peer File-sharing Networks}
\label{ch3:p2pnetworks}

\subsection{Napster}
\begin{figure}[htb]
\centering
\includegraphics[trim = 0mm 0mm 0mm 0mm, clip, width=\textwidth]{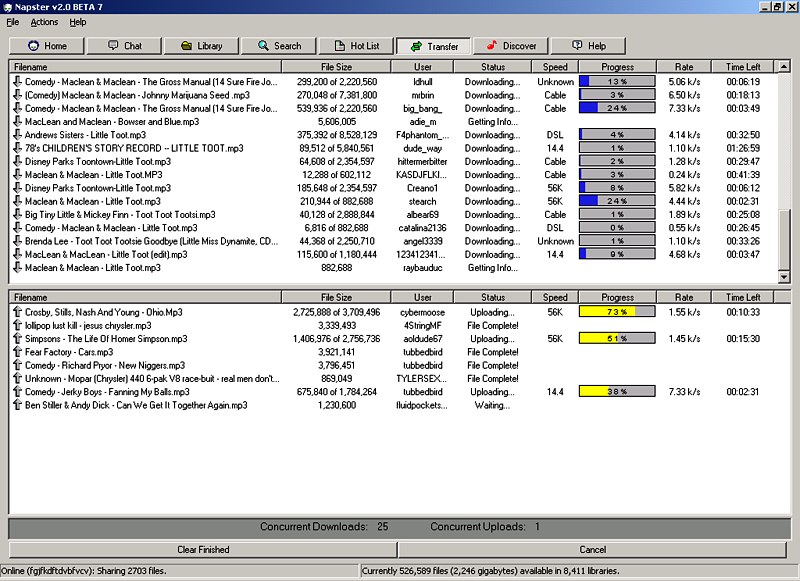}
\caption{Screenshot of Napster. Downloads can be seen at the top, with uploads at the bottom.}
\label{Figurenapster}
\end{figure}
In 1999, Napster pioneered the idea of global P2P file sharing. This early stage MP3 sharing network was supported by a centralised file search facility. Users could query this centralised database for desired content and their clients would be advised of other users currently sharing that piece of content. The client would then proceed to download this content directly off the user hosting the content. Due to its ``free'' usage model, Napster quickly grew to become a large global P2P network. By the end of 2000, Napster grew to over 75 million users sharing over 10,000 MP3 files every second \cite{fox2001peer}. The centralised index ultimately was the downfall of Napster's design. In July 2001, after losing a court case with the Recording Industry Association of America, Napster became the first P2P system to be ordered to shutdown in what has become known as the ``Napster Decision'' \cite{carrier2013sopa}.

\subsection{Gnutella}
\label{ch3:gnutella}

\begin{figure}[htb]
\centering
\includegraphics[width=1\textwidth]{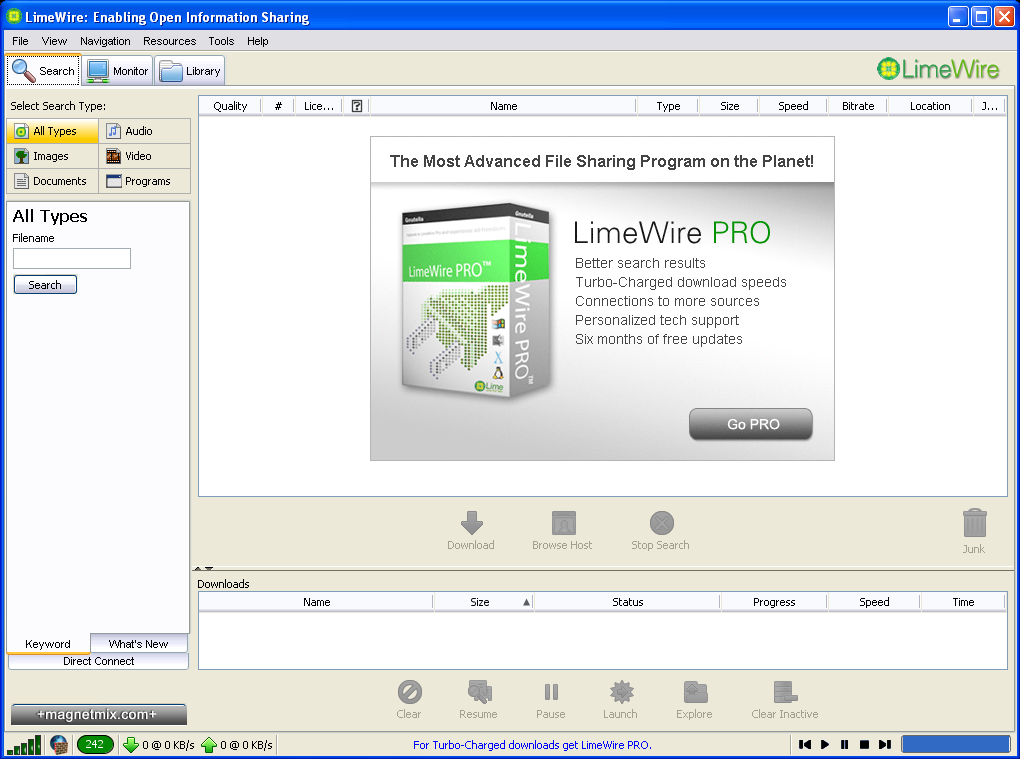}
\caption{Limewire Screenshot.}
\label{Figurelimewire}
\end{figure}

Gnutella was released in 2000 as an open source file-sharing protocol by Nullsoft. When the system was released online, Nullsoft's owners, America Online Inc., first learned of its existence and the company quickly ordered the removal of the release from the Internet. However, the protocol was already downloaded by numerous other developers who were able to publish the specification \cite{berkes2003decentralized}. This enabled numerous popular clients to be built on the protocol including BearShare, FrostWire, Morpheus, Shareaza and LimeWire (shown in Figure \ref{Figurelimewire}). Gnutella clients are generally capable of resuming any partially completed downloads by reestablishing connections to previously known or new peers sharing the same content \cite{Lewthwaite2013}.
 
\begin{figure}[h]
\centering
\includegraphics[width=0.5\textwidth]{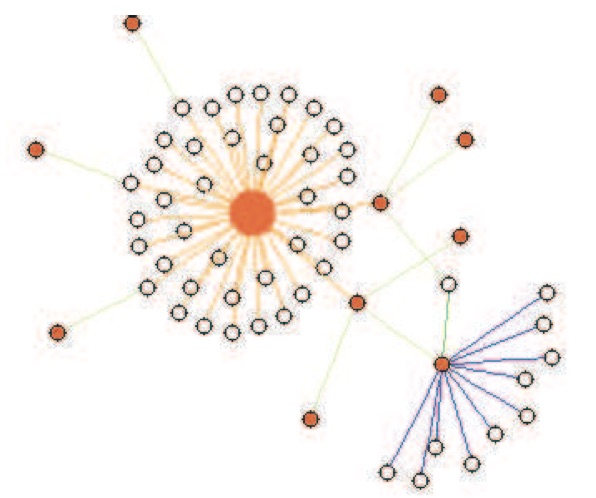}
\caption{Gnutella Node Map.}
\label{Figuregnutellamap}
\end{figure}

The Gnutella protocol adopts decentralised search algorithms which effectively eliminate the single point of failure of the centralised approach creating a much more robust network. This results in clients searching for content shared on Gnutella bouncing the query from node to node, with any hits being reported back through a reversed sequence of these network bounces \cite{hannaway}. A sample Gnutella node map is shown in Figure \ref{Figuregnutellamap} with supernodes, represented as solid dots, acting as servers for many leaf nodes (represented as hollow dots) \cite{berkes2003decentralized}.

The decentralisation of the network also left the network open and vulnerable to exploitation. While ``legitimate'' sharing of copyrighted content is the primary focus of the network, it was quickly exploited for malicious purposes by cybercriminals, e.g., the spread of viruses, worms and botnets. In 2008, Kalafut et al. found that 68\% of all downloadable responses in Limewire (Gnutella's largest client at the time) containing archives and executables contained malware \cite{kalafut}.

\subsection{eDonkey}
\label{ch3:edonkey}

eDonkey is one of the most successful P2P applications and operates on a hybrid P2P network of the same name. Alternative clients built on the protocol include eMule, Morpheus, Shareaza and MLDonkey. The systems uses a distributed network of servers running a specific server application. The servers do not share any files and only aid in the management of the distributed information through indexing which peers are sharing which files. The network gained significant popularity in Europe; with Germany, France and Austria topping the geographical overview accounting for 66.21\%, 6\% and 1\% respectively in 2004 \cite{heckmann2004edonkey}. Files are divided into chunks of 9,500kb which an MD4 checksum associated with each chunk.

\subsection{BitTorrent}

\begin{figure}[htb]
\centering
\includegraphics[width=1\textwidth]{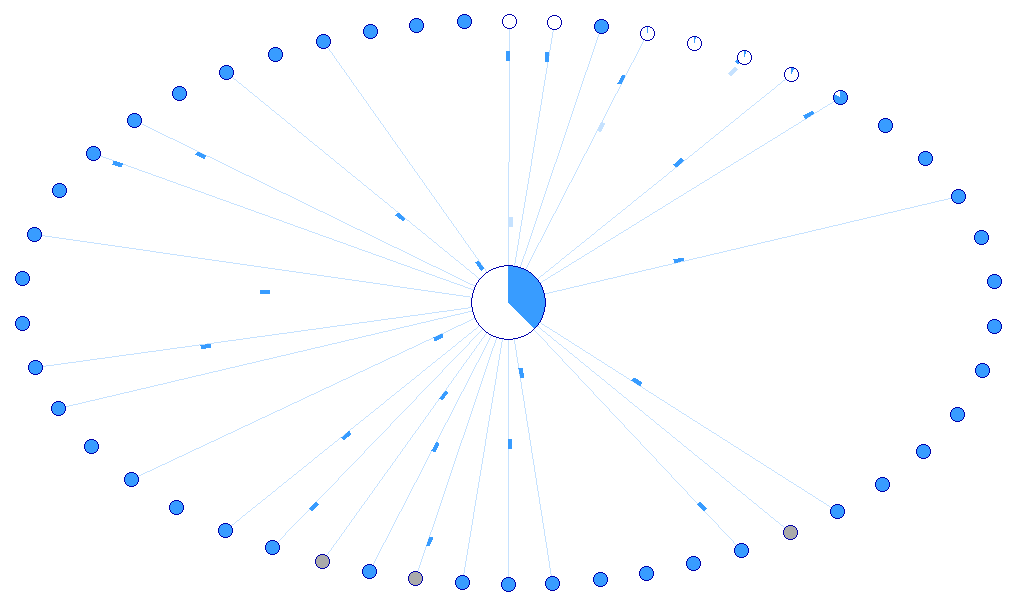}
\caption{Visualisation of a Typical BitTorrent Swarm}
\label{Figurebt}
\end{figure}

In July 2001, the first implementation of the BitTorrent protocol was released. The BitTorrent protocol is designed to easily facilitate the distribution of files to a very large number of downloaders with minimal load on the original file source \cite{btspec}. This is achieved through the downloaders uploading their completed parts of the entire file to other downloaders. A BitTorrent swarm is made up of seeders, i.e., peers with complete copies of the content shared in the swarm, and leechers, i.e., peers who are downloading the content. Due to BitTorrent's ease of use and minimal bandwidth requirements, it lends itself as an ideal platform for the unauthorised distribution of copyrighted material. This typically commences with a single source sharing large sized files to many downloaders.

Based on global bandwidth usage, BitTorrent is the most popular P2P network in use today. In 2005, D. Erman measured BitTorrent traffic to account for over 60\% of the world's bandwidth usage \cite{erman2005bittorrent}. The BitTorrent protocol is designed to easily facilitate the distribution of files to a potentially large number of interested parties, i.e., other peers, with a minimal load on the original file source, as outlined in the BitTorrent protocol specification. This is achieved through the following steps:

\begin{enumerate}

\item The file is split up into a number of uniformly sized pieces or chunks with typical chunk sizes generally ranging from 128kB to 4MB.
\item The initial source of the file creates a UTF-8 encoded ``.torrent" metadata file, which includes unique SHA-1 hash values for the entire file and each of the file chunks, along with other required file information, e.g., filenames, chunk size, total file size, path information, client information, comments etc.
\item This metadata file is then shared by the creator with other users interested in acquiring the original content either through direct distribution, e.g., email, instant messaging etc., or through the much more common method of uploading onto a torrent indexing website, such as ThePirateBay.org. Following the recent trend of maximising decentralisation of the BitTorrent eco-system, many indexing websites now only serve ``magnet'' \nom{URI}{Uniform Resource Identifier}s. These URIs enable the user to connect to a distributed hash table, as outlined below, and acquire the metadata file from other users.
\item Users interested in downloading the available content must then download this metadata file and open it using a BitTorrent client, such as Azureus/Vuze or  $\upmu$Torrent.
\item The BitTorrent client is then tasked with identifying other peers who are sharing the file uniquely identified in the metadata file, i.e., other peers in the swarm. This includes identifying seeders, i.e., peers with complete copies of the content shared in the swarm, and other leechers, i.e., peers who are currently downloading the content, but are sharing the completed chunks with others. This peer discovery is achieved through a variety of methods including tracker communication, distributed hash tables and peer exchange.

\end{enumerate}

The success of the BitTorrent protocol can be attributed to uploaders incurring no additional cost (besides their Internet connectivity costs) to share files with many users. In practice, the original uploader needs only to stay connected to the swarm until a sufficient number of leechers have one full copy of the file between them. This is made possible through the leechers uploading their completed chunks of the entire file to other downloaders. Due to BitTorrent's ease of use, minimal bandwidth requirements and perceived Internet anonymity, it lends itself well as an ideal platform for the unauthorised distribution of copyrighted material. Initially this distribution consists of a single original source for sharing a large sized file between many peers.

Each BitTorrent client must be able to identify a list of active peers in the same swarm who have at least one piece of the content and is willing to share it, i.e., that has an available open connection and has enough bandwidth available to upload. By the nature of the protocol, any peer that wishes to partake in a swarm, must be able to communicate and share files with other active peers. There are a number of methods that a client can attempt to discover new peers which are in the swarm:
\begin{enumerate}
\item Tracker Communication -- BitTorrent trackers maintain a list of seeders and leechers for each BitTorrent swarm they are currently tracking. Each BitTorrent client will contact the tracker intermittently throughout the download of a particular piece of content to report that they are still alive on the network and to download a short list of new peers on the network.
\item Peer Exchange (\nom{PEX}{Peer Exchange}) -- Peer Exchange is a BitTorrent Enhancement Proposal (\nom{BEP}{BitTorrent Enhancement Proposal}) whereby when two peers are communicating, a subset of their respective peer lists are shared during the communication.
\item Distributed Hash Tables -- Within the confounds of the standard BitTorrent specification, there is no intercommunication between peers of different BitTorrent swarms. Azureus/Vuze and $\upmu$Torrent contain mutually exclusive implementations of distributed hash tables as part of their standard client features. These DHTs maintain a list of all active peers using each client and enables cross-swarm communication between peers. Each peer in the DHT is associated with the swarm(s) in which it is currently an active participant.
\end{enumerate}

\begin{figure}[ht]
\centering
\includegraphics[width=1\textwidth]{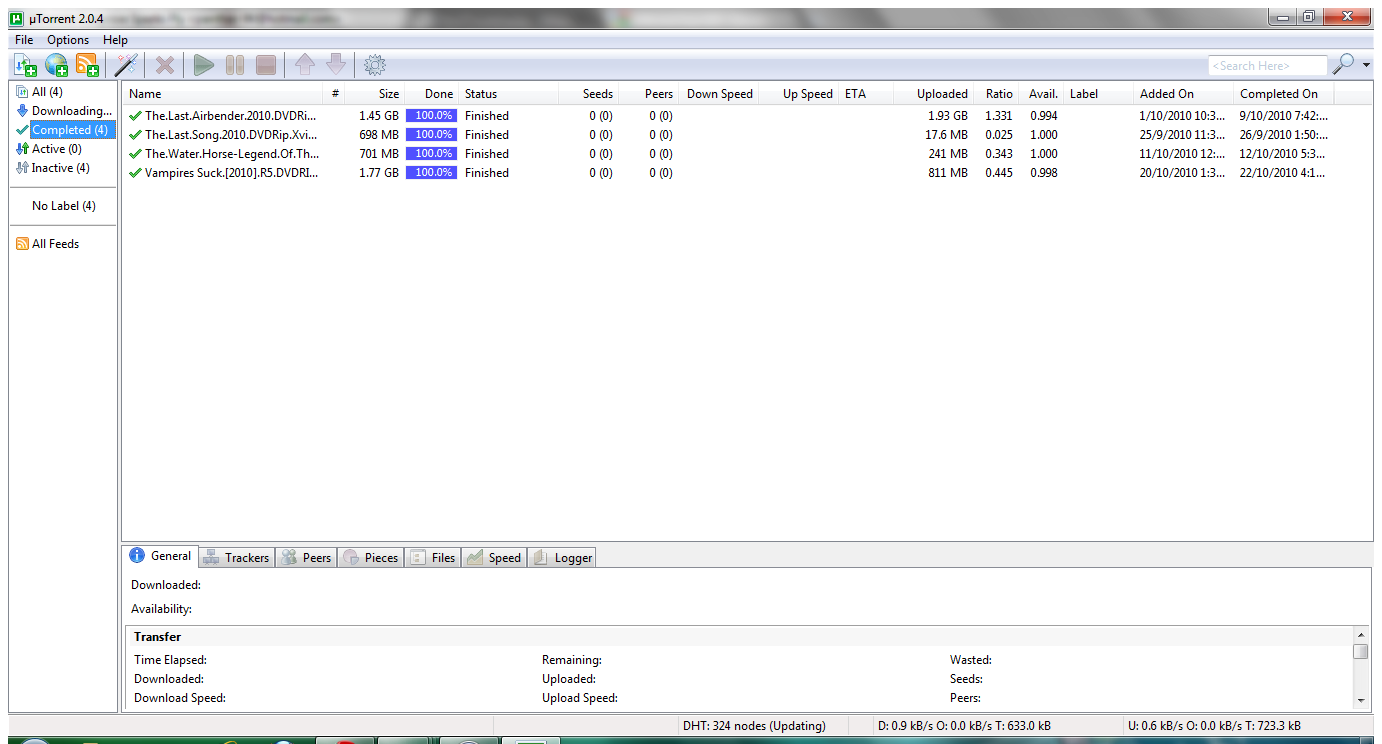}
\caption{$\upmu$Torrent Screenshot.}
\label{Figureutorrent}
\end{figure}

Due to the fact that the protocol is openly documented, numerous BitTorrent clients are available besides the official BitTorrent clients produced by BitTorrent Inc. \cite{bittorrentinc}, such as Azureus/Vuze, $\upmu$Torrent, Shareaza, BitLord, BitComet. Each client maintains its list of known active peers in a different manner. Different client parameters such as when the client decides to close an inactive connection and how the upload capacity is distributed results in some clients being more performant than others. In 2010 Iliofotou et al. conducted a large scale measurement study involving more than 11 million clients across over six thousand ISPs comparing the real world performance of the two most popular BitTorrent clients, Vuze and $\upmu$Torrent \cite{iliofotou2010comparing}. It was found that $\upmu$Torrent is on average 16\% faster than Vuze.

\section{Anti-Infringement Measures}
\label{ch3:antiinfringement}

With the popularity of acquiring copyrighted content illegally, many content producers are employing a number of technical solutions in an attempt to combat online piracy. In recent years, a number of anti-P2P companies have started to offer their services to the content producing industries.

\subsection{Attacks on Leechers}
\label{ch3:attacks}
In 2008, Dhungel et al. identified some of the techniques employed to interfere with unauthorised downloading and measured how successful attacks on BitTorrent leechers were \cite{dhungel2008measurement}. Two different attack vectors were identified:

\begin{enumerate}
\item Fake-Block Attack -- In this attack, the goal is to prolong the download time for a particular leecher. This is achieved by offering fake blocks of the desired content. While this is easily identified client-side through the hashing of the completed block, nonetheless time is wasted as the hashing can only occur once the 128kB to 4MB block is completely downloaded. The download could be further delayed if the peer decides to redownload this block, or any other block, from an attacker.
\item Uncooperative Peer Attack -- In this scenario, the attacker joins a swarm and establishes connections with as many peers as possible without ever sharing any blocks of the content. With each peer generally maintaining in the order of ten active connections at a time, taking up one or more of these valuable connections can have a significant impact on the performance of the user's download.
\end{enumerate}

\subsection{Pollution}
\label{ch3:pollution}

In 2005, Liang et al. identified that one way employed to combat unauthorised file-sharing of copyrighted content is to spam the network with large volumes of bogus or polluted files \cite{liang2005pollution}. With many P2P networks relying on a simple metadata (movie title, artist, song title, etc.) keyword search to locate desired content, polluting the P2P ecosystem with bogus files might be seen as a useful copyright infringement countermeasure to copyright holders. Due to each piece of content having as many as 50,000 different variations available, polluting the system with fake versions of the content can be a simple process merely requiring a user to rename a bogus file with the desired popular title.

\section{Forensic Process/State of the Art}
\label{ch3:p2pprocess}

\subsection{Network Crawling}
\label{ch3:crawling}

Crawling a P2P network attempts to discover each node participating in a given network. Depending on the network design, total peer enumeration may be possible, with decentralised networks generally proving easier to crawl. For example, Napster can only be crawled through the responses returned by querying the system for specified content whereas Gnutella can be crawled through the exploitation of the ping/pong peer discovery messages built into the protocol \cite{saroiu2003measuring}. Network crawling may attempt to find all the nodes sharing a specified piece of content, or attempt to enumerate the size and geolocation of the entire network.

\subsection{Deep Packet Inspection}
\label{ch3:dpi}
\begin{figure}[htb]
\centering
\includegraphics[width=0.8\textwidth]{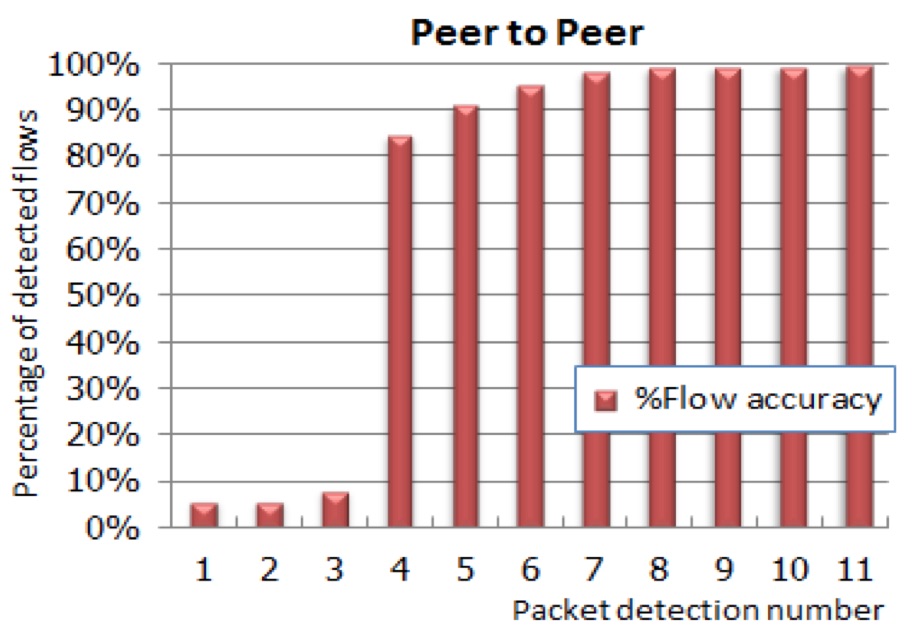}
\caption{Flow accuracy results for P2P traffic as a function of the packet detection number}
\label{Figureopendpi}
\end{figure}

Deep packet inspection attempts to classify packets as belonging to a given network. Traditionally this task was deemed relatively simple to implement as specific applications generally had a specific port number assigned to it. P2P traffic is becoming harder to identify due to port obfuscation, encryption and tunnelling. DPI has evolved to take packet flows into consideration when attempting to identify traffic. A combination of per-packet sampling and per-flow sampling is generally used to aid in identification. In 2013, Khalife et al., deployed an OpenDPI testbed in an attempt to identify encrypted P2P traffic \cite{khalife2013performance}. Incorporating packet flow analysis greatly improved the accuracy of the detection of P2P traffic, as can be seen in Figure \ref{Figureopendpi}.

\subsection{Identifying Copyrighted Content}
\label{ch3:identifyingcontent}
In 2008, Nasraoui et al. proposed a system to identify copyrighted content (movies, TV shows, ebooks, audio files, etc.) \cite{nasraoui2008node}. It was proposed that a database of known file hashes and metadata could be maintained by law enforcement agencies facilitating the identification of any content. In order for such a system to be complete, it would require cooperation from content producers and most likely a large number of volunteers in a wiki-style approach to maintenance. In this model, each audio or video file would have to be downloaded and manually verified to be a copy of the original. The hash of the verified file would then be added to the shared database.

An alternative to the per-file hashing approach has been developed by Audible Magic Corp. The company has patented an audio and video identification methodology for identifying any piece of content through an heuristic approach \cite{wold2012system}. This system creates a ``fingerprint'' of the content not based on the digital signature, but instead based on the recognition of the audio wave patterns produced by playing the file. This facilitates the identification of known content, irrespective of the codec, bitrate or metadata of the file.

\section{Forensic Counter-measures}
\label{ch3:countermeasures}

Due to privacy concerns from the monitoring of P2P file-sharing, some users attempt to circumvent detection through the use of forensic countermeasures. Many ISPs monitor their network traffic and perform throttling or ``traffic shaping" in order to curtail bandwidth hogging services, such as P2P file-sharing.

\subsection{Anonymous Proxies}
\label{ch3:anonymous}
Some users of P2P file-sharing employ anonymous proxy services, such as Tor (The Onion Router) \cite{tor} or I2P (Invisible Internet Project) \cite{i2p}. These services are distributed overlay networks designed to anonymise TCP-based applications. Each packet sent from a client operating on the network is encrypted and subsequently bounced through a random number of nodes before reaching its destination server. The response is then sent back through another random path \cite{heo2013study}. There are currently no methods available to reverse engineer a users actual IP address from the traffic originating from a Tor exit node. However, in 2012, Gilad et al. outlined a number a methods for detecting whether a given client is using Tor or not \cite{gilad2012spying}.

\subsection{Encrypted Traffic}
\label{ch3:encrypted}

The use of encryption helps to overcome some of the network forensic investigations utilising packet sniffing or deep packet inspection. An number of P2P networks employ encryption methods for all communication, e.g, \nom{SSL}{Secure Sockets Layer}. Many of the tools deployed by ISPs and network monitoring companies rely on DPI and payload heuristics to analyse network traffic and this encryption renders these approaches ineffective \cite{karagiannis2004p2p}.

\subsection{IP Blocking}
\label{ch3:blocking}

To avoid detection, some users employ custom firewall tools, such as PeerBlock, PeerGuardian, Moblock, etc., to ban any incoming or outgoing communication with specific other users \cite{6201810}. These tools accept a list of defined ``bad" IP addresses. Users can create their own lists or acquire lists of known bad IP addresses from numerous online services. One such service is iBlocklists \cite{iblocklists}, which allows users to download lists of known IP address ranges for a number of  content producing companies, governmental organisations, educational institutions, anti-piracy companies, etc. Such lists can contain over 222,000 IP address ranges and in total can cover over 796,128,149 IP addresses \cite{siganos2009monitoring}.

\section{Malware Risks on P2P Networks}
\label{ch3:p2pmalware}

\begin{figure}[htb]
\centering
\includegraphics[width=1\textwidth]{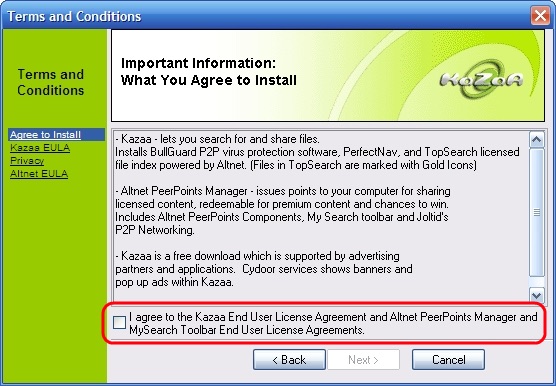}
\caption{Kazaa end user licence agreement}
\label{Figurekazaa}
\end{figure}

For almost as long as P2P networks have been used for file-sharing, they have been exploited for the propagation of viruses and malware. Keyword based metadata searching systems are most vulnerable to attack due to their simplified searching method. For example, a user may easily download and attempt to play an executable file as opposed to the music file he was attempting to download. Users can be easily tricked into downloading these files, by renaming the malware file with a desirable popular artist name, e.g., ``Katy Perry.exe''. In 2006, Shin et al. found that over 12\% of Kazaa client hosts were infected by over 40 different viruses, with 15-22\% of the total crawled data in their investigation containing viral files \cite{shin2006malware}. The Kazaa installation file also came bundled with malware. During the install process, and in order to complete the installation, the user has to agree to installing some third-party software alongside the Kazaa installation, as can be seen in Figure \ref{Figurekazaa}.

In an attempt to propagate itself, many malwares will connect to a popular P2P network and attempt to get other users to download and infect their machines. Many of these self-propagating malwares will cleverly respond with a dynamic filename based on whatever keywords the incoming request query contains \cite{kalafut2006study}.


\section{Summary and Discussion}
\label{ch3:summary}

In this chapter, P2P file-sharing systems and their differing design characteristics were introduced. Many of these popular networks are used mainly for the unauthorised distribution of copyrighted material which costs the content producing industry billions of dollars every year. Some of the methods used for investigating these networks were also introduced. These generally involve either building a bespoke network crawling tool or deploying a hardware/software network packet analysis system. 

\subsection{Weaknesses of Current Investigative Approaches}
\label{ch3:weaknesses}
If different investigative bodies, e.g., law enforcement from different countries, wish to investigate the same network, each body needs to start from scratch in the development of their own tool. As a result, from a P2P cybercrime investigation perspective, a significant amount of time is wasted globally in the duplication of developmental, investigate and analysis efforts in an attempt to reach a common goal. The universal P2P network investigation framework described in Chapter \ref{ch5designandarch} introduces solutions to some of these issues.



\chapter{Botnet Investigation}
\label{ch4botnetinvestigation}

\section{Introduction}
\label{ch4:introduction}

In the past, cyberattackers required high-end computer equipment coupled with high bandwidth Internet connections to accomplish their goals. In recent years, high bandwidth at home and workplace broadband Internet connections have become common-place. This has resulted in these computers being targeted by criminals to create large, global distributed systems, i.e., botnets, to perform their bidding. The software robots, or bots, which form these distributed systems are controlled remotely by the criminal attacker, or botmaster. The paradigm of modern botnet cybercrime involves enslaving compromised computers as a strategic criminal asset. Traditionally viruses were created with the intention to attack and destroy infected systems, but now malware has evolved to gain control of infected machines and use these machines to build a global network to perpetrate cybercrimes \cite{mielke2008botnets}.

Botnets have become the tool of choice to conduct a number of online attacks, e.g., DDoS, malware distribution, email spamming, phishing, advertisement click fraud, brute-force password attacks, etc. Criminals involved in conducting their craft online all share one common goal; not to get caught. Botnet design, as a result, has moved away from the traditional, more traceable and easily blocked client/server paradigm towards a decentralised P2P based communication system. P2P Internet communication technologies lend themselves well to use in the world of botnet propagation and control due to the level of anonymity they award the botmaster. For the cybercrime investigator, identifying the perpetrator of these P2P controlled crimes has become significantly more difficult. This chapter outlines the state-of-the-art in P2P botnet investigation.

\begin{figure}
\centering
\includegraphics[trim = 0mm 0mm 0mm 0mm, clip, width=0.8\textwidth]{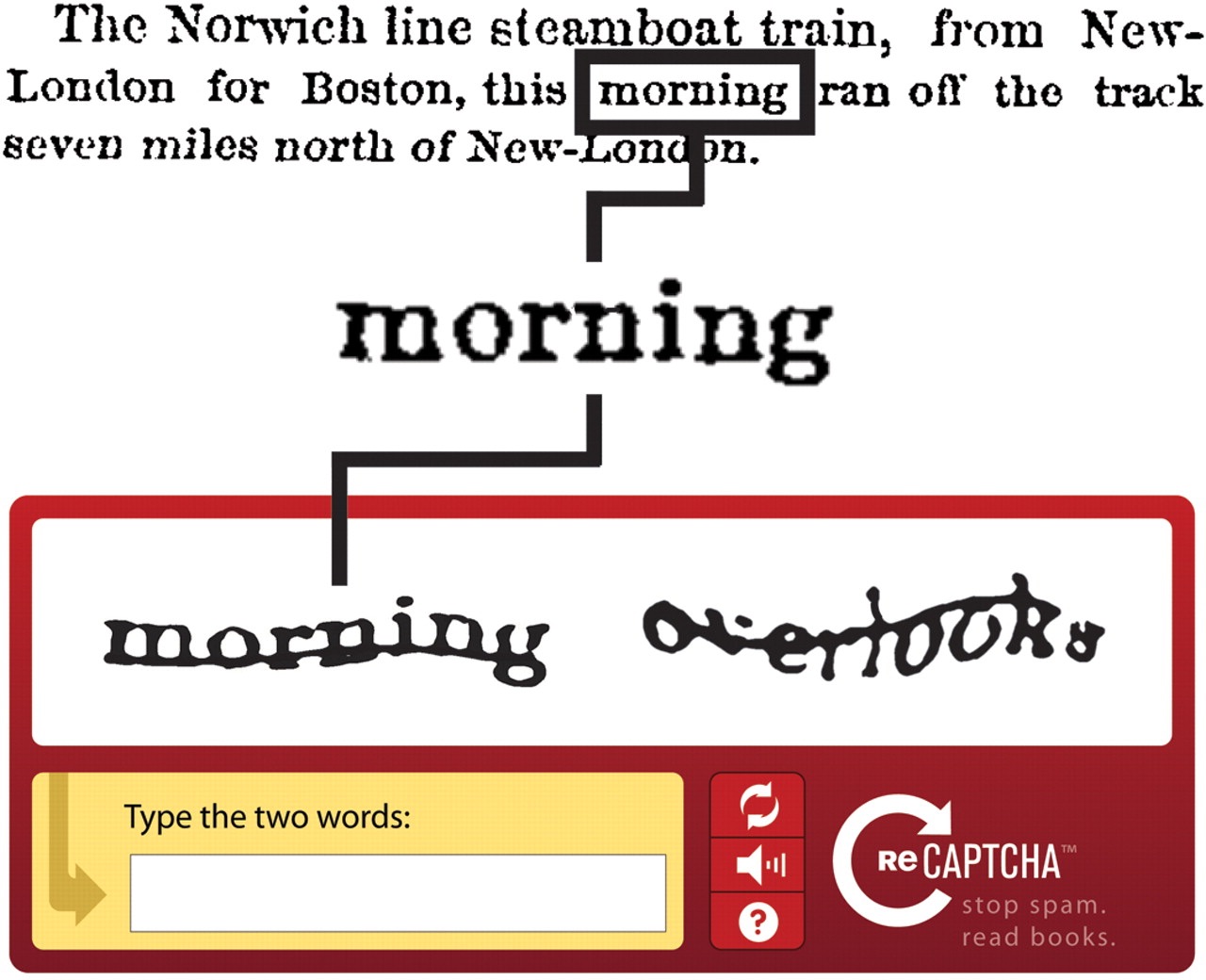}
\caption{Sample CAPTCHA from the reCAPTCHA online service and its automated book scanning source text}
\label{figurecaptcha}
\end{figure}

The prevalence of large, global botnets has resulted in many online services deploying human authentication systems to reduce or eliminate automated registration for forums, email accounts, social networks, etc. These systems aim to prevent automated botnet login attempts resulting in password cracking or spam. The most common test for telling whether any given visitor to a website is human or machine is to employ ``Completely Automated Public Turing test to tell Computers and Humans Apart'' (\nom{CAPTCHA}{Completely Automated Public Turing test to tell Computers and Humans Apart}) verification. The test involves presenting the user with a string of obfuscated characters and requires the user to identify the often scrambled words in order to proceed. Figure \ref{figurecaptcha} shows a sample word taken from a text with unreliable results from regular optical character recognition algorithms \cite{vonAhn2008recaptcha}. It has now become common to use these scanned words which are difficult automated recognition to verify human website usage. The human delivered results from these CAPTCHA systems can feed directly into text scanning and recognition projects, such as Google Books.


\section{Botnet Architectures}
\label{ch4:botnetarch}

\begin{figure}
\centering
\includegraphics[width=1\textwidth]{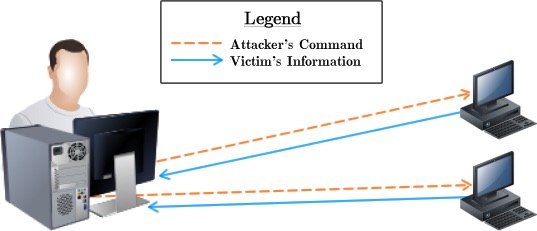}
\caption{Simple Trojan Horse Architecture Controlling Multiple Computers}
\label{FigureSimpleBotnet}
\end{figure}

Arguably, the simplest implementation of botnet technology involves an attacker manually taking control of each victim's computer using a remote trojan horse based attack, as shown in Figure \ref{FigureSimpleBotnet} \cite{schneier1999inside}. In this model, the attacker had one-to-one direct control of the victim's machine. The target of the attack would generally be the user of the infected machine as the attacker is capable of capturing keystrokes, executing any applications, intercepting print jobs, accessing local or network files and destroying the victim's operating system or system configuration. From an anonymity standpoint, this design has significant privacy issues for the attacker. Any investigation of the trojan's network traffic would easily reveal the attacker's IP address, which could subsequently be resolved back to reveal his identity.

\begin{figure}[t]
\centering
\includegraphics[trim = 0mm 0mm 0mm 0mm, clip, width=1\textwidth]{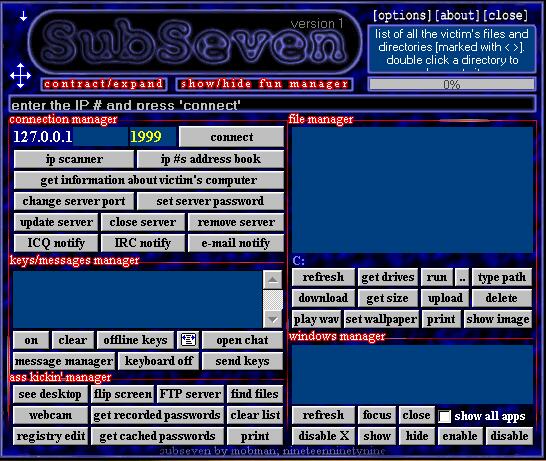}
\caption{Subseven Control Panel}
\label{Figuresub7}
\end{figure}

Subseven was one of the most popular trojan horse systems (alternatively referred to as Sub7 or Sub7Server). It consisted of a client trojan horse virus and a server or control panel to facilitate the attack. Released in 1999, it targeted vulnerable Windows machines \cite{abimbola2002subseven}. The types of commands the attacker could execute can be seen in Figure \ref{Figuresub7}.

\subsection{Client/Server Botnet Design}
\label{ch4:candc}

\begin{figure}[t]
\centering
\includegraphics[trim = 0mm 0mm 0mm 0mm, clip, width=1\textwidth]{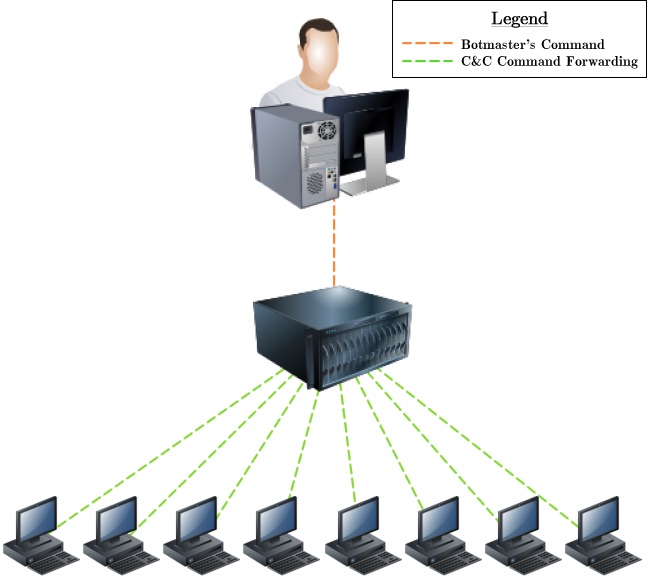}
\caption{Command and Control Server Botnet Network Architecture}
\label{Figurecandc1}
\end{figure}

The simplest botnet architecture relies on a central server to control all nodes on the network, often referred to as a centralised design. When a bot comes online, it registers its availability with the server, which in turn issues the bot with commands for the work it must complete. These Command and Control (\nom{C\&C}{Command and Control}) servers are directly controlled by the botmaster. The architecture of this C\&C based system contains a single controlling server with multiple compromised machines communicating with it, as outlined in Figure \ref{Figurecandc1}.

Traditional botnet design was centred on a client/server paradigm (see Figure \ref{Figurecandc1}). Using this model, the botmaster issues requests to the \nom{HTTP}{HyperText Transfer Protocol} (regular website based communication) or Internet Relay Chat (\nom{IRC}{Internet Relay Chat}) based C\&C server. In the case of IRC, each connecting bot will pick a randomised nickname for the chatroom and as a result, if necessary, the server has the ability to issue unique commands to each bot. The use of a C\&C server eliminates the need for the botmaster's computer to remain online in order to distribute the latest orders to the entire botnet. C\&C servers also award the botmaster an added level of anonymity from detection.

An advantage to using the HTTP based design over the IRC design, is that the bots themselves do not need to be continuously connected to the server \cite{jaiswal2013botnet}. Instead, the client-side HTTP bot software, which runs on the infected nodes, is programmed to periodically "check-in" with the C\&C server in order to get its latest commands. In the IRC based model, the bots remain connected to the IRC channel while online. This leaves the system vulnerable to IRC based investigation and manipulation by law enforcement or other forensic investigators.

The main concern with this simple client/server model is that it leaves the botnet vulnerable to a single point of failure. To counteract this, multiple C\&C servers may be used optionally in conjunction with a dynamic Domain Name System (\nom{DNS}{Domain Name System}) service, such as DynDNS \cite{dyndns} or No IP \cite{noip}. The dynamic hostnames required are hard-coded into the bot software, enabling the botmaster to quickly and easily swap in a new command and control server when needed. This is achieved by simply updating the IP addresses associated with the dynamic DNS provider ensuring no disruption of the botnet's regular operation. Cloud services lend themselves well to being exploited for running C\&C servers and offer the botmaster the ability to quickly and easily change not only the IP address, but the geographic location of the C\&C servers frequently. In 2009, Amazon discovered that its Elastic Compute Cloud (EC2) was being used by a new version of the Zeus botnet for its C\&C functionalities \cite{choo2010cloud}.

\begin{figure}
\centering
\includegraphics[trim = 0mm 0mm 0mm 0mm, clip, width=1\textwidth]{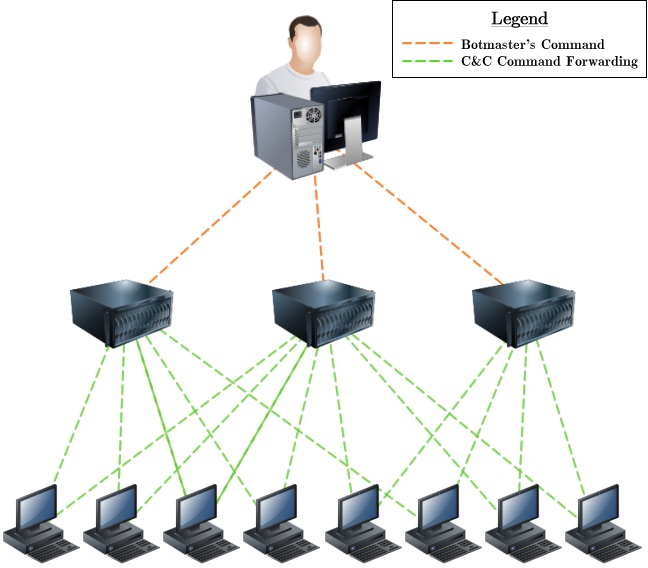}
\caption{Evolution of botnet architecture to eliminate single point of failure}
\label{Figurecandc2}
\end{figure}

The weak point of the original HTTP/IRC based command and control centre botnet design, as it can be seen in Figure \ref{Figurecandc1}, is that there is a single point of failure. If law enforcement or any other third party wished to destroy the botnet, the command and control server can be targeted and the botnet can be effectively destroyed, i.e., left without any commands or work to complete. The next evolution of botnet design incorporated multiple C\&C servers to attempt to alleviate the strain and weaknesses of a single server design, as it can be seen in Figure \ref{Figurecandc2}. In this model, each bot in the system will register and check-in with multiple C\&C servers.

\subsection{P2P Design}
\label{ch4:p2parch}

\begin{figure}
\centering
\includegraphics[trim = 0mm 0mm 0mm 0mm, clip, width=1\textwidth]{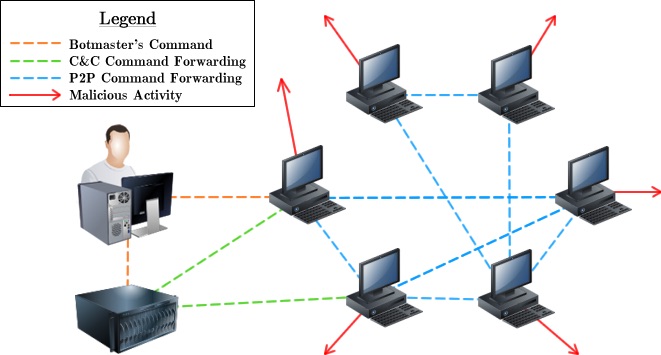}
\caption{Typical botnet topology with commands optionally routed through a C\&C server.}
\label{Figuremaliciousdecentralised}
\end{figure}

The design of botnet architecture has continuously evolved over the last number of years to improve the performance and attack resilience of the system, while awarding a greater degree of anonymity to the botmaster. A natural progression of the client/server design involved expansion to incorporate as many C\&C servers as possible. Further extension of this model involves utilising P2P technologies in the architecture of the systems. This effectively turns each bot in the network into a C\&C server. This results in every active node in the network having the ability to communicate with a subset of all the other nodes in the same botnet. Commands and updates are passed from peer to peer in a manner that eliminates the need for controlling servers. As a result, the single point of failure of the client/server botnet design is effectively eliminated \cite{gu}.

Decentralised botnet design relies on a DHT to record all the active nodes on the network \cite{scanlon2012peer}. In order for any new bot to participate in the network, it must have an avenue available to it to initially connect to this DHT. This is generally done through the hardcoding of an initial ``seed'' list or a list of bootstrapping servers. The seed list is a local cache of IP addresses which are more likely active nodes in the network. A node attempting to join the network can contact either a seed or one of the bootstrapping servers to connect to the DHT and begin regular operations. Without some initial hardcoded bootstrapping method to connect to the DHT, the only other completely decentralised option available to a P2P system is to employ random address probing . While this may initially appear an unlikely method, recall that discovering a single node connected to the DHT is sufficient to join the system and some of the DHTs involved can contain millions of active nodes. In practice, Dinger et al. \cite{dinger2009decentralized} found that limiting the randomised scan rate to 100 packets per second resulted in locating a BitTorrent DHT peer within ten minutes with a probability of $\ge$94\%.

Botnet developers are continuously updating and improving botnet design. In 2013, Memon et al. proposed a new botnet system, named Tsunami, which attempts to eliminate the bespoke bootstrapping server weakness of decentralised P2P botnet design by parasitically transmitting all botnet communications on the existing global Kad network \cite{memontsunami}. The Kad network is a DHT based P2P network with over 4 million users most commonly used by numerous P2P file-sharing applications, such as eMule and MLDonkey. Commands are sent from the botmaster to any active bots using the ``lookup'' messages in Kad. This facilitated a hidden payload instruction up to 106 bits in length. The lookup command in Kad is automatically passed from node-to-node in a similar method to that of querying the Gnutella P2P file-sharing network, as described in Section \ref{ch3:gnutella}. It was found that Tsunami could reach 75\% of its bots within 4 minutes and receive responses back from 99\% of these bots.

\subsection{Hybrid Design}
\label{ch4:p2phybrid}

A hybrid botnet topology builds upon the P2P design by promoting particular high uptime and high bandwidth nodes to ``supernode'' status, in a similar fashion to Gnutella outlined in Section \ref{ch3:gnutella}. This results in a two tiered model with client bots and servant bots \cite{jang}. The servant bots are the only ones receiving the commands directly from the botmaster and are responsible both for spreading those orders throughout the network and are responsible for the maintenance of the network itself \cite{dittrich}.

\section{Botnet Lifecycle}
\label{ch4:lifecycle}

For most botmasters, the botnet lifecycle starts with the configuration of the botnet client (infecting malware to run on victims' machines) and the botnet controlling server (responsible for the dissemination of the latest updates and commands). There are numerous software solutions available to criminals wishing to create their own botnet, requiring varying degrees of technical knowledge and costs:

\begin{enumerate}
\item Buy or rent an existing botnet -- There are numerous avenues available to the criminal to buy or rent partial or entire botnets. Botnets are rented or sold for differing prices depending on the associated ``quality'', i.e., size, node uptime, bandwidth, latency, geolocation, etc. This option requires minimal technical knowledge and the perpetration of the crime can commence almost instantaneously. This model shares many of the characteristics as renting cloud computing resources from legitimate providers. Minimal upfront time and money is required on behalf of the botmaster to get started.

\item Buy pre-developed botnet software -- This option will supply the criminal with developed software. This software is configured to spread the client malware via P2P networks, email attachments, instant messaging, etc. The cost of purchasing such a system increases with the decreased likelihood that the client executable will be discovered by anti-virus software and the broader the operating system compatibility. Choosing this option will allow the criminal to create numerous different botnets if desired, based on the purchased technology. Some technical knowledge and time is required in order to spread and infect the desired number of nodes. A significant downside to this option is that the same software will likely be sold to numerous criminals who may each create several botnets. The more prevalent the software is, the higher the likelihood that the malware will be detected or reported to anti-virus/malware detection providers. In this instance, all botnets based on this software may be rendered useless. However, in order for this to occur, all victims must update their virus definitions and remove the infection. 

\item Develop customised bespoke botnet software -- This is the most resilient option against detection, but obviously requires the highest amount of technical ability. Before any development can commence, a vulnerability must first be discovered in a popular operating system, browser, e-mail client or other common piece of software or hardware. This vulnerability facilitates the infection of the victims' machines, as outlined in greater detail in Section \ref{ch4:infection}. Assuming a functional botnet client can be developed, the botmaster still has the same task as described above in spreading the infection to as many victims as required.
\end{enumerate}

Once the botmaster has the required software (often a PHP/MySQL driven back-end installed on a command and control server), the configuration of the infecting malware must take place. Each newly infected machine must accept new orders, update the bot client software, send information back to the botmaster and potentially distribute orders with other known nodes in the P2P system. This configuration file will contain settings and required operational information, such as trusted C\&C servers, update servers, DHT information, communication frequencies, resource usage limits, etc., \cite{nazario2007blackenergy}. In a purchased or rented system, this configuration step will likely have been completed by the seller.

To ensure maximum flexibility, any hardcoded host information, such as the list of trusted C\&C servers, will generally be included using a dynamic hostname supplied by dynamic DNS providers such as DynDNS \cite{dyndns}. This allows the botmaster to regularly shutdown or move the C\&C servers without needing to update any of the bots on the network. To achieve this move, the botmaster merely changes the IP address associated with each dynamic domain name supplied in the configuration file.

\begin{figure}[h]
\centering
\includegraphics[trim = 0mm 0mm 0mm 0mm, clip, width=1\textwidth]{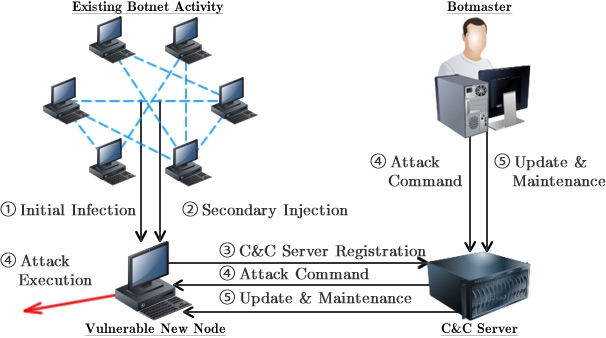}
\caption{Typical Botnet Lifecycle from a Victim's Point-of-View}
\label{Figurelifecycle}
\end{figure}

A typical botnet deployment lifecycle from a new vulnerable host's point-of-view can be seen in Figure \ref{Figurelifecycle} \cite{feily2009survey}. Each of the indicated steps are outlined in greater detail in Sections \ref{ch4:infection} to \ref{ch4:maintenance}. Most botnets aim to achieve, from the point-of-view of the users of the effected machines, no discernible performance reduction to the regular expected operational speed. The bot's client can be configured to intelligently use only available resources and as a result, potentially never get discovered. A user with what appears to be a fully functional computer may never decide to install or run anti-virus software.

\subsection{Spreading and Infection Phase}
\label{ch4:infection}
In order for a botmaster to execute his desired acts, the bot malware client must be running on a significant number of infected machines. In order for a machine to get infected with the malware, either some user interaction is required or a software vulnerability is exploited in order for the binary executable to run \cite{zhang2013conversation}.

\begin{figure}[h]
\centering
\includegraphics[trim = 0mm 0mm 0mm 0mm, clip, width=1\textwidth]{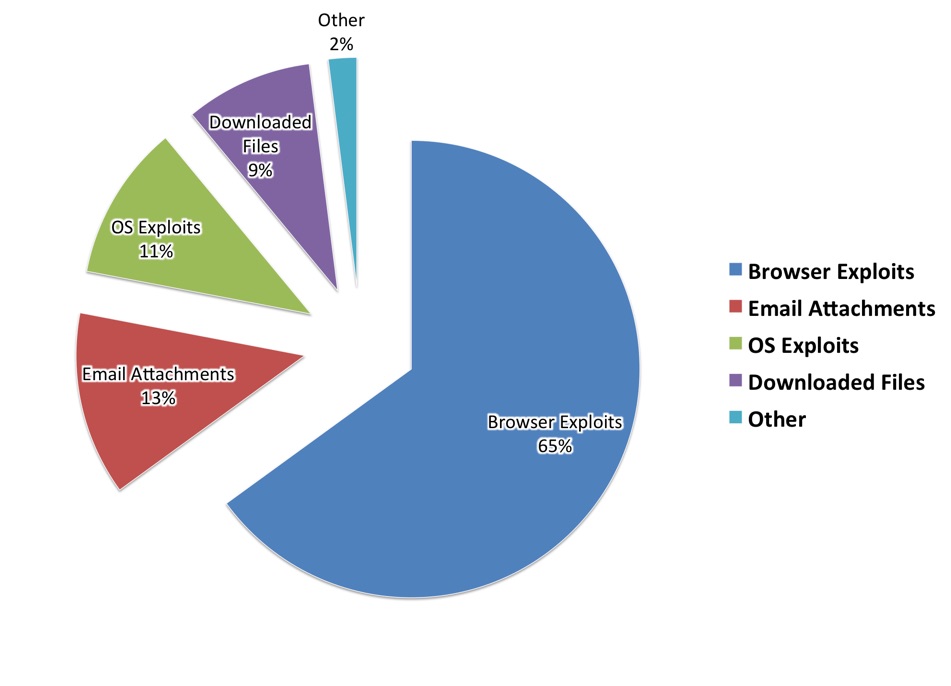}
\caption{Typical Malware Attack Vectors}
\label{Figureexploits}
\end{figure}

The infection or ``recruitment" phase is referred to as the phase in which a clean host is infected by a bot binary and, as a result, becomes a member of the botnet \cite{zhangthesis}. This generally involves some malicious executable compromising a host through any available means, e.g., taking advantage of a software or hardware exploit through social engineering, instant messaging, unprotected network shares, malicious email attachments or the mimicking of desirable content on download sites or on P2P file-sharing networks \cite{dittrich2008discovery}. In 2007, the European Network and Information Security Agency (\nom{ENISA}{European Network and Information Security Agency}) conducted research into the infection vectors of botnets and found that the most common infection method was through browser exploits \cite{barroso2007botnets}. The complete results of their findings are shown in Figure \ref{Figureexploits}. 

Once a new machine is infected, many bots attempt to self-propagate by sending emails to the victim's address book, instant message contacts, etc. Some bots will also connect to popular P2P file-sharing systems in an attempt to dupe users into downloading the infection.

\subsection{Secondary Code Injection Phase}
\label{ch4:secondaryinjection}

Many of the attack vectors for the malware involve a memory buffer overflow exploit in software. Due to the nature of the attack, it is common that the initial ``break-in'' or ``dropper'' binary might not contain the entire bot client. It may merely serve the function of gaining the required operating system access rights to facilitate the install of the client software \cite{langner2011stuxnet}. During this phase, the binary will download and install the latest version of the botnet client onto the infected machine. Seeing as a common method for malware distribution is to email the malware as an executable or to entice a download from a P2P file-sharing network, it is also common for the version of the malware distributed to be out of date. For example, the default behaviour of most P2P file-sharing systems is to automatically share all downloaded files. As a result of this potentially outdated software, it is also during this secondary phase that the bot software will update itself to the latest version from the C\&C server. 

\subsection{Command and Control Phase}
\label{ch4:commandandcontrolphase}

It is during the C\&C phase that a newly infected computer will become part of the botnet. A traditional client-server bot, once installed on a new machine, will immediately attempt to ``phone home'' through an IRC network or contacting a HTTP C\&C server. Decentralised P2P bots are distributed with a predefined bootstrapping method to connect to the relevant DHT. Once connected, the newly compromised machine will ask one of its peers for the latest command. Some of the P2P bots require that a specific port is open for the peers to be able to communicate with each other \cite{schoof2007detecting}. Through the deployment of a firewall, many of the unnecessarily open ports on any given machine will be blocked. Any new application that attempts to access the network for any reason can also be flagged to the user, e.g., immediately after a recent infection of the botnet malware. Newer versions of the bot client software will not use a predefined port number to aid in avoiding detection.

There are two different ways to spread a command in a botnet system, namely push and pull. IRC based bots belong to the push-based category as they sit in an IRC chat room listening for a new command. HTTP based bots periodically check with the server to verify if there is any new work. P2P bots can do both as they send and receive commands to/from other bots.


\subsection{Attack Phase}
\label{ch4:attackphase}

The attack phase is the most important in the botnet lifecycle from the botmaster's perspective. The purposes of designing, developing and spreading the botnet client malware onto as many machines as possible is to conduct whatever distributed illegal activities the botmaster has in mind. When a bot receives a command, so long as it remains online it will execute that command until one of the following two events occur:

\begin{enumerate}
\item A predefined stopping condition is met -- This condition may be when a specified execution time has elapsed or when the objective goal is accomplished, e.g., the taking down of a website or service, or a password having been cracked, etc.
\item A new order is received -- Any new order received will overwrite the current operation. Modern botnet design facilitates the execution of multiple orders simultaneously and each job may need to be manually ordered to cease.
\end{enumerate}

Examples of the types of attacks conducted by botmasters and real-world monetary rewards for each are outlined in greater detail in Section \ref{ch4:attacks}.

\subsection{Update and Maintenance Phase}
\label{ch4:maintenance}

\begin{figure}
\centering
\includegraphics[width=1\textwidth]{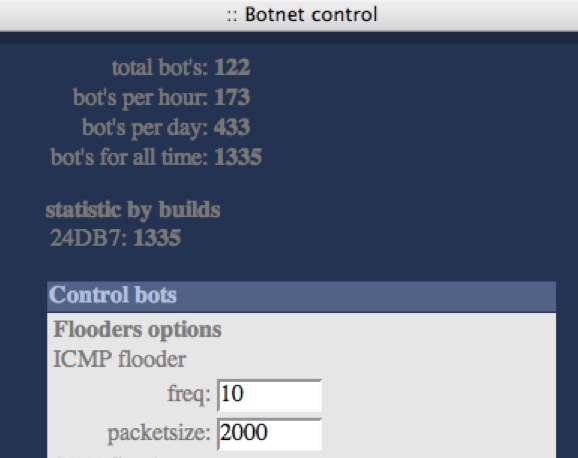}
\caption{Screenshot from the Blackenergy Botnet C\&C Server}
\label{Figureactivenodes}
\end{figure}

Botmasters have the ability to issue update commands to their entire system of slave machines. Due to the interest in botnet detection and investigation, the attackers need to have a facility to upgrade their tools. This stage allows the botmaster to update the existing binaries and/or configuration to make the entire system more resilient to new digital forensic techniques. 

The maintenance phase of the botnet lifecycle involves feeding execution, uptime and update information back to the botmaster. Figure \ref{Figureactivenodes} shows some of the information available to a botmaster including the number of currently active nodes, the churn rate per hour or day and the total number of detected bots \cite{nazario2007blackenergy}.

\section{Underground Economy}
\label{ch4:economy}

Given that financial gain is perhaps the biggest driving force in the growth of botnet technologies and volume of attacks, it was a natural progression that a large underground botnet enabled economy would develop. Numerous vendors, merchants, malware authors and customers are involved in the daily trade of personal information, attacks, spamming services and botnet technologies. Among the network of cybercriminals, a vibrant market has emerged trading in compromised credit cards and financial information. Much of this collected information would have been harvested by botnets. Those botmasters who target the collection of these financial details are usually unable to extract the funds directly from the accounts, so they usually sell them at a fraction of their value to experienced criminals or organisations who have a greater infrastructure available to them \cite{mielke2008botnets}.

Much research has been conducted into how this large underground market of trading criminal tools and technologies can be stopped, or at least hindered. Specifically to the trade of botnets, one method proposed to stem the profitability of this economy was proposed by Li et al. in 2009 \cite{li2009botnet}. This method suggests that by introducing virtual bots into the system, an uncertainty level in the performance of the network is introduced, e.g., a botmaster needs a specific amount of active nodes to perform a DDoS attack on a server and if many of the nodes currently active in the botnet are fake, the goal cannot be accomplished. This makes the task of achieving the optimal botnet attack size infeasible for botnet operators and will ultimately effect their profitability. 

\subsection{Valuation}
\label{ch4:valuation}

A significant underground economy of selling, trading or renting botnets has developed in recent years. A botnet with 10,000 infected machines can fetch approximately \$300-\$800, depending on the geolocation of the nodes and the quality of the nodes. This quality is determined by the nodes being infected solely by a single botnet client and the nodes' uptime and internet connection speeds. Botnets with infected nodes based in the United States are the most valuable at \$125 per 1,000, with European based botnets valued at \$35 per 1,000 and Asian botnets valued at just \$13 per 1,000 \cite{stone2011underground}. In 2010, Danchev found that the average price for renting a botnet is \$67 for 24 hours and \$9 for hourly access \cite{danchev2010study}. Often more money can be made through the renting of a botnet to multiple customers concurrently, referred to as ``Botnet as a Service (BaaS)''  \cite{stone2009your}. 

\subsection{Spamming}
\label{ch4:spamming}

In 2010, it was estimated that over 89\% of all emails sent were spam, resulting in over 262 billion spam emails being sent per day. In 2013, dealing with the volume of spam will cost over \$338 billion in network bandwidth and infrastructure costs in 2013. The majority of spam originates from botnets and it is estimated that 80-85\% of all spam is produced by 6-10 huge botnets \cite{hameed2012leveraging}. Almost all of this spam is illegally distributed under current laws in North America and Europe. Rao et al. estimate that the sending of spam is a \$200 million per year business for the botmasters \cite{rao2012economics}.

In 2011, Stone-Gross et al. analysed a popular underground web-based forum known as ``Spamdot.biz'' \cite{stone2011underground}. The forum required significant social engineering to gain access, with the authors requiring a reference from at least three existing members of the forum before they were granted access. It was found that this forum was used by almost 2,000 users to advertise their spamming services and to buy/sell information. E-mail address lists were worth between \$25 and \$50 per one million, contingent on the geolocation of the users and the proportion of addresses belonging to email providers with stronger spam filters, e.g., Gmail, Hotmail or Yahoo.

Kanich et al. found in one specific instance that a major spam campaign involving the sending of almost 350 million emails using the Storm botnet, only made \$2,731.88 in revenue for the advertiser \cite{kanich2008spamalytics}. The campaign required over 75,000 active bots in the network to send the emails and resulted in 28 purchases from the associated online pharmacy website, or a conversion rate of just 0.0000081\%. All but one of these purchases were for male pharmaceutical products, such as Viagra and Cialis, and the average purchase price was close to \$100. The authors continue to estimate that the cost of such a campaign would be in the order of \$25,000 and as a result, speculate that the botmasters of the Storm botnet may be the purveyors of the pharmacies advertised.

An obvious approach for ISPs and email providers to blocking spam emails is to refuse communication, or blacklist, known IP addresses that are found to be sending high volumes of messages. In the traditional spamming model, whereby a spammer hires a server or number of servers to send out the emails, this approach can prove very effective. However, blacklisting is not an efficient approach for blocking spam originating from infected nodes in a botnet. Any single node in the botnet is most likely a regular home or business user who is likely to get assigned a new IP address regularly, due to the common ISP practice of DHCP based IP address allocation. As a result, blocking every IP address that is found guilty of sending spam will result in the email service provider ultimately blocking a high number of legitimate users (who innocently may have been assigned an IP address previously used by an infected bot).

\subsection{Phishing}
\label{ch4:phishing}

Phishing generally involves an attempt to trick the user of an infected machine to enter personal or confidential information through faux webpages \cite{jaiswal2013botnet}. These often convincing data harvesting webpages are injected into the user's regular browsing habits, popped up in their browser or OS or sent as links in emails. Phishing attacks are also commonly distributed by email spam. Webpages emulating those of a bank, online payment provider or lottery are common themes used in email based phishing attacks to entice the unsuspecting user into parting with valuable personal information.

Links have been found between distributed online phishing attacks collecting credit card information and the funding of terrorism. In one example, three men were arrested in the United Kingdom in 2008 and were found guilty of funding the terrorist organisation, al-Qaeda \cite{mielke2008botnets}. The trio were found to be in the possession of over 37,000 stolen credit card numbers, along with associated personal information from victims. They had made over \$3.5 million in fraudulent charges and had purchased over 250 airline tickets.

\subsection{Scamming the Scammers}
\label{ch4:scamming}

In 2010, Herley et al. documented that a large proportion of the underground economy has evolved offering bogus botnet software, email addresses, botnets, etc., to unsuspecting criminals \cite{herley2010nobody}. The presence of these scammers ultimately represents a tax on every ``honest'' transaction, where neither party might be familiar with the other. It was also found that a two-tier underground economy now exists. The top tier consists of elite cybercriminals where their organisation, alliances and trust is established. At this tier, transactions take place between known or ``reputable'' criminals. The lower tier, generally conducted on IRC marketplaces, is occupied by criminal newcomers without any experiential skills or alliances and are easily cheated out of their money by fake sellers or ``rippers'', who have no intention of providing the goods or services offered.

\section{Botnet Powered Attacks}
\label{ch4:attacks}

As proven by the Anonymous ``hacktivist'' attacks in recent years, innocent victims' infected machines do not solely contribute to the the distributed power of botnet attacks. Regular Internet users with shared political or activist views can voluntarily decide to contribute their processing power to a collaborative cause. Partaking in the Anonymous attacks involved users downloading and configuring an open source network stress testing tool called ``Low Orbit Ion Cannon'' (\nom{LOIC}{Low Orbit Ion Cannon}). Alternatively, users can donate their computational power via a JavaScript-based version facilitating anyone who visits the site to participate in the attack \cite{mansfield2011anonymous}. Regular P2P file-sharing networks, such as BitTorrent, can also be manipulated by malicious users to aid in the execution of a DDoS attack through the exploitation of vulnerabilities in the protocol and operation of the network \cite{el2007bottorrent}. In a cyberwarfare scenario, it is conceivable that citizens of countries with limited computational infrastructure or supporters of terrorist organisations could similarly be called upon to donate their systems to aid in a collaborated attack on an enemy's infrastructure.

The potential for attacks originating from a growing number of sources is a concern for the security of many nations across the globe. In 2012, Amoroso et. al defined five possible motivations behind cyberattacks \cite{amoroso2012cyber}:

\begin{enumerate}
\item Country-sponsored warfare -- This is whereby national infrastructure is attacked by enemy cyber forces in an attempt to disable critical resources of the opposing country in a similar manner to traditional physical warfare. The intensity of this attack is only limited by the resources and devotion of the attacking nation. In a P2P botnet facilitated attack, citizens could voluntarily donate their computing power towards the national goals in a similar manner to the Anonymous attacks outlined above.
\item Terrorist Attack -- Groups driven by terrorist motivations could quickly gain sufficient funding and expertise to conduct their attacks. Also in this scenario, regular Internet users could partake in a terrorist operation without requiring any skill or expertise by donating their regular computer equipment to the attack.
\item Commercially motivated attack -- Competing companies might target their competitors' e-commerce infrastructure in order to prevent regular users from purchasing anything from their online stores. A popular e-commerce site being taken offline has the added effect of harming the victim company's reputation in their customers' eyes.
\item Financially driven criminal attack -- These types of attacks could target individual computer users by recording their Internet banking details, online payment services and other financial services. Companies can also be targeted with extortion threats against their online infrastructure.
\item Hacking -- This scenario generally involves an individual or group of hackers attacking targets motivated by little more than mischief or attaining online recognition of their achievements.
\end{enumerate}

A number of popular botnet powered attacks are outlined in the following subsections:

\subsection{Infection}
\label{ch4:victiminfection}

Often botnet developers integrate some method of self-propagation into their design. Bots can scan for and infect more vulnerable computers with network or browser based exploits. Increasingly, web-based infection mechanisms have been observed online, whereby drone machines infect legitimate websites with a drive-by exploit and consequently, visitors to those websites can become infected \cite{mielke2008botnets}. These website based attacks are often targeted towards popular websites to try to infect as many web users as possible.

\begin{figure}[htb]
\centering
\includegraphics[width=1\textwidth]{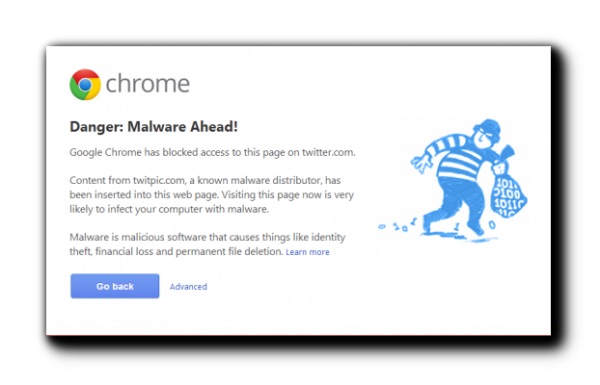}
\caption{Google Chrome Malware Warning}
\label{Figurechromemalware}
\end{figure}

The prevalence of web based malware distribution has resulted in most of the major web browsers implementing malware detection to aid in the protection of their users. A sample browser warning is shown in Figure \ref{Figurechromemalware}.

\subsection{Distributed Denial of Service Attacks (DDoS)}
\label{ch4:ddos}

Distributed Denial of Service attacks are attempts to overload or monopolise a machine or web service resulting in the resource becoming unavailable for its intended purpose. The first DDoS attack occurred in the summer of 1999 and ranged from several hundred to more than two thousand computers \cite{dittrich2010case}. This type of attack can result in significant financial losses for their targets, as their web services are rendered useless for the duration of the attack. While many DDoS attacks result in the target being effectively offline for a certain period of time, ``kinetic world'' cyberattacks also exist through firmware code injection into physical hardware resulting in destroyed routers, firewalls, motherboards, etc.

The first DDoS attacks were motivated by petty online fights on IRC channels but soon shifted to monetary motivations with extortion of online gambling, e-commerce and pornography websites in 2003. These attacks evolved into politically motivated attacks against national infrastructures in 2007 \cite{dittrich2010case}. 

In December 2010, the Anonymous ``hacktivist'' group launched ``Operation Payback'' in response to the controversy of the attempted take-down of the whistle-blowing website, Wikileaks \cite{pras2010attacks}. The operation involved a coordinated DDoS attack on the financial organisations (e.g. MasterCard, Visa, Paypal), Wikileaks's DNS provider and political and legal websites involved in the controversy. LOIC is the network stress testing application that was used by Anonymous to accomplish its DDoS attacks. LOIC was utilised as a botnet whereby individual users download the client application and voluntarily contribute their computing resources to the collective attack of targets instigated by Anonymous.
\\
\begin{lstlisting}[caption=Example Low Orbit Ion Cannon Instruction, label=snip:anon]
!lazor default targethost=www.moneybookers.com subsite=/ speed=3 
threads=15 method=tcp wait=false random=true checked=false 
message=Sweet_dreams_from_AnonOPs port=80 start
\end{lstlisting}

One example attack command sent by Anonymous to the LOIC bots targeting the online payment provider Moneybookers.com and the included parameters can be seen in Snippet \ref{snip:anon} \cite{mansfield2011anonymous}. This particular attack was focused on overloading the Moneybookers.com web server running on TCP port 80, with each bot running 15 concurrent threads to the server and repeating the request every 3 seconds without waiting for any reply from the server.

\begin{figure}[htb]
\centering
\includegraphics[width=\textwidth]{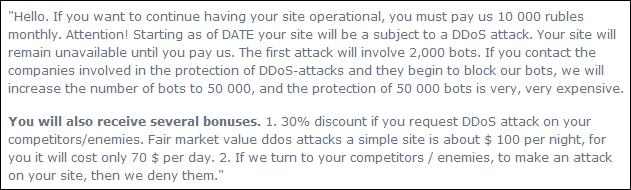}
\caption{DDoS Extortion Example}
\label{Figureddosransom}
\end{figure}

Botnet driven DDoS attacks are also commonly used for extortion. In this scenario, a website or service is sent a threatening demand and must pay extortion or else be faced with their website being taken offline. A sample extortion threat is shown in Figure \ref{Figureddosransom} \cite{danchev2010study}.

\subsection{Espionage}
\label{ch4:espionage}

An infected machine is in complete control of the botmaster and often the intended targets of the criminal activity are the home or corporate owners of the compromised computer. Local documents, passwords, keystrokes, financial and personal information are all desirable material that is capable of being sent back to the controller of the bot. In a similar vein to the original trojan horse viruses described above, desktop screenshots and webcam snapshots have been observed getting relayed to the botmaster \cite{mielke2008botnets}.

\subsection{Proxies}
\label{ch4:proxies}

Many botnet systems have proxy capabilities which effectively let the botmaster tunnel their internet traffic through one or more of the zombie machines. This can help facilitate anonymity for the botmaster and aid in the evasion of capture from the authorities in a similar vein to anonymous proxies, as outlined in Section \ref{ch3:anonymous}.

\subsection{Clickthrough Fraud}
\label{ch4:clickthrough}

``Clickthrough fraud'', often referred to as just ``click fraud'', involves automatically gathering hits, advertising impressions and advertising clicks on specific websites operated by the cybercriminal. As each bot on the P2P network has a unique IP address, each infected machine appears to the advertising network provider (e.g. Google Adwords) as though it is any other unique visitor to the website. Generally a subset of the entire botnet will ``choose'' to click on the advertising available on the site, emulating regular visitor usage in an attempt to avoid detection. Advertisers are forced to trust that the advertising engine providers detect and prevent clickthrough fraud even though the engines still get paid for every undetected fraudulent click \cite{wilbur2009click}.

\subsection{Cyber Warfare}
\label{ch4:cyberwarfare}

Cyberattacks on critical domestic or wartime infrastructure, e.g. power, water, communication and emergency systems, could bring a country and much of its military coordination to its knees during a time of war. Next generation warfare will involve coordinated attacks on two fronts; both ``traditional'' or ``kinetic'' ground, sea and air based fighting and attempting to remotely destroy a nation's cyber-infrastructure.

In July 2009, several government and business websites in the United States and South Korea were reportedly invaded. Initial suspicion was focused on North Korea as a source of these attacks, though no conclusive evidence was discovered \cite{zeichick2009reality}.  The rules and principles that govern physical warfare are largely dictated by easy to comprehend physical laws and limitations. Due to the fact that the majority of cyberwarfare attacks are conducted online, physical limitations, such as distance between warring nations, becomes almost irrelevant \cite{parks2011principles}. 

\section{Existing Detection Methods}
\label{ch4:existingdetection}

As quickly as botnet technology is evolving, so too must the methodologies attempting to keep up to date with the latest botnet advancements. Primarily, the objective of any botnet investigation is to attempt to decipher the methods of communication used by the system. This is in order to eavesdrop on the botnet chatter in an attempt to record the manner with which the botnet propagates itself, what commands the botnet is executing, what systems are at risk and how many machines are infected. There are three main entry points to P2P botnet investigation \cite{dittrich}:

\begin{enumerate}
\item Deliberately infect a host and participate in the botnet. This is the most realistic scenario as a real machine is infected and, as a result, no flags should be raised to either the bot client or any other peers that an investigation is taking place. In this instance, the network traffic of the machine can be monitored and analysed.
\item Deliberately infect a virtual host (or multiples thereof). This allows multiple bot clients to run on the same physical machine allowing much more network traffic to be gathered in a shorter period of time. However, many modern bots have the ability to detect if their host is a virtual machine and may adjust their behaviour accordingly.
\item Create a crawler and mimic the protocol used by the botnet. In order for a crawler to be built, the bot itself will need to be completely reverse engineered. The crawler can then act as though it were a regular bot on the network to every other peer. This method awards the investigator much control over the network, from enumeration to forwarding bogus commands and potentially destroying the botnet.
\end{enumerate}

\begin{figure}
\centering
\includegraphics[width=\textwidth]{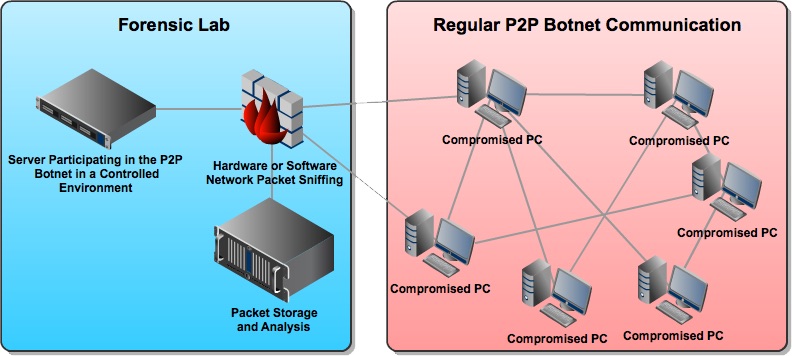}
\caption{Typical Investigation Topology}
\label{FigureP2PInvestigation}
\end{figure}

Irrespective of the method used, the investigation will appear similar to that outlined in Figure \ref{FigureP2PInvestigation}. A client machine in a controlled, forensically sound environment will attempt to partake in the botnet. In order not to raise any flags to any built-in, counter-forensic measures to either the botnet client or any other peers on the network, this client machine must appear as any other regular infected machine. All network communication from that client machine can then be monitored, recorded and analysed.

In 2009, Feily et al. highlighted four main categories of detection available in botnet investigation \cite{feily2009survey}:

\begin{enumerate}
\item Signature based detection -- This can only be used to aid in the detection of known bots and operates in a similar fashion to regular anti-virus signature detection, besides being applied to the identification of network traffic streams using an Intrusion Detection System.
\item Anomaly based detection -- This attempts to identify botnet activity based on network anomalies such as high network latency, high volumes of traffic, suspicious port usage and other unusual system behaviour.
\item DNS based detection -- Due to the prevalence of dynamic DNS (\nom{DDNS}{Dynamic Domain Name Server}) providers being employed to avoid hardcoding C\&C server IP addresses, the unusual querying of a DDNS provider may trigger detection.
\item Mining based detection --  Identifying C\&C based traffic may prove difficult, as they generally operate on commonly used ports, e.g., HTTP traffic on TCP port 80. C\&C traffic is also generally quite infrequent and of low volume. As a result, data mining techniques, such as classification and clustering can be used efficiently to detect C\&C traffic.
\end{enumerate}

\begin{table}[htbp]
  \centering
\scalebox{1}{
    \begin{tabular}{|p{1.7cm}|p{1.8cm}|p{2.5cm}|p{1.8cm}|p{1.9cm}|p{2cm}|}
    \cline{1-6}
   \centering \textbf{Detection Type}
& \centering \textbf{Unknown Bot \\ Detection} 
& \centering \textbf{Protocol \& Structure Independent} 
& \centering \textbf{Encrypted Bot \\ Detection} 
& \centering \textbf{Real-time Detection}
& \textbf{Low False Positive} \\
    \cline{1-6}
     Signature  & No    & No    & No    & No    & Yes  \\
\hline
     Anomaly  & Yes   & No    & Yes   & No    & No  \\
\hline
     DNS  & Yes   & Yes   & No    & No    & No  \\
\hline
     Mining  & Yes   & Yes   & Yes   & No    & Yes  \\
    \hline
    \end{tabular}%
}
  \caption{Comparison of Botnet Detection Techniques}
  \label{tab:detectiontechniques}%
\end{table}%

In 2013, Vania et al. published the Table \ref{tab:detectiontechniques} which presents the most up to date facts about the various detection methods outlined above \cite{vania4review}. From the data, it is clear that no single detection method is perfect and as a result, multiple methods should be deployed in any detection system.

\subsection{Host Based Approach}
\label{ch4:hostbased}

In 2007, Nummipuro outlined the three main methods available for detecting and identifying P2P botnets on an infected host machine \cite{nummipuro2007detecting}:

\begin{enumerate}
\item Tracking Network Data -- This involves tracking the remote machines that a specific process is in communication with. Communication with known C\&C servers can aid in identifying a botnet system.
\item Analysing Network Data -- This involves the analysis of the payload of individual packets to identify common botnet communication patterns.
\item Behaviour Based Identification -- When a specific piece of malware is running on an infected machine, it calls on specific Microsoft Windows \nom{API}{Application Programming Interface} functions. The behaviour of these calls, in combination with one or both of the above methods, can help identify the infection.
\end{enumerate}


\subsection{Hardware Based Approach}
\label{ch4:hardware}

Honeypots are a common system used to detect security threats, collect malware and to understand the behaviours of malware and their perpetrators \cite{jaiswal2013botnet}. Honeypots are specially constructed computers or network traps which attract malicious attacks. However, advancements in botnet technology has resulted in more intelligent honeypot aware, self-destructing bots. From 2006, forensic researchers began documenting ways that a bot could detect that it was running in a honeypot \cite{zou2006honeypot}. The earliest detection methods were based on the assumption that security and forensic professionals have liability constraints, such that they cannot allow their infected honeypots to participate in real (or too many real) attacks. Subsequently honeypot detection methods expanded to include firewall, anti-virus and virtual machine detection.

Deep packet inspection (as outlined in Section \ref{ch3:dpi}) is another common hardware based approach and may be used in conjunction with honeypots.


\section{Investigation Types}
\label{ch4:investigation}

\subsection{Anatomy}
\label{ch4:anatomy}

Investigating the anatomy of a particular botnet includes analysis of the behaviour of the bot binary and analysis of the network communication patterns. This type of investigation attempts to classify the botnet as centralised/decentralised, client-server or P2P based command and control. The classification can continue past the architecture of the system to cover some of its counter-detection and anti-forensic techniques. For example, Goel et al. discovered that ``Agobot'' had a built in defence mechanism to kill an upgradable list of over 610 anti-virus programs \cite{goel2006botnets}.

\subsection{Wide-Area Measurement}
\label{ch4:widearea}

Wide-area measurement investigations concentrates on attempting to enumerate the population of the botnet, the bandwidth usage, the computational capabilities as well as the commands being issued. Gathering the population of a botnet is a non-trivial task, as the number of nodes connecting to a C\&C server may only ever count for a small proportion of the total infected nodes. There are two definitions of a botnet's size, as specified by Rejab et al. \cite{rajab}:

\begin{enumerate}
\item Footprint -- This indicates the aggregated total number of machines that have been compromised over time.
\item Live Population -- This measure denotes the number of compromised machines that are concurrently in communication with the C\&C server.
\end{enumerate}

A relatively straightforward method for measuring the size of a botnet is to run a bot on a deliberately infected machine and monitor the resultant traffic. The number of IP addresses the infected node is in communication with can be easily counted, having eliminated all non-botnet related network traffic. While it would be unsafe to assume that a single node will ultimately communicate with every other node over any reasonable timeframe; increasing the number of infected machines (physically or virtually) and amalgamating the results should lead to a more accurate representation.

Byung et al. proposed in 2009 a methodology for improving botnet size estimates through the implementation of a botnet crawler, called Passive P2P Monitor (PPM) \cite{kang2009towards}. PPM acts as though it were the same as any other node on the network by implementing the ``Overnet Protocol", as explained below. This method involves mimicking the functionality of a regular bot with regards to maintaining the DHT. For each peer the crawler connects to, it can ask for a list of all known peers. In this manner, a list of all known peers on the network can be compiled. This approach closely resembles that employed in the crawling of P2P file-sharing networks described in Section \ref{ch3:crawling}.

\subsection{Takeover}
\label{ch4:takeover}

Botnet takeover involves a third party gaining control of a botnet from its owner. This third party could be law enforcement, researchers or another botmaster. Once control of the botnet has been gained, the new botmaster is able to issue commands, update configurations and operate the botnet as desired. In 2009, Stone-Gross et al. successfully took over the Torpig botnet for 10 days \cite{stone2009your}. During this time, the researchers identified more than 180,000 compromised machines and were sent over 70GB of automatically harvested personal information.

\begin{figure}
\centering
\includegraphics[trim = 0mm 0mm 0mm 0mm, clip, width=1\textwidth]{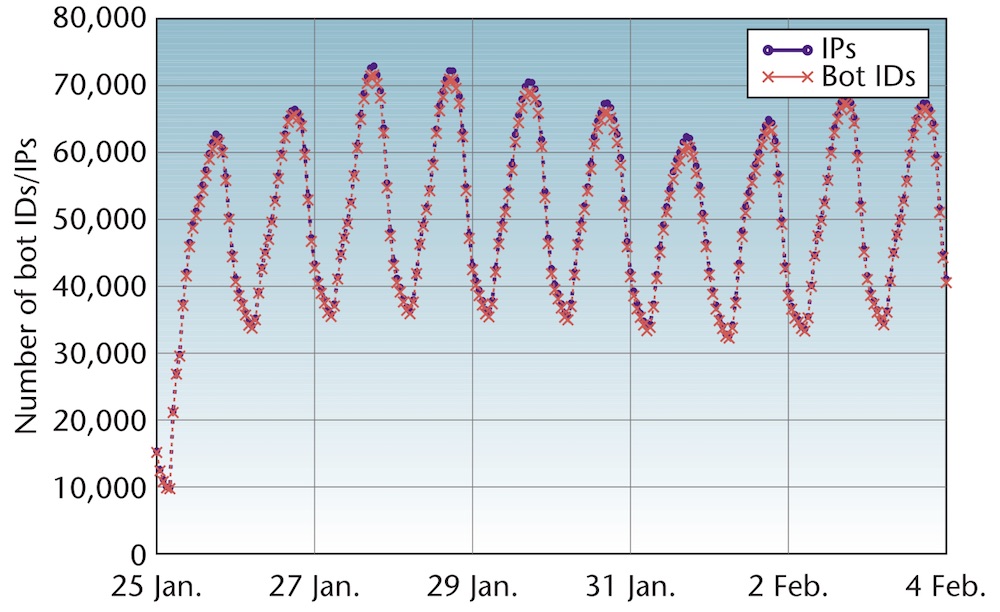}
\caption{Unique Bot IDs and IP Addresses per Hour}
\label{Figuretorpig}
\end{figure}

The number of discovered unique Torpig bot IDs and corresponding number of IP addresses can be seen in Figure \ref{Figuretorpig} \cite{stone2011analysis}. The discrepancy between the number of bots and IP addresses found is accountable by network effects such as DHCP churn and \nom{NAT}{Network Address Translation}. Stone-Gross et al. discovered 182,914 different bot IDs originating from 1,247,642 distinct IP addresses over the ten day controlling window.

\subsection{Investigation Obstacles}
\label{ch4:obstacles}

Many of the obstacles facing an investigation on P2P botnets are shared by the investigation of any P2P network, documented or undocumented \cite{scanlon2010week}:

\begin{enumerate}
\item DHCP -- Due to a typical lease from an Internet service provider lasting in the order of 2-7 days, dynamic reallocation of the same IP address may result in two or more infected machines participating in the network appearing as a single peer.
\item Proxy servers -- Similar to the issue caused by DHCP, any bots that access the Internet through a transparent or anonymous proxy server will also appear as a single bot.
\item NAT -- Numerous machines behind a shared router may appear to the outside world as a single machine, as a result of sharing a single external IP address.
\item Encrypted Communication -- Should the bot employ encrypted communication, the only method available for investigation is to attempt to reverse engineer the bot. The decryption key for any incoming commands must be stored within the bot's client.
\item Difficulty in Take Down -- Fighting back against botnets is often a matter of discovering a vulnerability in the design. Traditionally, this has meant attempting to take down their centralised C\&C server \cite{schoof}. However, with the popularity of employing a fully decentralised network design, the ability to take down a botnet has been made considerably more difficult. Should the bot be reverse engineered, it is possible that the botnet could be destroyed or ``imploded'', i.e., through the issuing of an uninstall command to each infected node.
\end{enumerate}

\section{Case Studies}
\label{ch4:casestudies}

\subsection{Nugache}
\label{ch4:nugache}
Nugache used a list of 22 hardcoded IP addresses which each newly infected host attempted to connect to \cite{dittrich2008discovery}. These 22 hosts maintained a list of active nodes, which they shared with each new node. The list of active nodes that any given peer maintained always contained the initial 22 hosts, along with any newly shared active IP addresses. The weakness of this design is that once these 22 hardcoded nodes are taken down, no newly connecting peer will be able to gather its initial list of active peers to communicate with. The Nugache botnet communicates across its own bespoke network protocol. The communication between each node is not encrypted, but there is a degree of obfuscation employed \cite{schoof2007detecting}. In June 2007, Dittrich et al, discovered that there were at least 6,000 active IP:port pairs in the Nugache botnet at any given time and a total infected footprint of nearly 11,000 IP:port pairs \cite{dittrich}. It was also discovered that Nugache propagated itself through two remotely accessible exploits in the Window's LSASS and RPC-DCOM services, emailing copies of itself to targets found in the Windows Address Book and via instant messenger clients, such as AIM and MSN Messenger.

\subsection{Storm}
\label{ch4:storm}
The ``Storm" botnet, first discovered in January 2007 \cite{grizzard}, was the first botnet discovered that utilised a P2P protocol. It spread through a mixture of social engineering and exploiting vulnerabilities in Windows XP and Windows 2000. The social engineering aspect of the worm was realised through the sending of topical, newsworthy emails with attachments or links to videos and pictures, which were in fact executables to infect the user's machine. When it infected any given machine, it would disable the Windows firewall and open a number of TCP and UDP ports. Communication in the Storm botnet relies on the ``Overnet Protocol". Once the malware was installed and the host machine was configured, it would then bootstrap onto the Overnet network and start listening for commands. The worm was also engineered to aggressively attack anyone who attempted to reverse engineer it \cite{mukamurenzi}.

The Overnet Protocol utilises a DHT, storing the IP addresses and unique IDs of each active peer in the network \cite{byung}. It is based on the Kademlia algorithm, similarly to BitTorrent \cite{mukamurenzi}. Kademlia assigns a 160-bit hash ID to each participating peer on the network. Each peer maintains a local routing table consisting of the binding values for other peers that are "close" to their own ID. In order to bootstrap onto the DHT, the Storm bot has a hardcoded list of over one hundred peers it can connect to \cite{nummipuro2007detecting}.

\subsection{Waledec}
\label{ch4:waledec}

The Waledec botnet has striking similarities to the Storm botnet, while simultaneously exhibiting unique refinements to aid in network uptime and performance, but in part more vulnerable to attack. Waledec follows a hierarchical architecture design. The lowest level were the spammer nodes, which, as their name implies, were responsible for sending spam emails. These spammer nodes communicated exclusively with repeater nodes or super-nodes. These super-nodes, in turn, were in control of the communication with the spammer nodes and would receive their commands from the next level up, known as the sub-controllers \cite{sinclair}. The highest level in the hierarchy, the C\&C server, only communicated directly with these sub-controllers.

Similar to the Storm botnet, the Waledec binary contains a list of hardcoded nodes to use to bootstrap onto the network. In the event of all of these hardcoded nodes being offline, a dynamic URL is also included in the binary to fall back on HTTP to receive commands. Due to this HTTP fall-back, this category of botnet can be referred to as a ``HTTP2P'' botnet \cite{jang}. Communication between nodes is encrypted, initially using a constant key for all nodes, which later evolved into a frequently changing key, which would be created at the C\&C server and passed down the hierarchy \cite{sinclair}.

\subsection{Zeus}
\label{ch4:zeus}

The Zeus botnet is one of the largest botnets in the world \cite{lu2012p2p}. Zeus uses an entirely decentralised P2P architecture and, like the majority of other botnets, it originally only operated on Microsoft Windows, but variants have been discovered infecting Blackberry and Android mobile phones \cite{scmag}. An infected machine attempts to connect to its C\&C channel by bootstrapping on to any one of hundreds of predefined nodes. The command and control channel consists of many thousands of server nodes \cite{rossow2012large}. Its purpose is primarily to spy on the users of infected machines, with the intent of gaining financial benefits for the botmaster \cite{5593240}. It has the ability to log any information entered by the unsuspecting user, as well as injecting data displayed on visited web pages. The targeted information includes email addresses, passwords, online banking accounts, credit card details and transaction authentication numbers.

\subsection{Stuxnet}
\label{ch4:stuxnet}

\begin{figure}
\centering
\includegraphics[trim = 0mm 0mm 0mm 0mm, clip, width=1\textwidth]{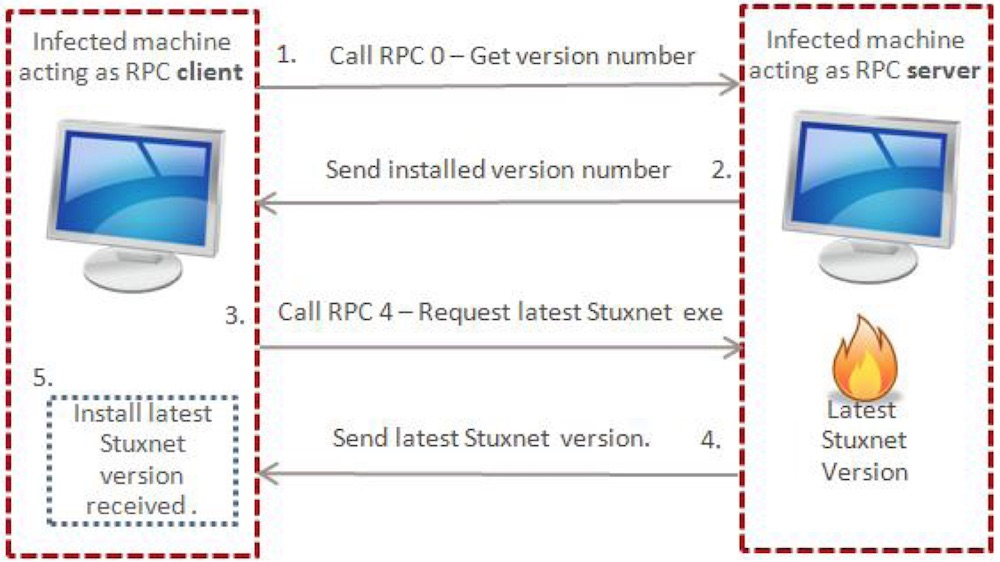}
\caption{Example of an Old Client Requesting Latest Version of Stuxnet via P2P}
\label{Figuresymantec}
\end{figure}

Discovered in June 2010, Stuxnet was the first cyberwarfare weapon targeting physical infrastructure \cite{langner2011stuxnet}. It is believed to have been developed by the United States and Israel in an attack against a nuclear power plant and processing facility in Natanz, Iran, although no conclusive evidence has been discovered about who lies responsible \cite{sanger2012obama}. Stuxnet was not remotely controlled; it was completely stand-alone and spread itself without any further interaction. The C\&C servers that Stuxnet contacted while in operation appear to have primarily been used for recording evidence of compromise. It spread via a Microsoft Windows vulnerability and targeted Siemens industrial software and equipment. This equipment included electronic controllers for pumps, valves, thermometers, motors and tachometers used in the nuclear facility. During the attacks in 2010, Stuxnet temporarily shut down nearly 1,000 of the 5,000 centrifuges Natanz had in operation purifying uranium \cite{sanger2012obama}.

The Stuxnet dropper client was designed to spread itself to as many machines as possible and spread through network shares, infecting removable storage devices and exploitation of software vulnerabilities. It utilised P2P communication for updating itself, as can be seen in Figure \ref {Figuresymantec} \cite{falliere2011w32}. The P2P component had two parts, namely an \nom{RPC}{Remote Procedure Call} server and client. When the malicious code compromises any machine it starts the RPC server. Through P2P chatter, any other infected machines on the network can update themselves from any peers that are running code with a higher version number.

\section{Ethics of Botnet Mitigation/Takeover}
\label{ch4:ethics}

With a lack of precise legal guidance in the investigation of botnets, much of the decisions required in botnet investigation are left in the hands of investigators to make the right ethical choices. When Stone-Gross et al. took over the Torpig botnet in 2009 as described in Section \ref{ch4:takeover}, they used two principles to guide their investigation \cite{stone2009your}:

\begin{enumerate}
\item The compromised botnet should be operated so that any harm and/or damage to victims and targets of attacks would be minimised.
\item The compromised botnet should collect enough information to enable notification and remediation of affected parties.
\end{enumerate}

\section{Summary and Discussion}
\label{ch4:summary}

\begin{table}[htbp]
  \centering
    \begin{tabular}{|r|c|c|c|}
   \hline
    \textbf{Factors } & \textbf{Centralised (IRC,HTTP) } & \textbf{Hybrid DDNS } & \textbf{Peer-to-Peer P2P } \\
   \hline
    Detection  & Easy  & Medium  & Hard  \\
    Resilience  & Low   & Fairly High  & Very High  \\
    Latency  & Low   & Medium  & Fairly Hard  \\
    Traceback  & Fairly Hard  & Hard  & Very Hard  \\
    Complexity  & Easy  & High  & Medium  \\
    Experience  & Very High  & None  & Medium  \\
   \hline
    \end{tabular}%
  \caption{Comparison of Botnet C\&C Architectures}
  \label{tab:investigativecomparison}%
\end{table}%

In this chapter, the evolution of botnet design culminating in Peer-to-Peer architectures was introduced. Table \ref{tab:investigativecomparison} presents a summary of the various C\&C botnet architectures available and the corresponding difficulties associated with each of the considerations in a botnet developer's design decisions \cite{vania4review}.

Ultimately, the P2P botnet topology is a desirable option to choose for botmasters, as it affords them an additional level of anonymity when conducting their crimes. The ideal design for a P2P botnet is one that is completely decentralised, utilises unique encryption methods and operates on a bespoke network protocol for communication. Investigation of such a botnet may prove particularly difficult. However, a combination of research, network monitoring, deep packet inspection and network crawling should result in successful, albeit more labour intensive, investigations. The fundamental requirement for any newly infected node (or a node coming online) to have a starting point to bootstrap onto the P2P network and discover other active nodes will always leave an avenue of investigation open for the digital investigator.

\include{ch5designandarch}

\include{ch6forensicinvestigationofbittorrent}


\chapter{Conclusion and Discussion}
\label{ch7conclusion}

Currently a need exists with law enforcement for a universal P2P investigative tool capable of identifying the crimes and the criminals behind some of the world's largest P2P networks. This thesis proposes a solution to this problem incorporating the individual investigative techniques required and a methodology for completing them. A proof of concept tool was developed and tested on BitTorrent, the world's largest documented P2P network. The future plan for expansion upon this work is to produce an intelligent P2P monitoring tool. Such a collaborative, investigative tool would be of significant benefit to law enforcement in investigating cybercrimes that utilise P2P communications.

In total, over 4TB of evidence was gathered using the prototyped system. This consisted of evidence specific to a number of peers in the order of tens of millions. A precise number is unattainable due to resource contraints for data storage throughout the project. The latest prototype of the system processes the results into a NoSQL database (based on MongoDB) capable of quickly performing cross-swarm and cross-investigation queries.

\section{Analysis of Outlined Approach}
\label{ch7:analysis}

The approach discussed in Chapter \ref{ch5designandarch} outlines a novel modular universal P2P network investigation framework. As of the date of this thesis, no other collaborative P2P network investigation system exists. The primary benefit of this collaborative approach is that it can be easily expanded upon and updated to create a leading tool in the arsenal of the forensic investigator. Through shared resources and expertise, many wasted man hours could be reallocated to the analysis of the evidence and the prosecution of those responsible for P2P based cybercrimes. It is envisioned that this framework will be made available for collaboration to law enforcement. This should help to eliminate some of the redundancy of efforts by local law enforcement agencies in an attempt to combat P2P based cybercrimes.

\subsection{Enhancements}
\label{enhancements}

Due to the aforementioned commonality in design and implementation of P2P networks, it is envisioned that the proof-of-concept P2P network investigation framework should be expanded to handle any P2P network. 

Given the gathered network traffic from any active node of a new P2P network, the system should be enhanced to automatically determine the networks topology, protocols and available commands. This would greatly speed up the first two steps, as shown in Figure \ref{steps}, and enable the monitoring procedures to commence as early as possible in the investigation.

\section{Further Ideas}
\label{ch7:furtherwork}

While the objectives of the research outlined in this thesis were met, there are some ideas and features which could be added to (or used in conjunction with the existing system) to improve the overall level of functionality. Potential modifications to the current system include implementing automated P2P network traffic pattern recognition, creation of a comprehensive database of P2P network signatures and automated result processing.

The framework developed was designed and prototyped in such a manner as to easily facilitate the expansion of the tool to deal with any P2P network. It is hoped that in the future, numerous botnet investigation bodies will contribute to the maintenance and development of the framework. 

\subsection{Bespoke Hardware Device}
\label{ch7:hardware}
A specific hardware device could be created to piggyback between an infected machine and its Internet connection. When in operation, this device would automatically acquire network evidence from the suspect computer's communication. This device could subsequently perform on-the-fly network identification and processing of the live communications.

\subsection{P2P Audio/Video Reconstruction}
\label{ch7:avreconstruction}
With P2P technology being increasingly utilised for VOIP communication, the reconstruction of captured audio or video content could be crucial to forensic event reconstruction activities. Through the analysis of captured UDP packets, the voice/video call should be capable of being reconstructed. Using pattern analysis, collected evidence could be reconstructed to potentially better quality than the original call, i.e., patching collected packets together in the correct order.

With the popularity of P2P based file-sharing, this reconstruction could also be used the verification of suspected content as being a true copy of the original. In this scenario, a partial sample of the entire content could be used to verify the infringement of copyright.

\subsection{Usability Test}
\label{ch7:usability}

As outlined as part of the technical requirements of the UP2PNIF system in Section \ref{ch5:techrequirements}, the framework should be relatively easy to use for regular law enforcement officers and should require minimal training. In order to measure this requirement, a usability test should be conducted. This test should invite law enforcement officers and digital forensic investigators to take part. The groups should be randomly divided into two teams, each given the same task of collecting digital evidence from known P2P networks. The two teams would be divided as follows:

\begin{enumerate}
\item One team would not be given any instruction on how to use the framework.
\item The second team would be given a short introduction to using the framework, how it operates and the best practices while using the tool.
\end{enumerate}

Should both teams achieve their task in a similar time frame, the ease of use of the tool would be proven. This result would also prove the reduced level required of digital forensic expertise to use the tool. Feedback received from the usability testing could be useful in building upon the current system.

\subsection{NIST Computer Forensics Tool Testing}
\label{ch7:cftt}

Computer Forensics Tool Testing (\nom{CFTT}{Computer Forensics Tool Testing}) is a standardised set of tests procedures, criteria and hardware compatibility checking performed by NIST to validate computer forensic tools for use by law enforcement. When a sufficient number of P2P networks are added to the system, the tool should be sent for independent, third-party verification.

\section{Future Vision}
\label{ch7:futurework}

\subsection{P2P in the Cloud}
\label{ch7:cloud}

With many everyday services being pushed to the cloud in recent years, one could assume that P2P networks themselves might become redundant in the future. However, alongside the push for cloud based services and storage, there has also been a significant rise in P2P anonymity services and P2P-aided, cloud driven services, e.g., P2P based streaming services such as Spotify and BitTorrent Live. Controlling a botnet from the cloud could easily facilitate criminals in adding an additional, often temporary, layer of removal from the botnet itself. This potentially could aid the botmaster in avoiding detection completely.

\subsection{Mobile P2P}
\label{ch7:mobile}

The vast majority of botnets existing today are developed to be executed on desktop computers worldwide. However, in the future, it is envisioned that mobile botnets will become commonplace. Smartphones and 3G-enabled tablets are an ideal ``next target'' device for botnet developers as they are difficult to trace solely based on the data connection. The mobile devices themselves are becoming more powerful with each device having its own always-on Internet connection \cite{cole2012all}.

\section{Conclusion}
\label{ch7:conclusion}

The phenomenon of the ever increasing number of crimes being aided by P2P networks is set to continue into the future due to the level of anonymity provided to cybercriminals. As a result of this inevitable increase in P2P based cybercrimes, digital forensic investigators' workload is set to drastically increase. Any saving of the investigators' time that can be allocated to performing the analysis of captured evidence will help to aid the turn around time for investigations. 

This thesis proposed and validated the viability of a forensically sound, P2P evidence acquisition framework. This framework processes the network evidence into an ``investigation-ready'' state for the forensic laboratory as early into the investigative process as possible. The existing model for P2P network evidence acquisition generally requires a digital investigator to first develop a bespoke tool capable of deciphering the captured packets from a compromised machine. The use of the UP2PNIF system can significantly improve on this traditional model by fast-tracking the investigation.


\newpage
\appendix

\chapter{Graphical Results}
\label{appendix}

\begin{figure}[!h]
\centering
\includegraphics[width=1\textwidth]{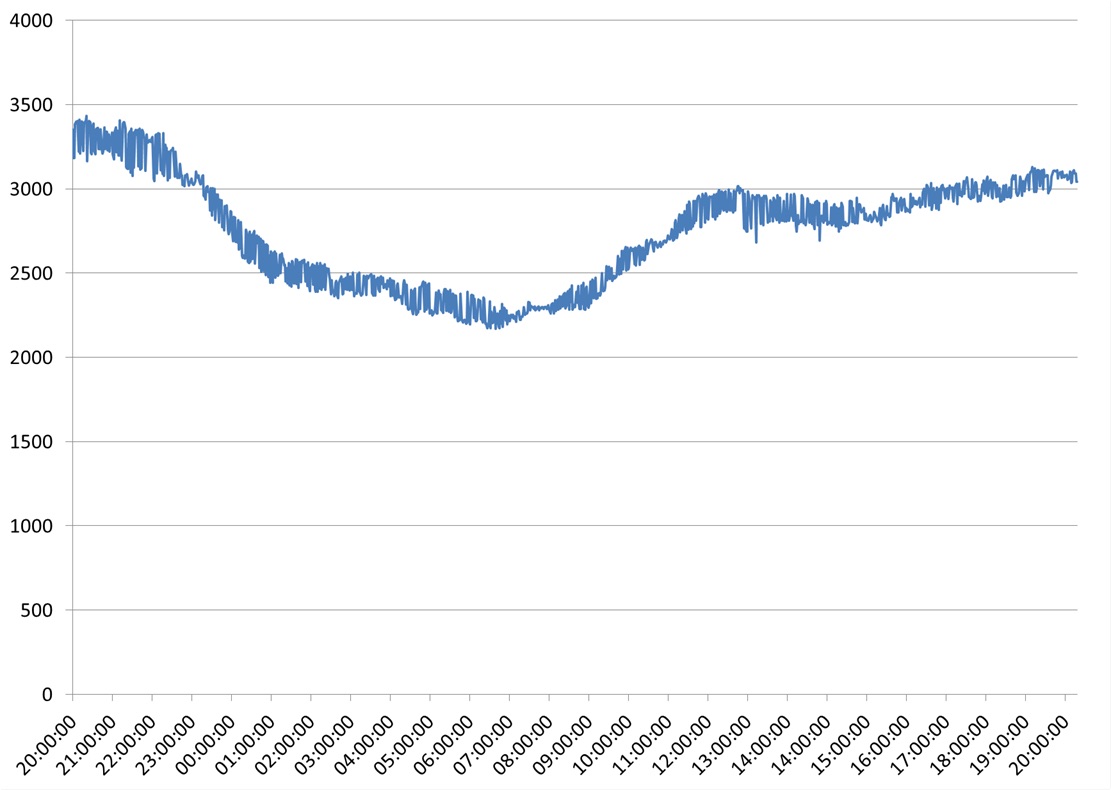}
\caption{Daft Punk: Active Swarm Size over 24 Hours}	
\label{Figuredpswarmsize}
\end{figure}

\begin{figure}
\centering
\includegraphics[width=1\textwidth]{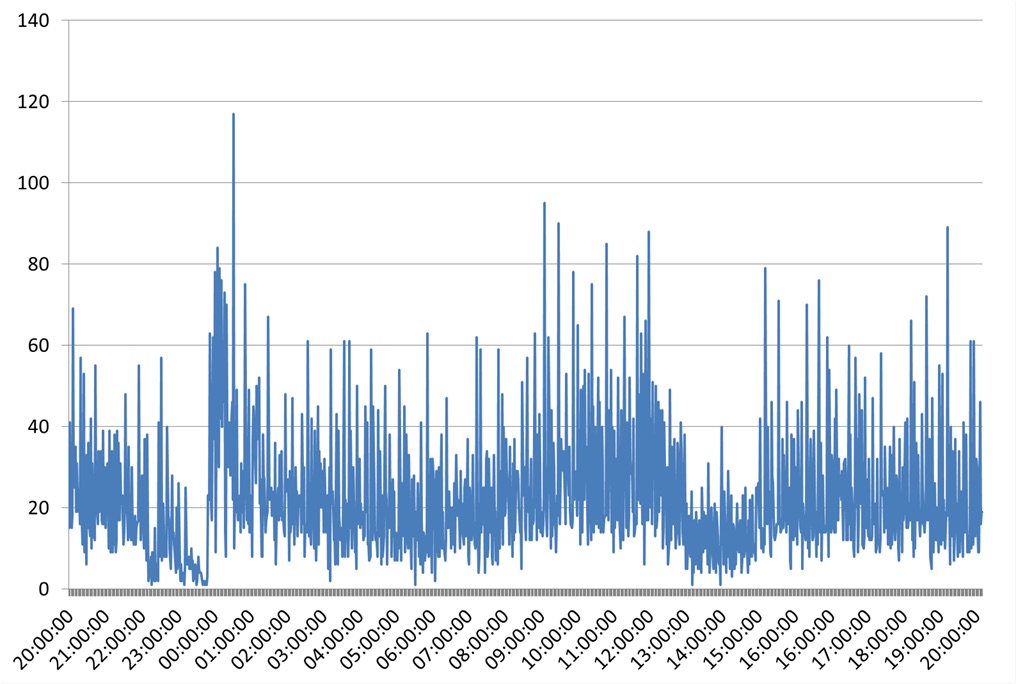}
\caption{Daft Punk: Newly Discovered Peers Identified per Crawl (Excluding the Initial Crawl)}
\label{Figuredpnewpeers}
\end{figure}

\begin{figure}
\centering
\includegraphics[width=1\textwidth]{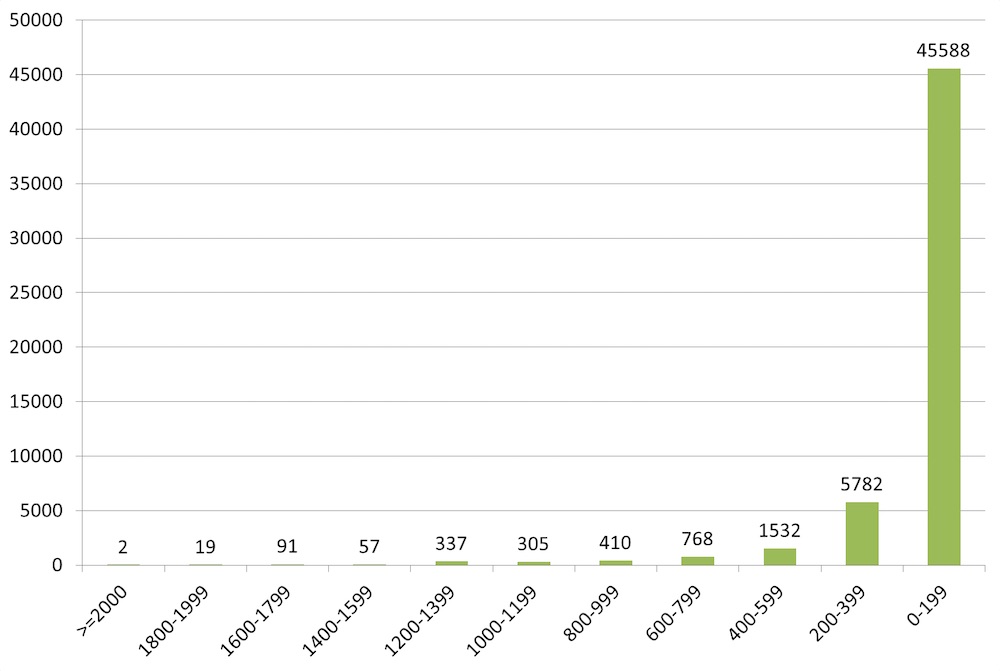}
\caption{Daft Punk: Overall Average Peer Crawl Count}
\label{Figuredpconnectiontime}
\end{figure}

\begin{figure}
\centering
\includegraphics[width=1\textwidth]{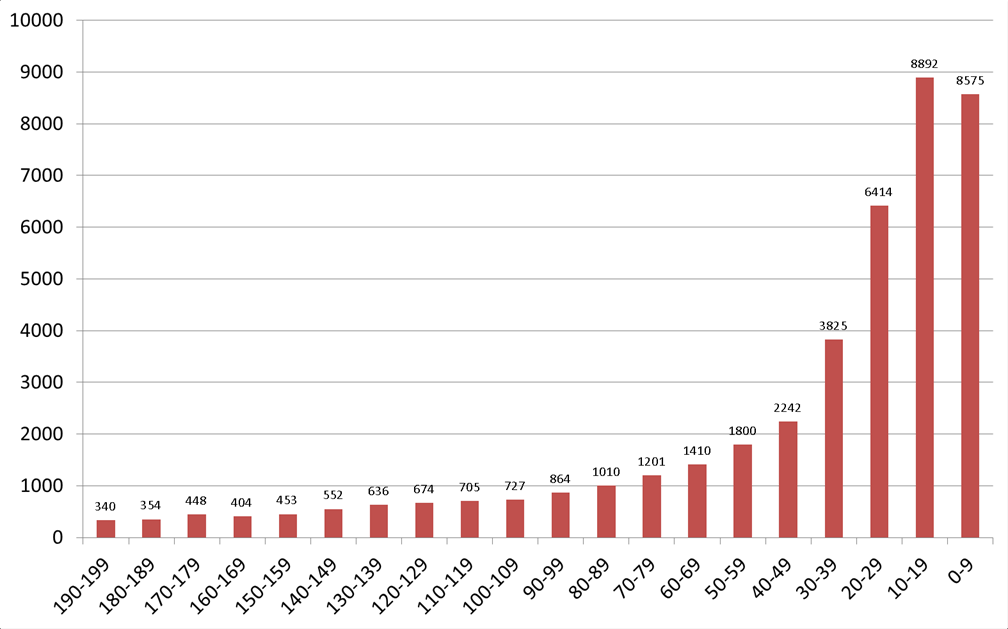}
\caption{Daft Punk: Average Peer Connection Time for 0-200 Crawl Count}
\label{Figuredpconnectiontime0to200}
\end{figure}

\begin{figure}
\centering
\includegraphics[width=1\textwidth]{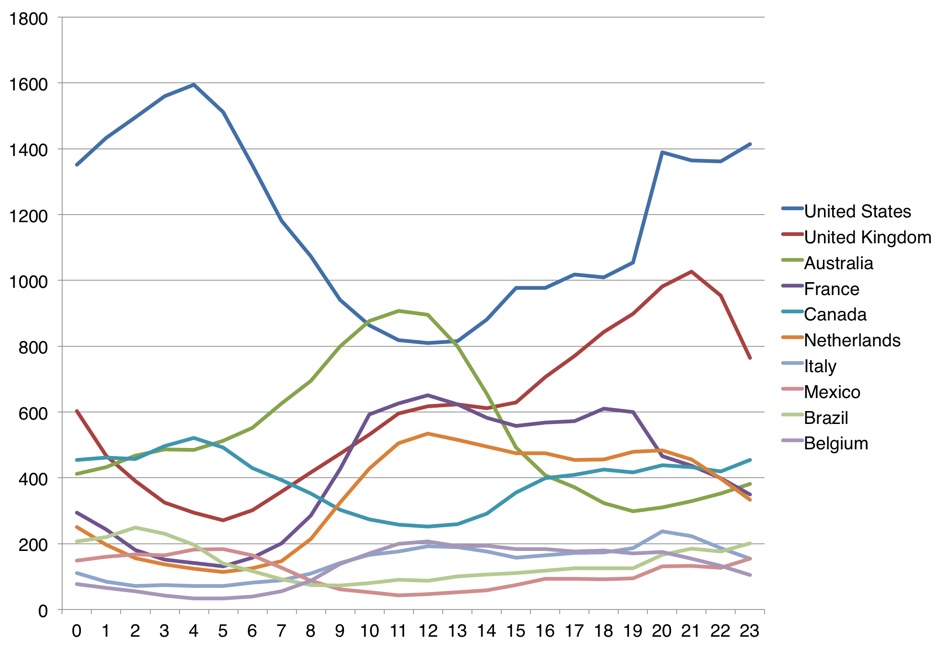}
\caption{Daft Punk: Top 10 Countries Hourly Activity (GMT)}
\label{Figuredptop10countries}
\end{figure}

\begin{landscape}
\begin{figure}
\centering
\includegraphics[width=1.47\textwidth]{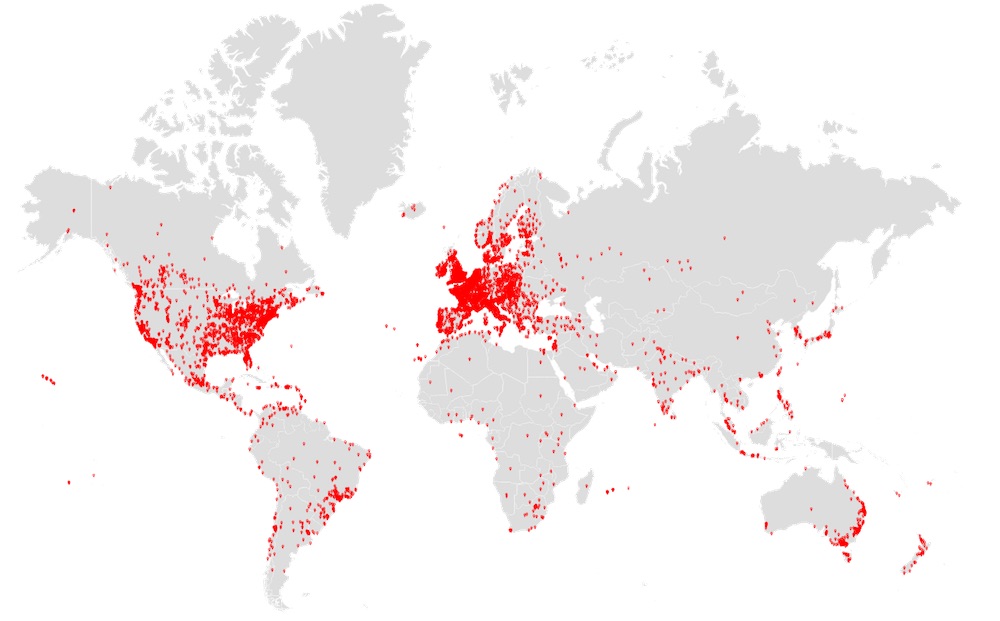}
\caption{Daft Punk: Geolocation for Worldwide Cities}
\label{Figuredpworldwidecities}
\end{figure}
\end{landscape}

\begin{figure}
\centering
\includegraphics[width=1\textwidth]{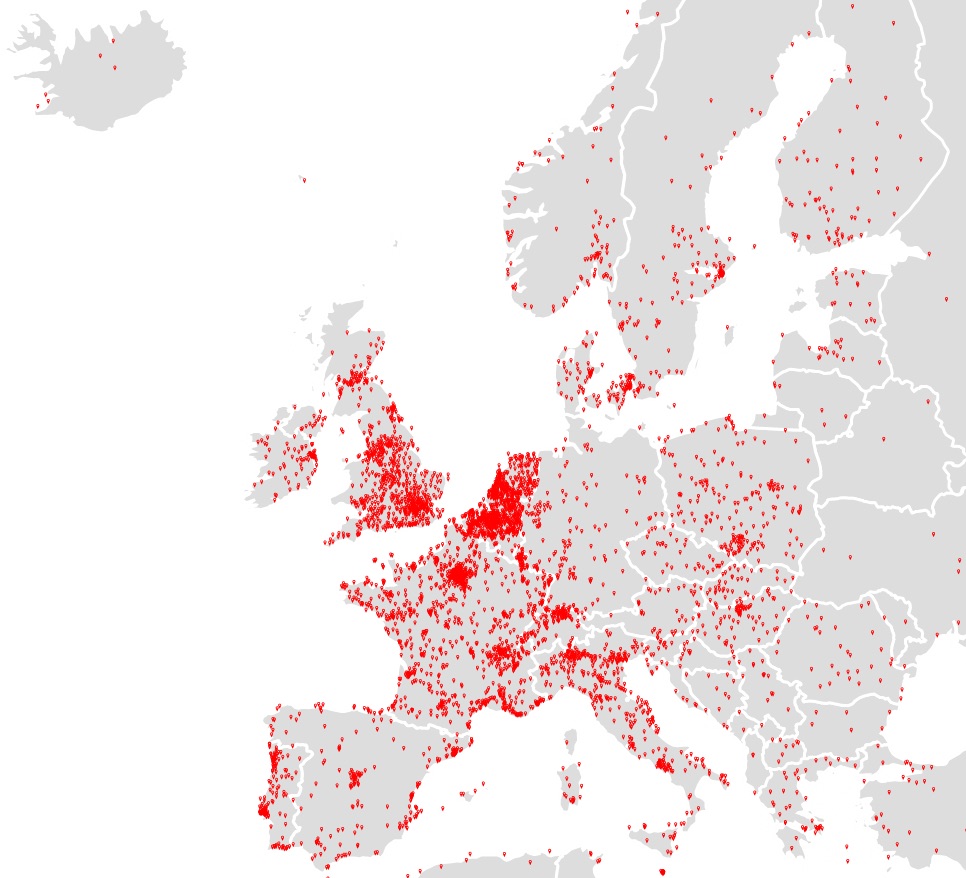}
\caption{Daft Punk: Geolocation for Mainland Europe}
\label{Figuredpmainlandeurope}
\end{figure}

\begin{landscape}
\begin{figure}
\centering
\includegraphics[width=1.45\textwidth]{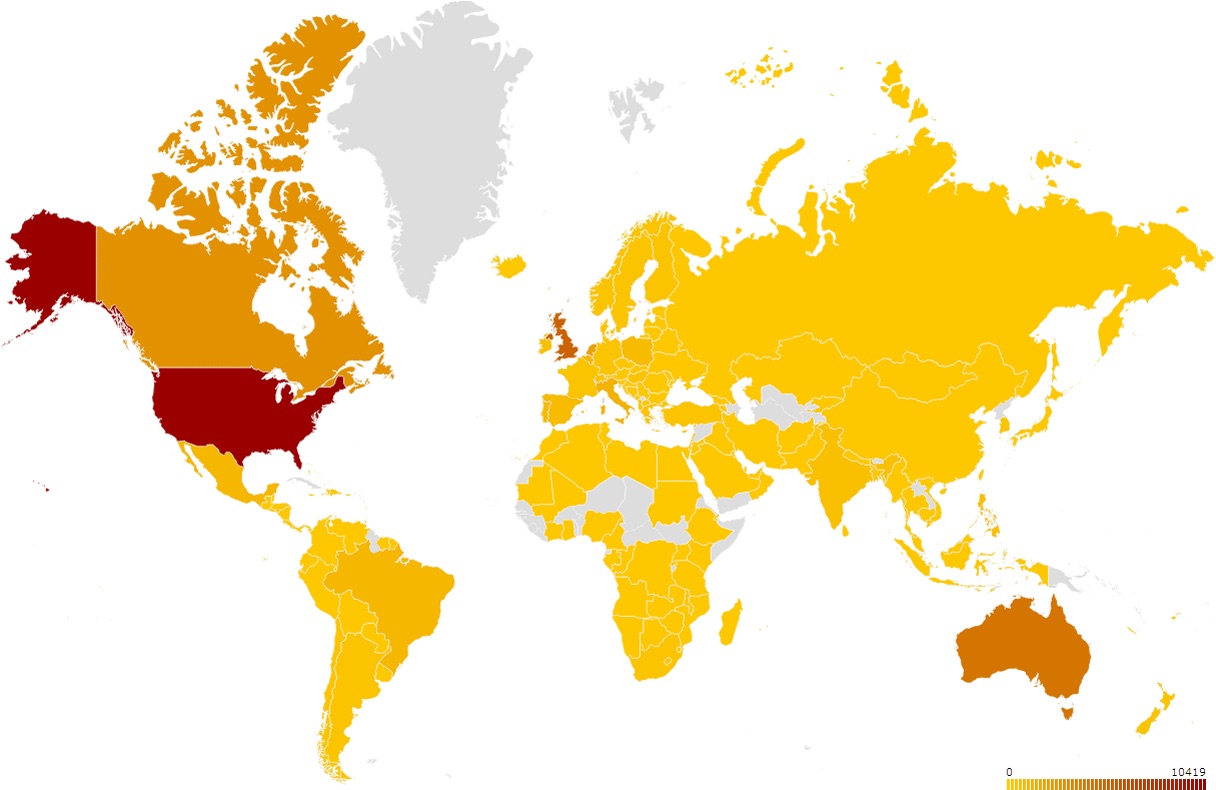}
\caption{Daft Punk: Global Heatmap}
\label{Figuredpglobalheatmap}
\end{figure}
\end{landscape}


\begin{figure}
\centering
\includegraphics[width=1\textwidth]{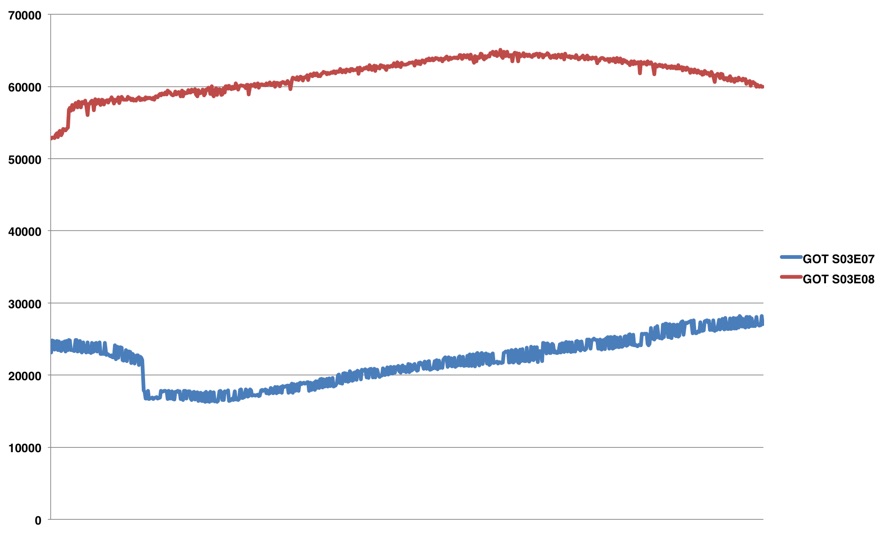}
\caption{Game of Thrones S03E07/S03E08: Swarm Sizes over 24 hours}
\label{Figuregotcombinedswarmsize}
\end{figure}

\begin{figure}
\centering
\includegraphics[width=1\textwidth]{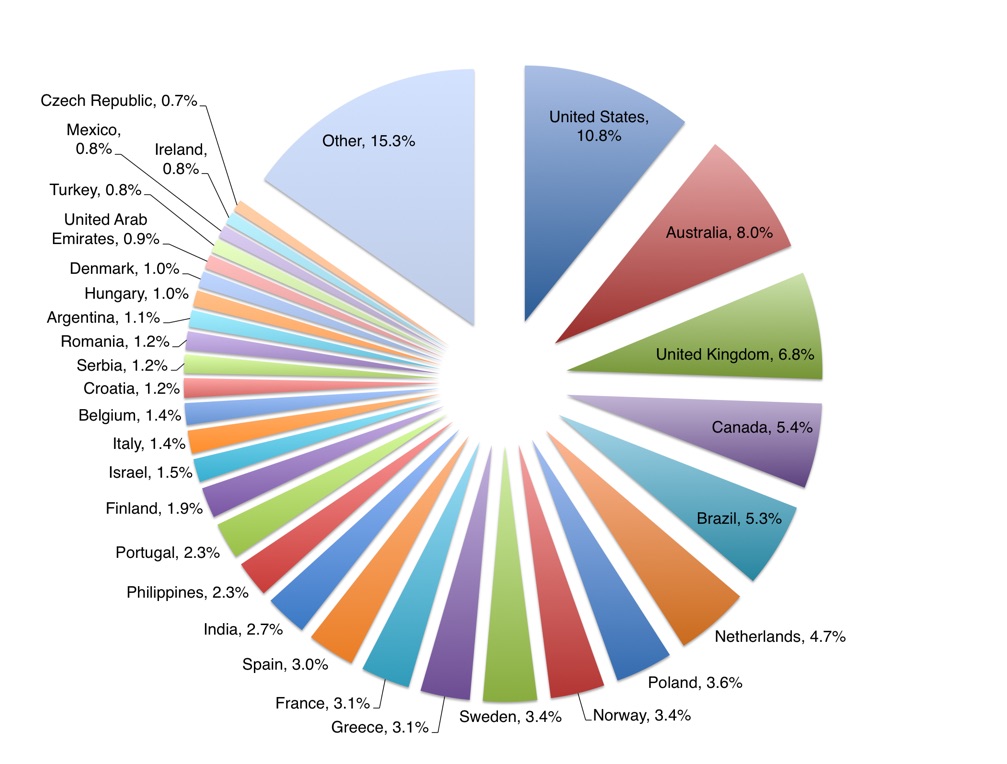}
\caption{Game of Thrones: Top 30 Countries}
\label{Figuregotcountrypie}
\end{figure}

\begin{figure}
\centering
\includegraphics[width=1\textwidth]{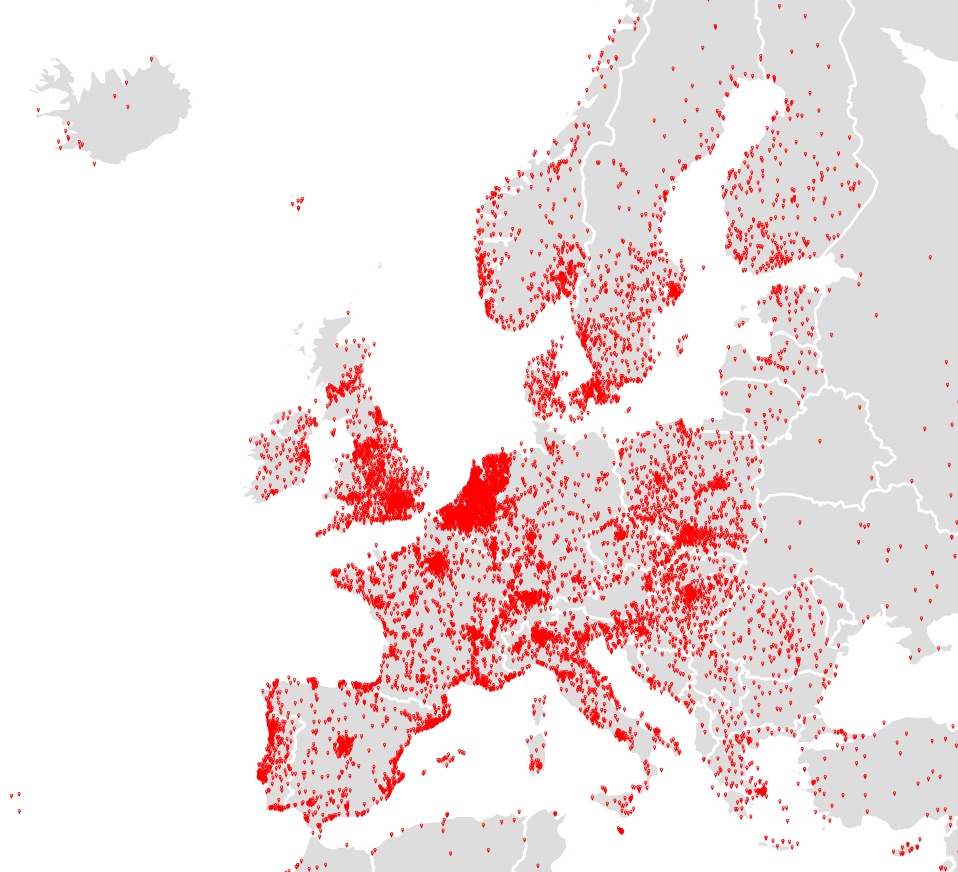}
\caption{Game of Thrones S03E07: Mainland Europe Activity}
\label{Figuregots03e07mainlandeurope}
\end{figure}

\begin{landscape}
\begin{figure}
\centering
\includegraphics[width=1.45\textwidth]{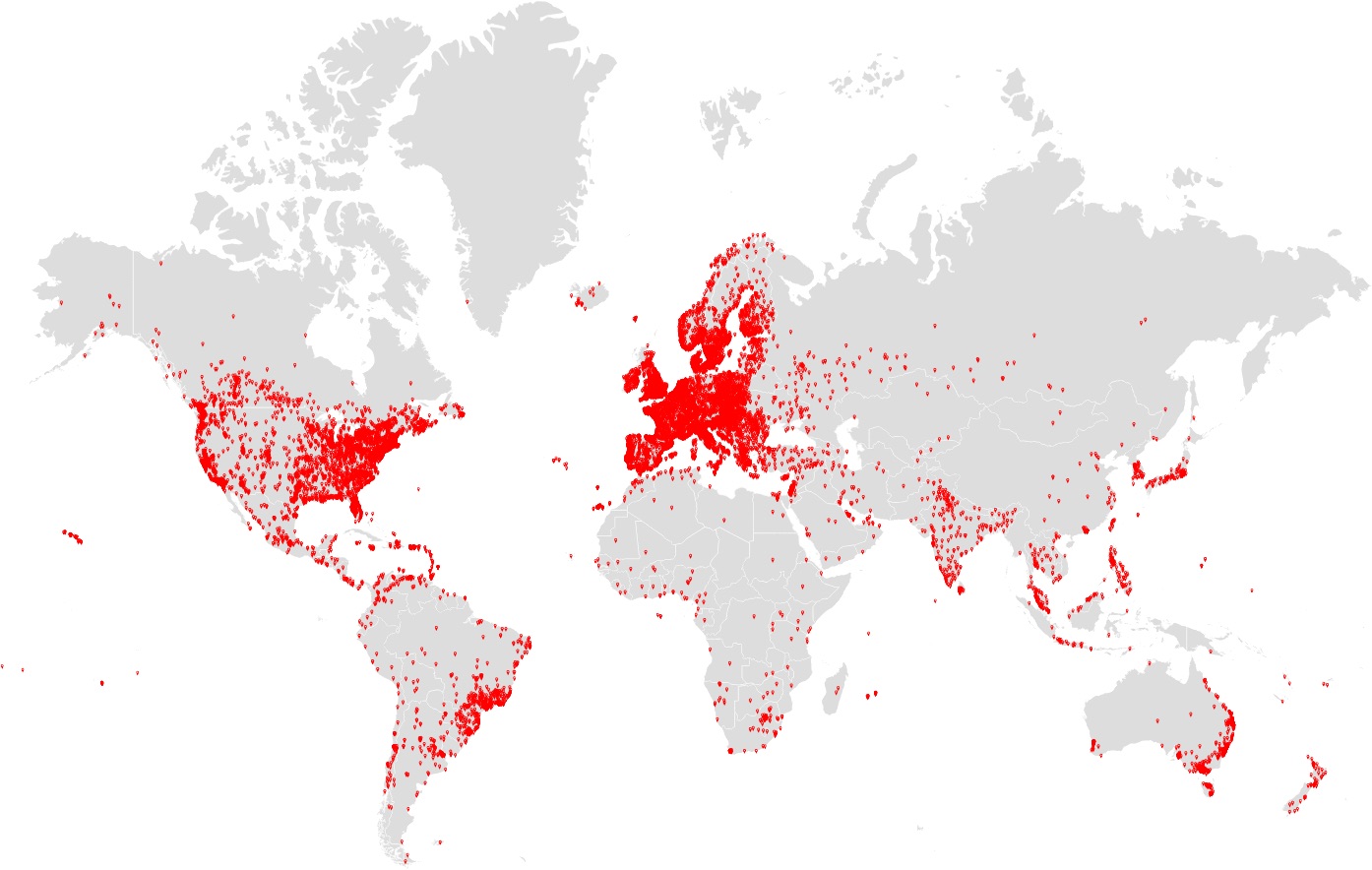}
\caption{Game of Thrones S03E07: Global City Level Activity}
\label{Figuregots03e07worlddots}
\end{figure}
\end{landscape}

\begin{landscape}
\begin{figure}
\centering
\includegraphics[width=1.45\textwidth]{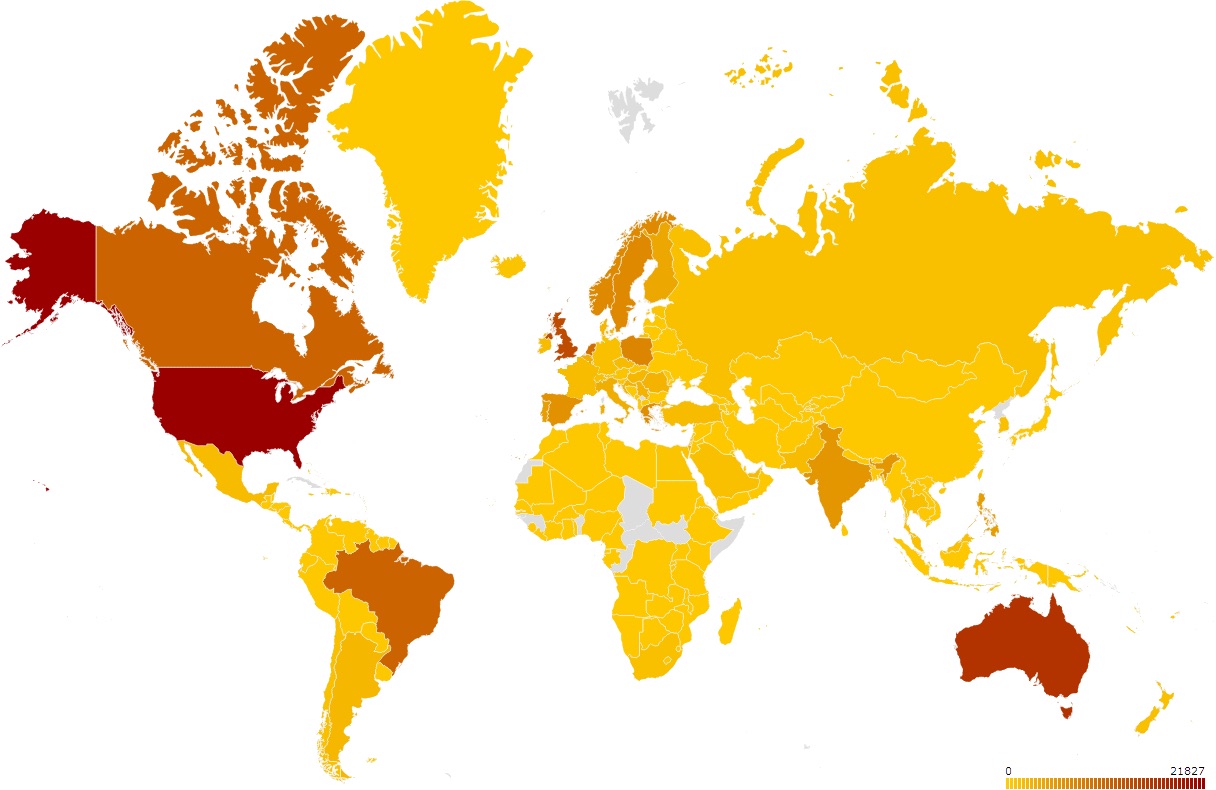}
\caption{Game of Thrones S03E07: Global Heatmap}
\label{Figuregots03e07heatmap}
\end{figure}
\end{landscape}

\begin{figure}
\centering
\includegraphics[width=1\textwidth]{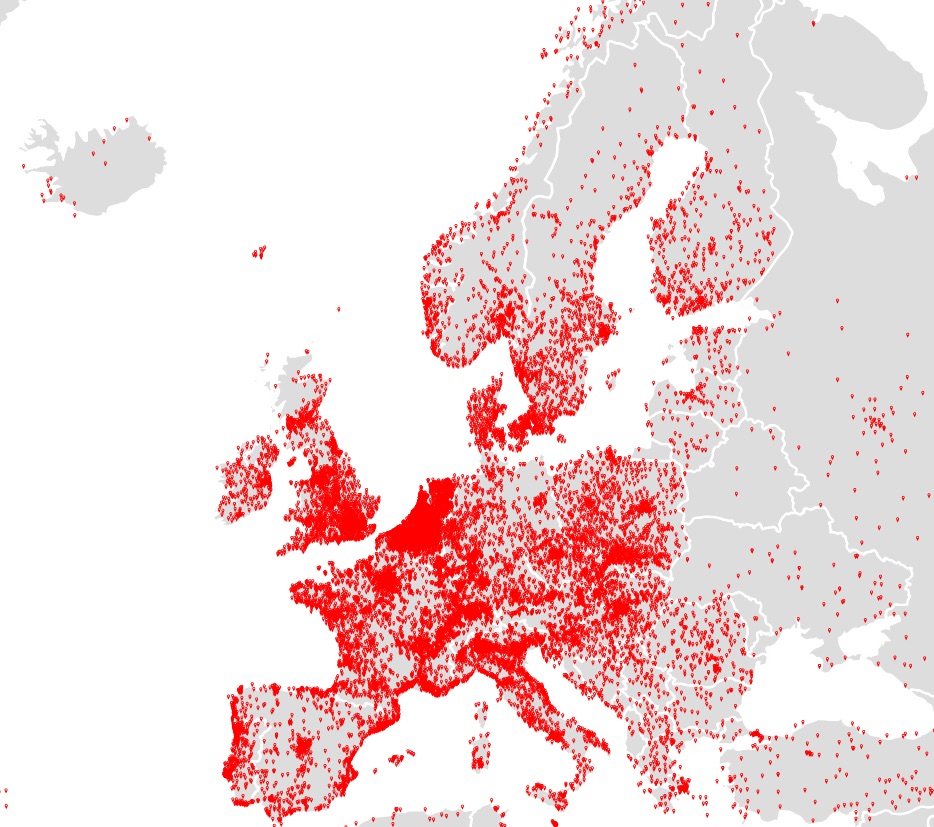}
\caption{Game of Thrones S03E08: Mainland Europe Activity}
\label{Figuregots03e08mainlandeurope}
\end{figure}

\begin{landscape}
\begin{figure}
\centering
\includegraphics[width=1.47\textwidth]{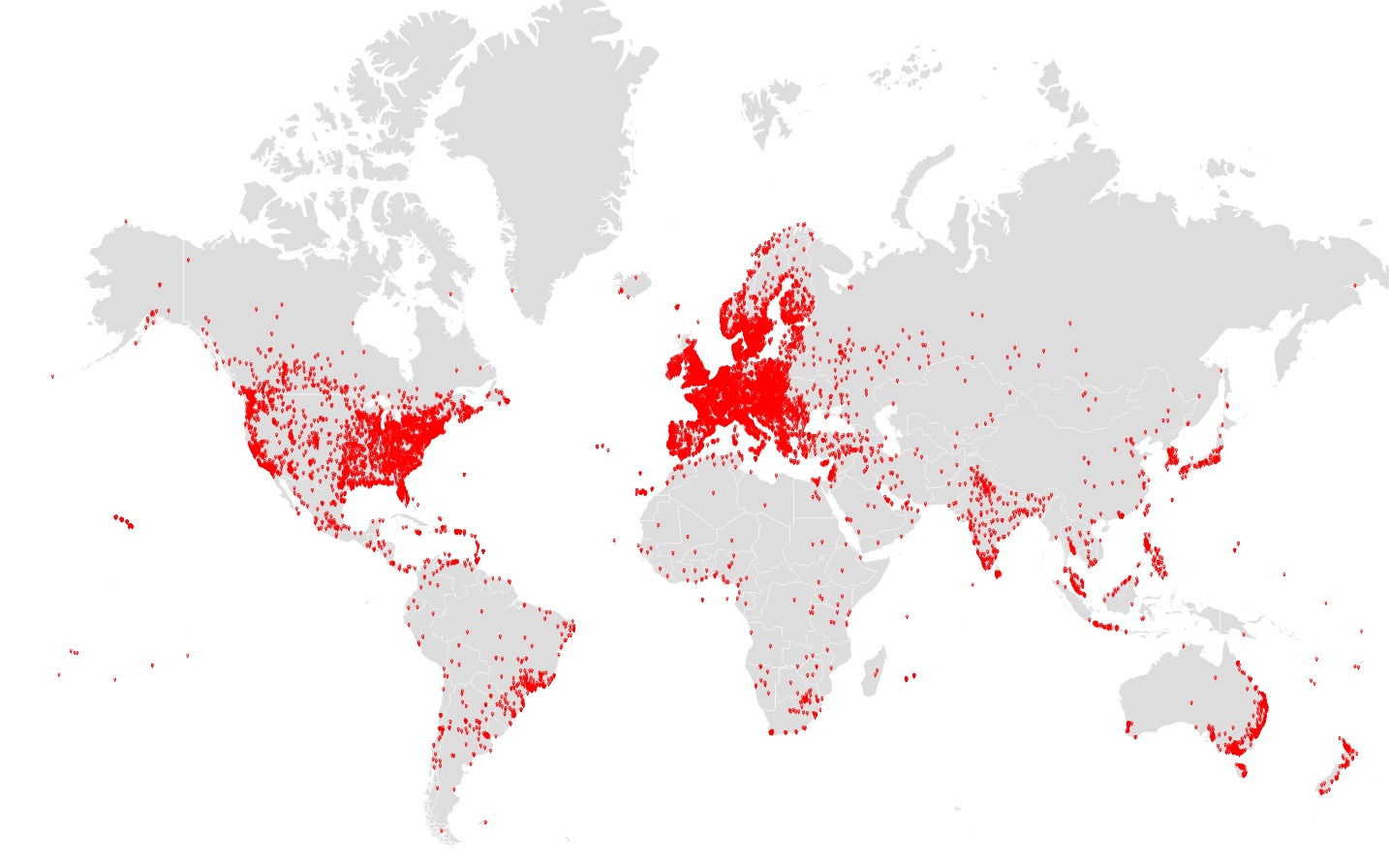}
\caption{Game of Thrones S03E08:  Global City Level Activity}
\label{Figuregots03e08worlddots}
\end{figure}
\end{landscape}

\begin{landscape}
\begin{figure}
\centering
\includegraphics[width=1.45\textwidth]{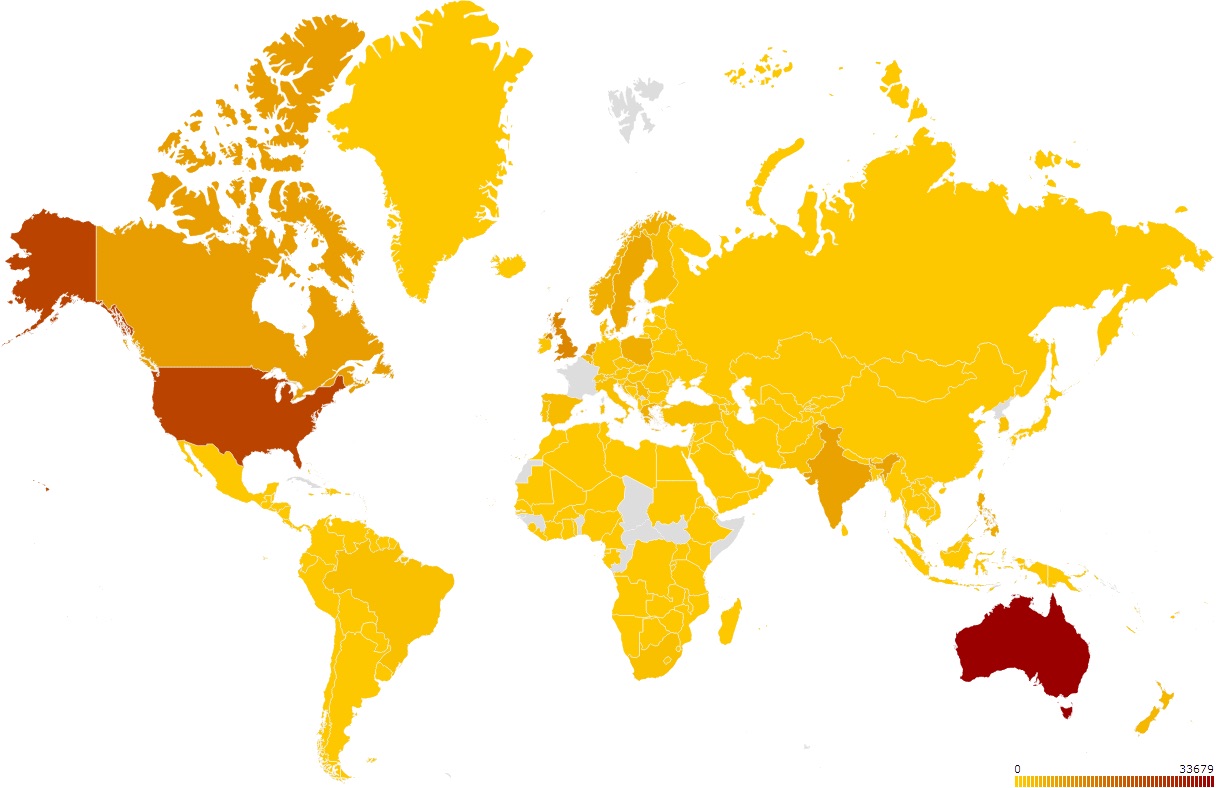}
\caption{Game of Thrones S03E08: Global Heatmap}
\label{Figuregots03e08heatmap}
\end{figure}
\end{landscape}

\begin{figure}
\centering
\includegraphics[width=1\textwidth]{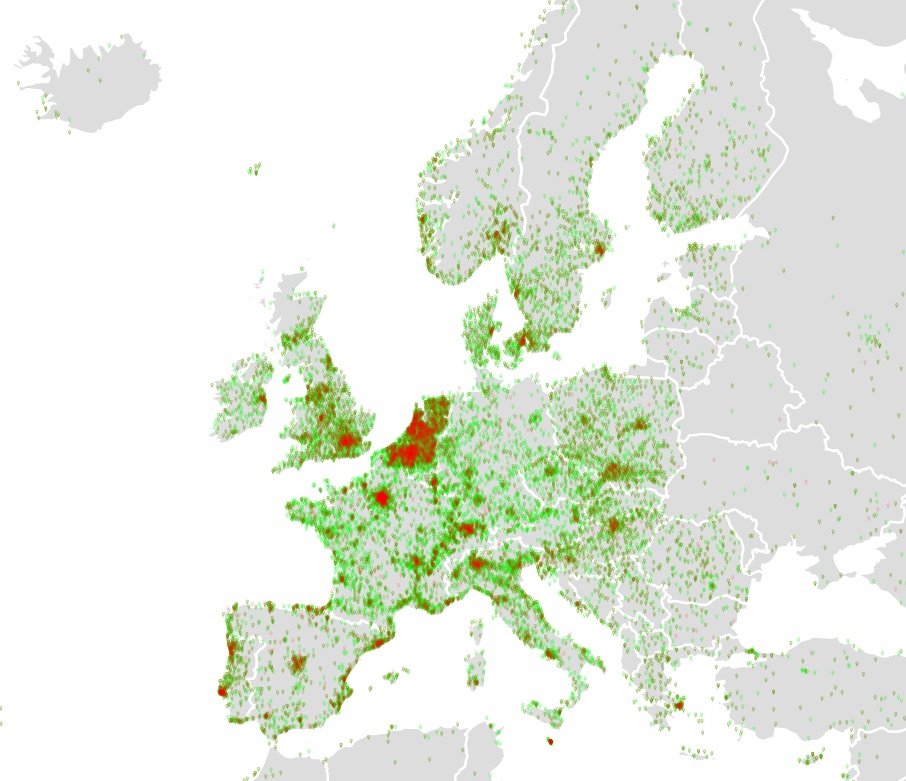}
\caption{Game of Thrones: Collated Results for S03E07 (Red) and S03E08 (Green) in Mainland Europe}
\label{Figuregotbotheurope}
\end{figure}

\begin{landscape}
\begin{figure}[htb]
\centering
\includegraphics[width=1.45\textwidth]{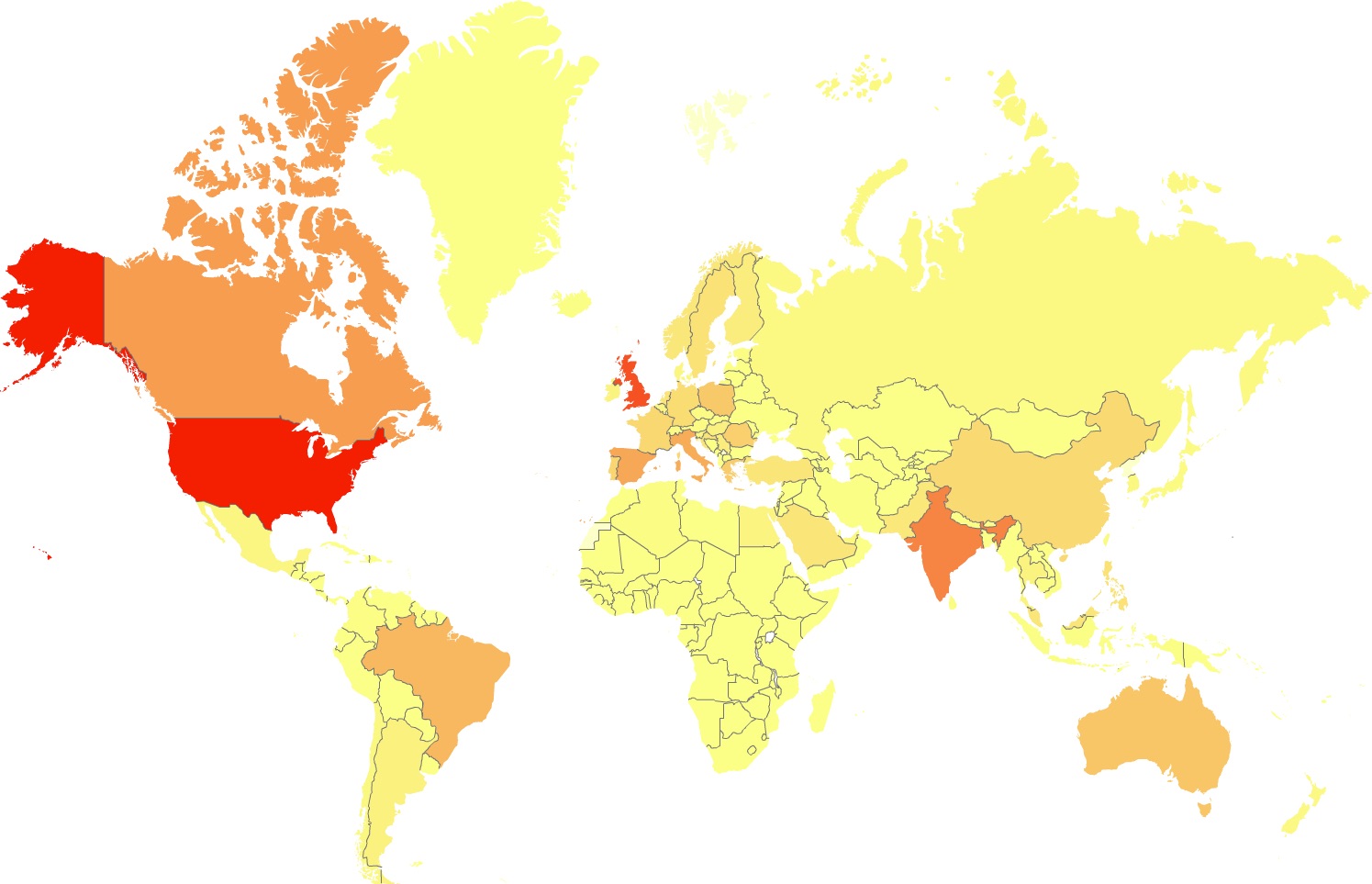}
\caption{Top 100 Swarms: Heatmap showing the worldwide distribution of peers discovered}
\label{Figureheatmap}
\end{figure}
\end{landscape}

\begin{figure}
\centering
\includegraphics[trim = 0mm 5mm 0mm 0mm, clip, width=1\textwidth]{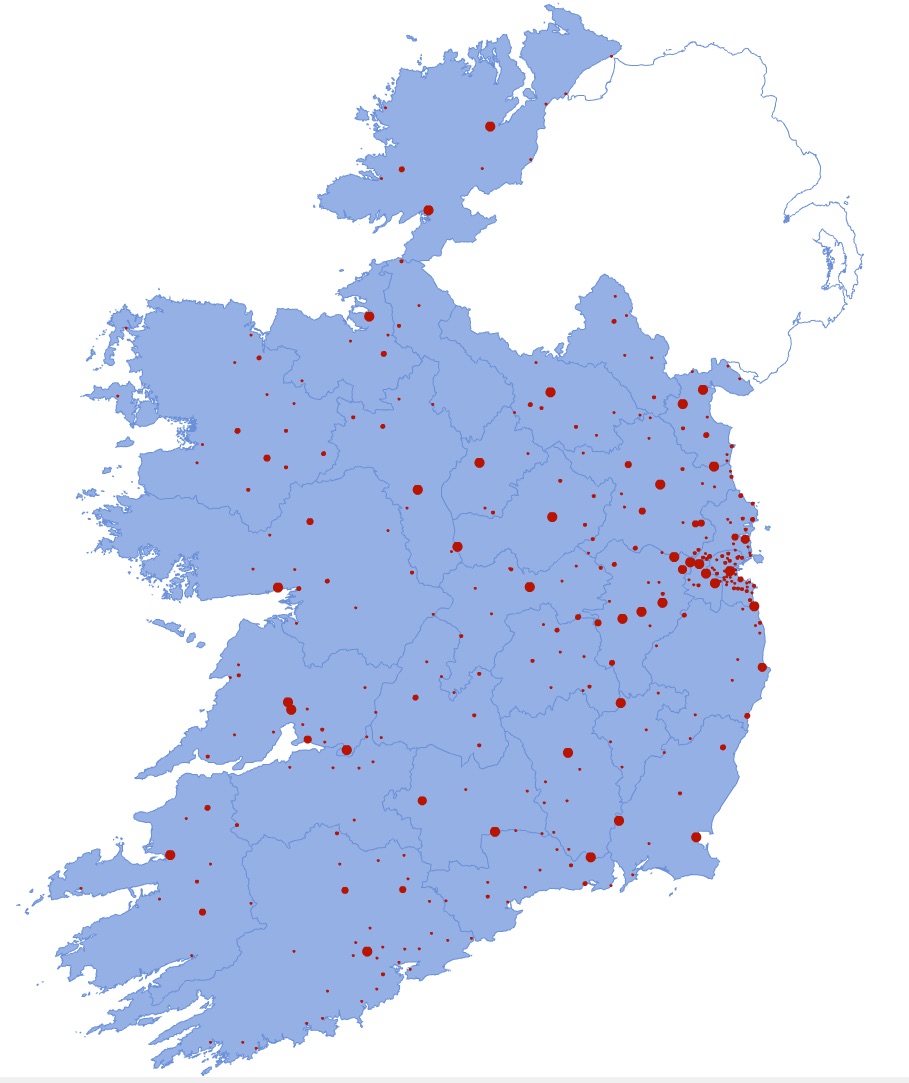}
\caption{Top 100 Swarms: Geolocation of the peers found across Ireland}
\label{Figureireland}
\end{figure}

\begin{figure}
\centering
\includegraphics[width=0.9\textwidth]{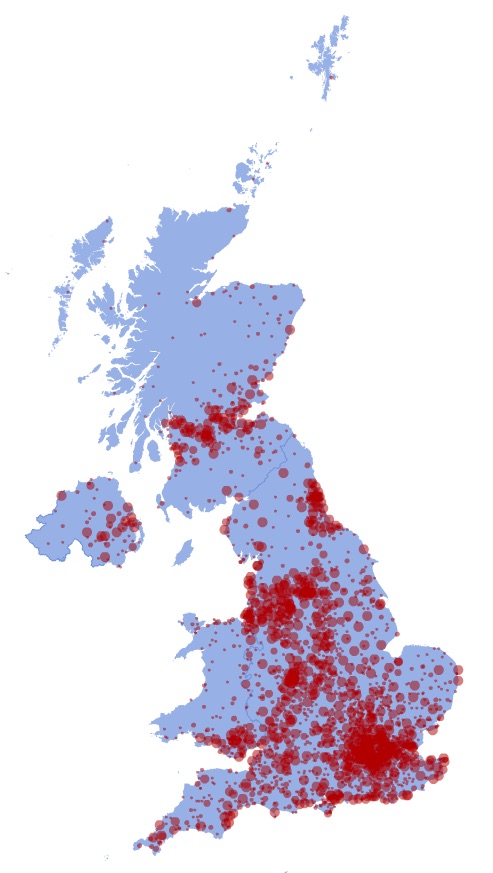}
\caption{Top 100 Swarms: Geolocation of the peers found across the United Kingdom}
\label{Figureuk}
\end{figure}

\begin{landscape}
\begin{figure}
\centering
\includegraphics[width=1.55\textwidth]{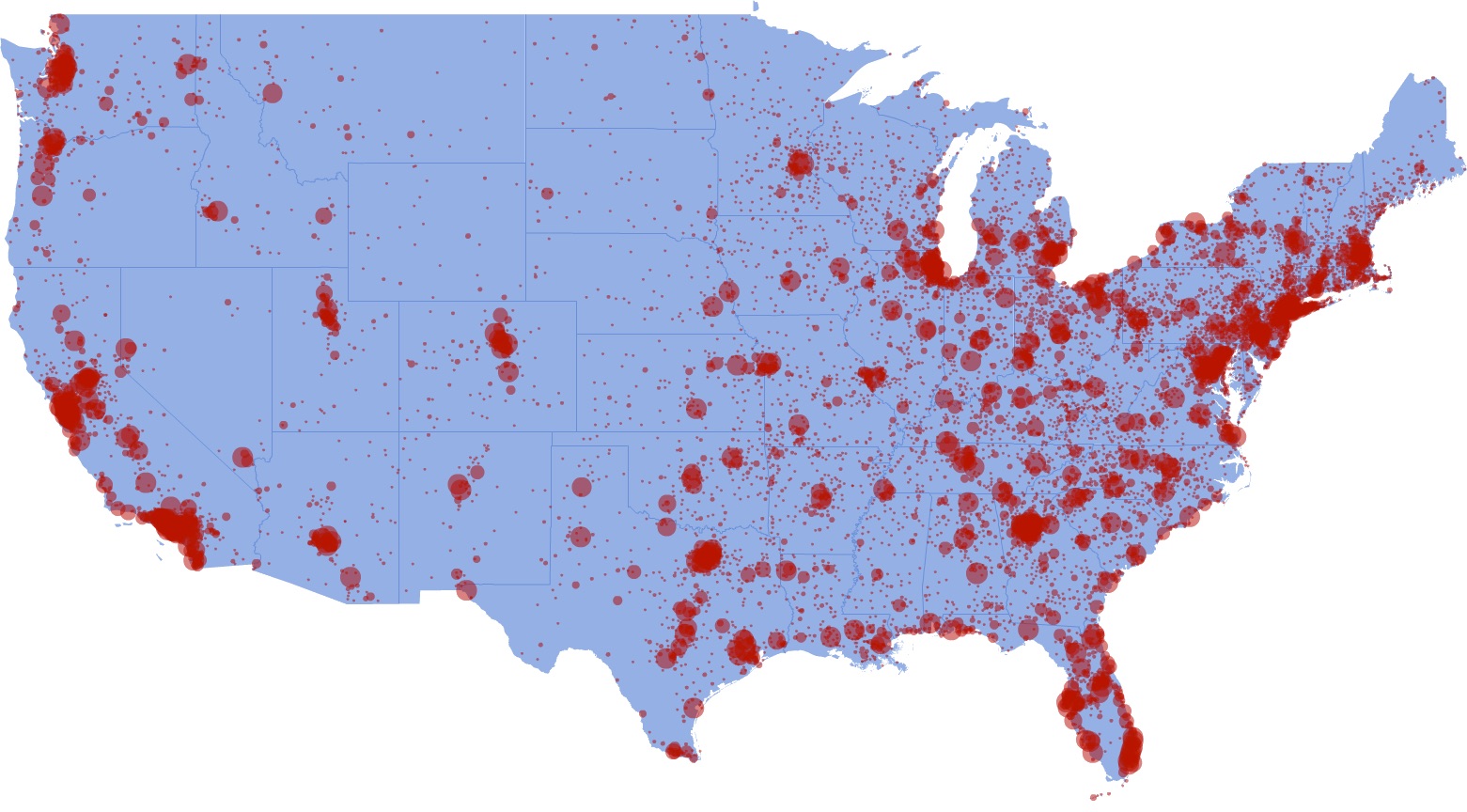}
\caption{Top 100 Swarms: Geolocation of the peers found across mainland USA}
\label{Figuremainlandus}
\end{figure}
\end{landscape}

\singlespacing

\bibliographystyle{unsrt}
\bibliography{bibfile}
\end{document}